\documentclass[prx,twocolumn,aps,epsf,showpacs,superscriptaddress,longbibliography,nofootinbib,notitlepage]{revtex4-2}

\usepackage{amsmath, amssymb, bm, braket, dsfont, amsthm,comment}
\usepackage{graphicx,xfrac}
\usepackage[dvipsnames]{xcolor}
\usepackage[breaklinks,colorlinks,allcolors=blue]{hyperref}
\usepackage{tikz-cd}
\usetikzlibrary{calc}
\tikzset{curve/.style={settings={#1},to path={(\tikztostart)
    .. controls ($(\tikztostart)!\pv{pos}!(\tikztotarget)!\pv{height}!270:(\tikztotarget)$)
    and ($(\tikztostart)!1-\pv{pos}!(\tikztotarget)!\pv{height}!270:(\tikztotarget)$)
    .. (\tikztotarget)\tikztonodes}},
    settings/.code={\tikzset{quiver/.cd,#1}
        \def\pv##1{\pgfkeysvalueof{/tikz/quiver/##1}}},
    quiver/.cd,pos/.initial=0.35,height/.initial=0}

\usepackage{mdframed}
\usepackage{array}
\usepackage[margin=1in]{geometry}
\newcolumntype{M}[1]{>{\centering\arraybackslash}m{#1}}
\renewcommand{\>}{\rangle}
\newcommand{\<}{\langle}
\newcommand{\tr}{\mathrm{Tr}\,}

\newcommand{\vc}{\sf{Vec}}

\newcommand{\rep}{\text{Rep}}
\newcommand{\A}{\mathcal{A}}

\newcommand{\C}{\mathcal{C}}

\newcommand{\E}{\mathcal{E}}
\newcommand{\F}{\mathcal{F}}

\renewcommand{\H}{\mathcal{H}}

\newcommand{\M}{\mathcal{M}}

\newcommand{\T}{\mathcal{T}}
\newcommand{\Z}{\mathcal{Z}}
\newcommand{\JW}{\mathcal{JW}}
\newcommand{\arf}{\text{Arf}}
\usepackage{enumitem}
\usepackage[mathscr]{euscript}
\newcommand\EA  {\EuScript{A}}
\newcommand\EC  {\EuScript{C}}

\newcommand\EF  {\EuScript{F}}

\newcommand\bC   {\mathbb{C}}

\newcommand\bR  {\mathbb{R}}
\newcommand\bZ  {\mathbb{Z}}
\newcommand\bc  {\overline{c}}
\renewcommand{\d}{\partial}

\usepackage[normalem]{ulem}

\definecolor{junglegreen}{rgb}{0.16, 0.67, 0.53}

\newcommand{\drew}[1]{\textcolor{RedOrange}{#1}}

\newcommand{\SymTO}{SymTFT} 
\newcommand{\oned}{$1+1D$}
\newcommand{\twod}{$2+1D$}



\begin{document}
\title{Topological holography for fermions}
\author{Rui Wen} \author{Weicheng Ye} \author{Andrew C. Potter} 
\affiliation{Department of Physics and Astronomy, and Stewart Blusson Quantum Matter Institute, University of British Columbia, Vancouver, BC, Canada V6T 1Z1}

\begin{abstract}
Topological holography is a conjectured correspondence between the symmetry charges and defects of a $D$-dimensional system with the anyons in a $(D+1)$-dimensional topological order: the symmetry topological field theory (SymTFT). Topological holography is conjectured to capture the topological aspects of symmetry in gapped and gapless systems, with different phases corresponding to different gapped boundaries (anyon condensations) of the SymTFT. 
This correspondence was previously considered primarily for bosonic systems, excluding many phases of condensed matter systems involving fermionic electrons.
In this work, we extend the SymTFT framework to establish a topological holography correspondence for fermionic systems. 
We demonstrate that this fermionic SymTFT framework captures the known properties of $1+1D$ fermion gapped phases and critical points, including the classification, edge-modes, and stacking rules of fermionic symmetry-protected topological phases (SPTs), and computation of partition functions of fermionic conformal field theories (CFTs). Beyond merely reproducing known properties, we show that the SymTFT approach can additionally serve as a practical tool for discovering new physics, and use this framework to construct a new example of a fermionic intrinsically gapless SPT phase characterized by an emergent fermionic anomaly.
\end{abstract}
\maketitle

\tableofcontents

\section{Introduction}
Symmetry provides a powerful set of organizing principles for understanding phases of matter. Beginning from the traditional Landau spontaneous symmetry-breaking paradigm, to the modern theory of generalized symmetries that provide a unifying framework for numerous phenomena from symmetry-protected topological orders (SPTs) with non-local (e.g. string) order parameters, to topological orders that can be understood as spontaneously broken higher-form symmetries, to non-perturbative dualities at critical points that are captured by non-group like, non-invertible symmetries~\cite{Gaiotto_2015,Bhardwaj_2018,thorngren2019fusion,luo2023lecture,Gomes_2023,shao2023whats,Kong_2020,Chatterjee_2023,ji2022unified,Chatterjee_2023_2,moradi2022topological,Lin_2023,freed2023topological}. 
A key lesson from this progress is that one must generalize the notion of symmetry beyond global unitary operators with group-like structure. Somewhat unexpectedly, when all such generalized symmetries are taken into account, the structure of a higher-dimensional topological order suddenly appears~\cite{Kong_2020,Chatterjee_2023,ji2022unified,Chatterjee_2023_2,Lichtman_2021,freed2023topological,Lichtman_2021,moradi2022topological,Lin_2023,inamura202321d,bhardwaj2023generalized,bhardwaj2024gapped,bhardwaj2023categorical}. 
For example, as discussed in~\cite{Kong_2020, Chatterjee_2023,inamura202321d}, a \oned{} Ising spin chain has both an ordinary $\bZ_2$ symmetry associated with spin-flip, and also a ``dual symmetry'' associated with the conservation (modulo two) of domain walls. The local action of both symmetries restricted to finite spatial ``patches"  unveils a collection of line operators that have precisely the same structure as the $e$ and $m$ anyons of a \twod{} toric code.

These observations have led to a conjectured \emph{topological holography} between symmetries of a quantum system and a topological field theory that is one-dimensional higher, which is referred to as symmetry topological field theory (\SymTO{}) \cite{Lin_2023,kaidi2023symmetry,kaidi2023symmetry2,bhardwaj2023generalized,bhardwaj2024gapped,bhardwaj2023categorical} or symmetry topological order (SymTO) ~\cite{inamura202321d,Chatterjee_2023,Chatterjee_2023_2}\footnote{In this paper, we primarily focus on \twod{} \SymTO{}, and the \SymTO{} is always described by some \twod{} topological order. Hence we will freely employ the terminology of \twod{} topological order like anyons, anyon condensation, etc.} Accordingly, different phases of a quantum system with the given symmetry correspond to different boundary conditions of the \SymTO{}. For a finite, unitary symmetry $G$ of a bosonic system, the corresponding \SymTO{} is the $G$-gauge theory in \twod{}, which can be described by the quantum double theory $D(G)$ in the categorical language. In this case, the electric and magnetic anyon excitations of the \SymTO{} correspond to charged local operators, and topological defects (domain walls) of the \SymTO{}, respectively.

The big advantage of \SymTO{} is that it disentangles topological aspects of the symmetry (e.g. `t Hooft anomalies, conservation of topological defects, the symmetry quantum numbers of topological defect operators), from dynamical aspects (e.g. the scaling dimensions at a critical point\footnote{Still, there are attempts in the literature to derive the scaling dimensions of \oned{} systems from \SymTO{}. See e.g. \cite{Chatterjee2022}.}). Moreover, it makes certain ``topological manipulations'', including gauging, stacking with invertible phases and duality transformation manifest by simply considering different boundary conditions of the same \SymTO{}. Hence \SymTO{} makes the connections of these theories with different appearances manifest.

In gapped systems, all universal aspects of symmetry are topological, and the topological holography correspondence states that different gapped phases of matter correspond to distinct gapped boundaries of the \SymTO{}, which are in turn characterized by different patterns of anyon condensations.
For gapless systems, the \SymTO{} captures only topological rather than dynamical aspects (e.g. it can constrain the possible symmetry actions on topological defect operators, but does not fully determine the spectrum of scaling operators at a critical point). This raises the intriguing opportunity to establish a non-perturbative, topological (partial) characterization of gapless systems -- which are notoriously difficult to study except in low-dimensions. 

The topological holographic correspondence has been systematically established~\cite{Kong_2020,Chatterjee_2023,Chatterjee_2023_2,ji2022unified,Lin_2023,inamura202321d,bhardwaj2024gapped,bhardwaj2023categorical,bhardwaj2023generalized}, in \oned{} bosonic systems, with finite, internal (non-spacetime) symmetries, including both ordinary group-like symmetries and non-invertible symmetries described by a general fusion category. In the group-like symmetry case, it has been shown recently that the SymTO can be derived from analyzing the algebra of patch operators associated with the symmetry and the dual symmetry~\cite{inamura202321d}.
The correspondence can also be pictorially understood via a dimensional reduction procedure called the ``sandwich construction" (see Fig.~\ref{fig:setup1}), in which the partition function for a $G$-symmetric \oned{} quantum system is written in terms of that of a thin- slab of \twod{} \SymTO{} sandwiched between two boundaries: 1) a ``trivial" reference boundary in which all the gauge-electric charges are condensed (corresponding to Dirichlet boundary conditions in the field theory language), and 2) a ``physical" boundary that determines the topological properties of the \oned{} quantum system.
Different boundary conditions of the \SymTO{} correspond to different patterns of anyon condensations at the boundary~\cite{Chatterjee_2023,Chatterjee_2023_2,kong2014anyon,Kong_2017}. Gapped systems correspond to patterns of anyon condensation on the physical boundary that fully confine the \SymTO{} (such that all particles are either condensed or confined in that boundary condensate). 
Anyons of the \SymTO{} together define a categorical version of symmetry of the \oned{} quantum system~\cite{Kong_2020,ji2022unified,Chatterjee_2023,Chatterjee_2023_2}.
These symmetries include both ordinary on-site unitary symmetries, but can also capture generalized symmetries such as string-orders of symmetry-protected topological phases (SPTs), and non-invertible symmetries~\cite{Bhardwaj_2018,thorngren2019fusion,kaidi2023symmetry,kaidi2023symmetry2,zhang2023anomalies,Lin_2023,bhardwaj2023generalized,bhardwaj2024gapped,bhardwaj2023categorical,cordova2024particlesoliton,cordova2023anomalies,antinucci2023anomalies} such as the Kramers-Wannier duality of the \oned{} Ising CFT (for which the corresponding \SymTO{} is the double Ising topological order). 

Recent work has made progress in extending these ideas to gapless systems such as gapless SPTs~\cite{wen2023classification,huang2023topological,bhardwaj2023club,bhardwaj2024hasse}, continuous symmetries and Goldstone phases~\cite{brennan2024symtft,apruzzi2024symth}, and has made forays into higher numbers of spatial dimensions~\cite{Kong_2020,cordova2023anomalies,antinucci2023anomalies}.
These advances have, so far, been developed mainly for \emph{bosonic} systems, that is, systems where all local operators are bosonic, except for some general discussions in a few pioneering works \cite{Gaiotto2020,freed2023topological} and very recently in \cite{smith2024backfiring,KantaroLecture}. This has limited the relevance of the framework to condensed matter problems involving electronic degrees of freedom. 

This work aims to fill this gap, and develop a topological holographic correspondence and \SymTO{} formalism for \emph{fermionic} systems enriched by symmetry. 
We focus on \oned{} fermionic systems with a (possibly anomalous)\footnote{In this work, we do not consider systems with pure gravitational anomaly, and systems with nonzero chiral central charge.} symmetry $G^F$ , which is a finite unitary symmetry that contains the $\bZ^F_2$ fermion parity conservation as a subgroup.
The central structures of our construction include:
\begin{itemize}
    \item For a non-anomalous fermionic symmetry $G^F$, the \twod{} \SymTO{} is given by the $G^F$ gauge theory.
    \item The \SymTO{} for a generic \oned{} fermionic symmetry is given by a ``gauged fermionic SPT" in \twod{}.
    \item The reference boundary of the \SymTO{} is defined by condensing a set of anyons of the \SymTO{} that have the same fusion and braiding structure as local symmetry charges of $G^F$. This involves ``condensing" a set of fermionic anyons, $f_R$, which is done by introducing local fermions, $c$, on the reference boundary, and condensing the bosonic composite $cf_R$. The reference boundary is therefore fermionic due to these local fermions $c$.
    \item The dependence on spin structure (or $G^F$-spin structure in the general case) of a \oned{} fermionic system is manifested in the \SymTO{} by the spin structure dependence of the fermionic reference boundary.
    \item Bosonization of the \oned{} system is manifested in the \SymTO{} simply by changing the reference boundary from ``fermionic'' to ``bosonic''.
\end{itemize}
We explain how this ``fermion condensed" boundary condition \cite{Aasen_2019,Lou_2021} can reproduce universal properties of various known \oned{} phases with- and without- additional symmetries, 
such as the Kitaev chain, the Majorana CFT, and arbitrary fermionic SPTs. 
These fermionic phases have many distinctive features absent in bosonic phases. For instance, the edge modes of the Kitaev chain are Majorana defects with radical quantum dimension $\sqrt{2}$, the fermion parity is an unbreakable symmetry, and the partition function of a fermionic system depends on the spin structure of the manifold. We find that all these features can find their places in a properly formulated \SymTO{}. We show that one can obtain the full classification of \oned{} fermionic SPTs from \SymTO{}. 
This includes not only the set of SPT phases, but also the stacking rules of these phases. We also show that bosonization/Jordan-Wigner transformation has a simple realization in the \SymTO{}: it amounts to a change of reference boundary condition of the \SymTO{}. We recover many bosonization-related results by simple \SymTO{} considerations.  

Beyond providing a fresh perspective on familiar phases, the fermionic \SymTO{} framework provides a convenient tool to theoretically discover new phases of matter. As an example, we use the \SymTO{} framework to construct an example of an intrinsically-fermionic and intrinsically-gapless SPT (igSPT) state, which exhibits an emergent anomalous fermionic symmetry (a discrete version of the \oned{ }chiral anomaly) that protects a gapless Luttinger liquid bulk and fractionally charged fermionic edge states. This fermionic igSPT is both intrinsically-fermionic, i.e., its topological properties cannot arise in a bosonic system, and intrinsically-gapless, i.e., the fractionally charged edge modes are not equivalent to those of any gapped fermion SPT state.
We show that the topological aspects of this fermion igSPT can be readily deduced from the \SymTO{} construction, which also directly facilitates a simple bosonized field-theory description of the fermion igSPT and its edge states.

This work is organized as follows. In Section~\ref{sec:bg} we review the theory of \SymTO{} for bosonic symmetries. Next we construct and analyze the \SymTO{} for the simplest fermionic symmetry--the fermion parity symmetry $\bZ_2^F$ in Section~\ref{sec:Z2FSymTO}. We discuss the Kitaev chain and the Majorana CFT within the framework of \SymTO{}. We also discuss how spin structure dependence is realized in \SymTO{}.  We construct and study an example of fermionic igSPT in Section~\ref{sec:gSPT} and reveal fascinating properties of the igSPT via \SymTO{} methods. In Section~\ref{sec: FSymTO general} we provide the general construction for a fermionic symmetry with no anomaly, analyzing the structure of the relevant fermionic condensation and the dependence on twisted spin structures. We show in Section~\ref{sec: SPT} how partition function of \oned{} fermionic SPTs may be derived from \SymTO{} methods. In Section~\ref{sec: stacking} we address the issue of stacking rules in \SymTO{}. We define the stacking of \SymTO{}s via a product structure of condensable algebras and show with examples that it matches with the stacking of the corresponding phases. 
In Section~\ref{sec: bz}, we explain how the standard procedure of bosonization/gauging fermion parity (or the Jordan-Wigner transformation for \oned{} fermion chains) manifests in the \SymTO{} construction, and use this to identify non-invertible symmetry transformations (dualities)  in Section~\ref{sec: bza}.


\section{Background: Bosonic \SymTO{}\label{sec:bg}}
We begin by briefly reviewing the \SymTO{} description of a \oned{} bosonic system $\sf{T}$ with a finite, unitary, internal (non-spacetime), group-like symmetry $G$, for which the corresponding \SymTO{} is simply a \twod{} $G$-gauge theory. For anomalous symmetry charaterized by cocycle $\omega\in H^3[G,U(1)]$, the \SymTO{} is a twisted (Dijkgraaf-Witten) gauge theory twisted by the same cocycle $\omega$, or the so-called gauged SPT in the condensed matter language, and corresponds to the quantum double topological order $D(G,\omega)$.\footnote{We focus mostly on group-like symmetries in this work, 
except in ~\ref{sec: bza} where we also use the \SymTO{} for non-invertible fusion categorical symmetry to discuss bosonization of anomalous fermionic symmetry. For a general fusion categorical symmetry $\EA$, the \SymTO{} is given by the Drinfeld center $\sf{Z}[\EA]$.  In the group-like symmetry case, the symmetry fusion category is $\vc_G^\omega$, and we recover the quantum double \SymTO{} due to $\sf{Z}[\vc_G^\omega]\simeq D(G,\omega)$.} The original \oned{} bosonic system $\sf{T}$ can be reconstructed from the \twod{} \SymTO{} according to the ``slab'' or ``sandwich'' construction illustrated in  Fig.~\ref{fig:setup1}: 
\begin{enumerate}
\item The \SymTO{} lives on the manifold $M\times I$, where $M$ is a \twod{} (closed, orientable) manifold and $I$ is the interval $[0,1]$.
\item The upper boundary $M\times \{0\}$ is a fixed \emph{reference boundary} $\sf{B_{ref}}$ that encodes \emph{topological aspects} of the symmetry.
\item The lower boundary $M\times \{1\}$ is a \emph{physical boundary} $\sf{B_{phys}}$ that can be varied to describe different possible $G$-symmetric phases. 
\item The original \oned{} bosonic system on $M$ is obtained by ``dimensional reduction'' of the interval $[0,1]$.
\end{enumerate}

\subsection{Reference boundary}

\begin{figure*}[t]
    \includegraphics[width=1.5\columnwidth]{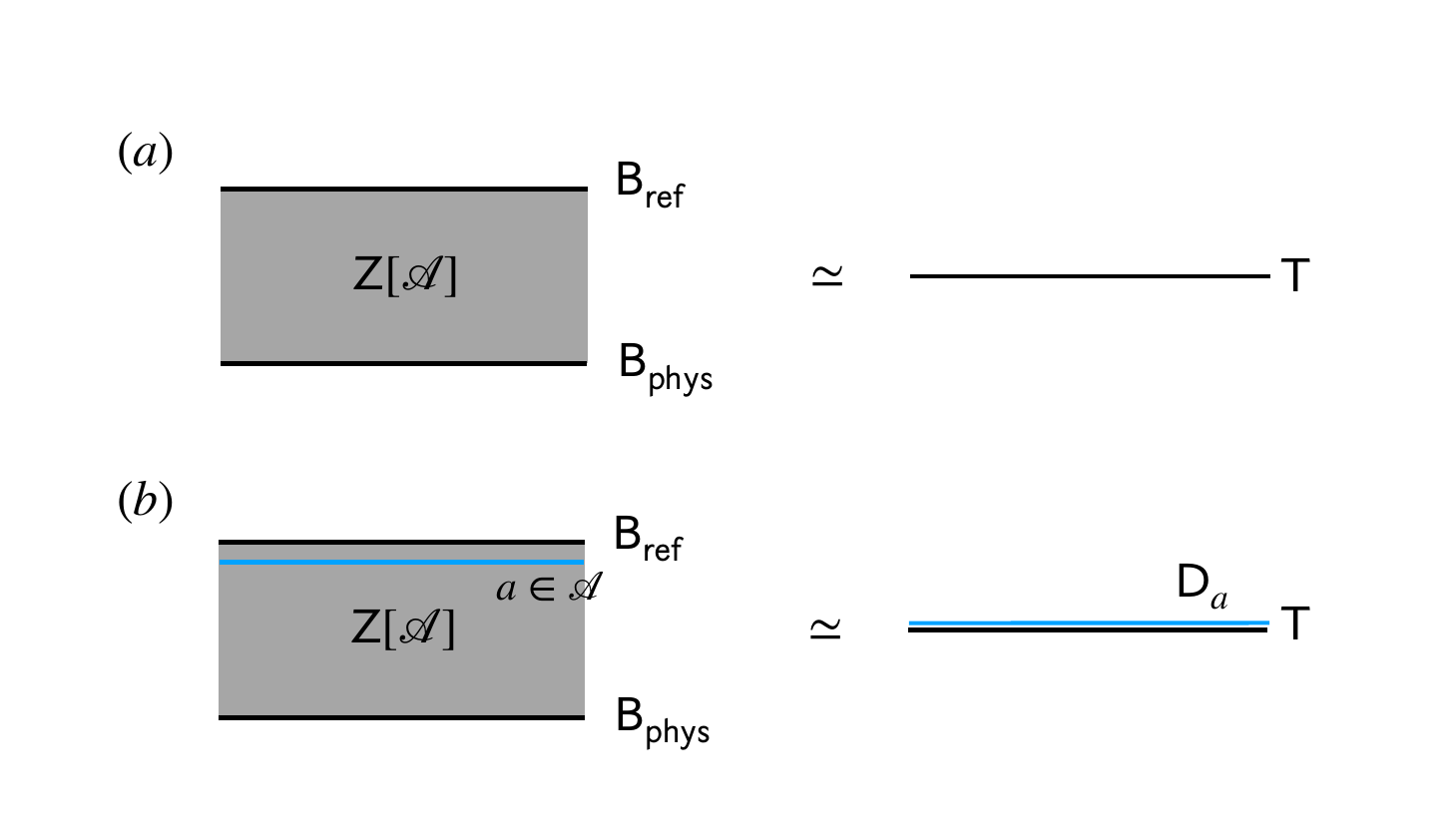}
    \vspace{-0.2in}
    \caption{\textbf{The \SymTO{} setup.} (a). The \SymTO{} for a general symmetry category $\A$ is the Drinfeld center $\sf{Z}[\A]$. The sandwich (left) reduces to the original system $\sf{T}$ (right) when viewed as an effective \oned{} system. (b). A non-trivial topological defect line near the reference boundary becomes a symmetry defect operator $\sf{D}_a$ of $\sf{T}$ after the dimensional reduction.}
    \label{fig:setup1}
\end{figure*}

For group-like symmetry $G$, the reference boundary is one where all of the gauge charges of the $G$ gauge theory are condensed (this corresponds to Dirichlet boundary conditions in the field theory language). This is substantiated by the following two observations. First, one defining property of the reference boundary is that the non-trivial anyon lines that cannot be absorbed on the reference boundary correspond to the symmetry defects of the original \oned{} system $\sf{T}$, as illustrated in Fig.~\ref{fig:setup1}. Indeed, by choosing to condense all gauge charges on $\sf{B_{phys}}$, the remaining non-trivial anyon lines that cannot simply be absorbed into $\sf{B_{ref}}$ are the flux lines. These flux lines exactly form the symmetry category $\vc_G^\omega$. Namely, the fluxes are labeled by elements in $G$ with the fusion rules from the group structure of $G$, and the $F$-symbol of the flux lines matches with the anomaly $\omega$. 

\begin{figure*}[t]
    \includegraphics[width=1.5\columnwidth]{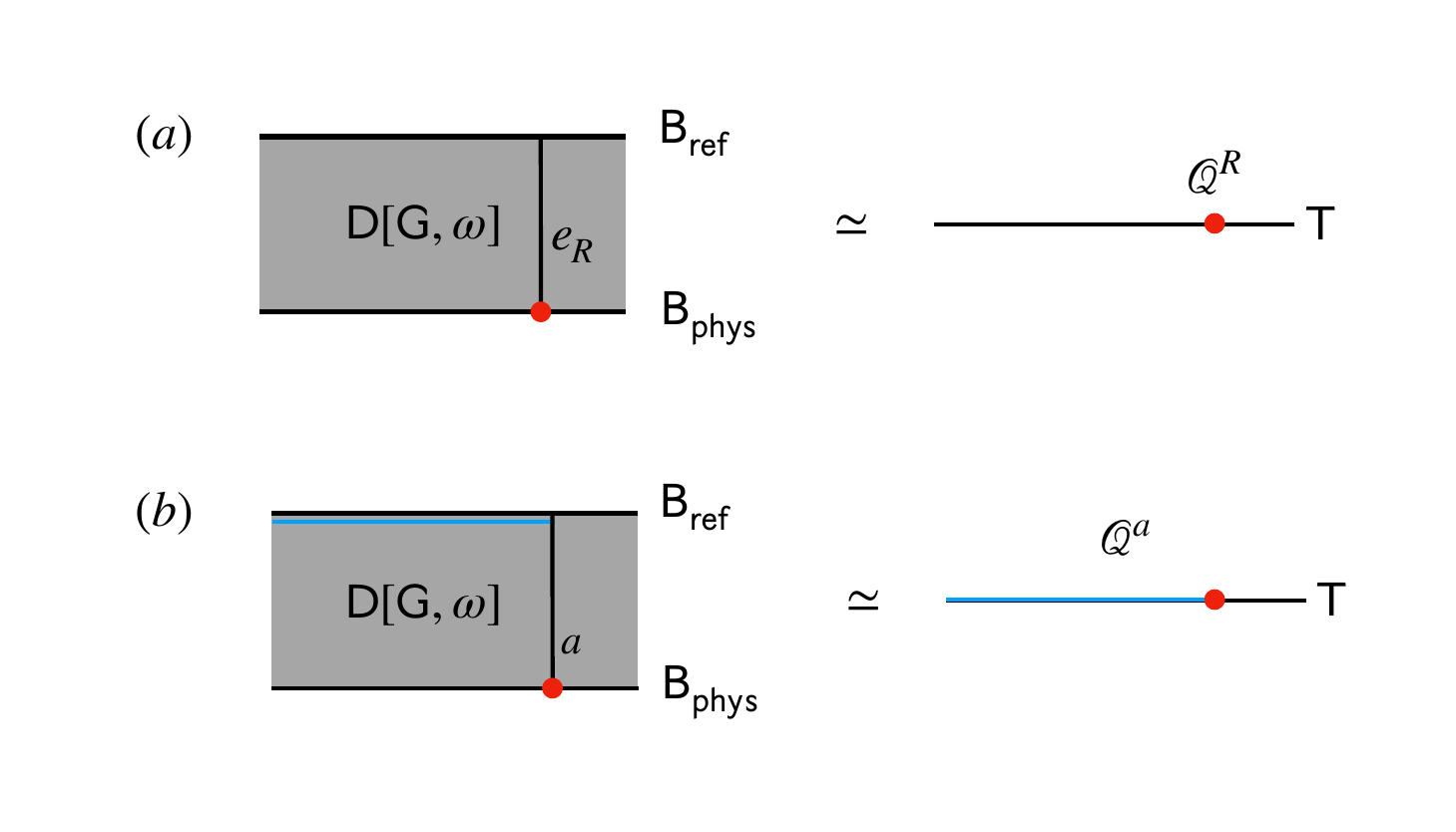}
    \vspace{-0.2in}
    \caption{\textbf{Local and non-local charges in \SymTO{}.} (a). A gauge charge line in the \SymTO{} can be absorbed by the reference boundary. This corresponds to a local symmetry charge of $\sf{T}$ after dimensional reduction. The red dot represents an excitation created by the anyon line at $\sf{B_{phys}}$. (b). A general anyon of the \SymTO{} is a dyon, the associated defect line can not be fully absorbed by $\sf{B_{ref}}$, instead it becomes a flux line on $\sf{B_{ref}}$(blue). After dimensional reduction this becomes a charge that lives on the end of a symmetry defect, i.e. a non-local charge or a string order parameter. }
    \label{fig:setup2}
\end{figure*} 
Secondly, the anyons condensed on the reference boundary correspond to local charges of the \oned{} system. Concretely, a short anyon string that is absorbed by the reference boundary corresponds to a local charged operator of the original system in the sandwich construction, also illustrated in Fig.~\ref{fig:setup2}. Therefore, anyons condensed on the reference boundary should have the same structure (fusion rule, $F$-symbols) as that of local charges of the symmetry. This is another defining property of the reference boundary. Indeed, the gauge charges of the twisted gauge theory $D(G,\omega)$ have the same fusion structure as local charges of the symmetry $G$, namely the fusion category of representations of $G$, $\rep(G)$. 

Besides fluxes and charges, there are dyons in the \SymTO{}. When a dyon line reaches the reference boundary, it cannot be fully absorbed but becomes a flux line living on the reference boundary. This configuration corresponds to a charge of $\sf{T}$ that lives at the end of a symmetry defect. These string-like charges are known as non-local charges of the symmetry. If the dyon is bosonic, its anyon string also corresponds to string order parameters of the symmetry $G$, whose vacuum expectation value differentiates different gapped or gapless SPTs. See Fig.~\ref{fig:setup2}. Therefore the \SymTO{} provides a unified description for all topological aspects of the symmetry, namely the anyons of the \SymTO{} correspond to the symmetry defects and symmetry charges, and the braiding of anyons encodes the algebraic relations among these symmetry defects and charges.

From the dimension-reduction construction, the \SymTO{} sandwich computes the partition function of the physical \oned{} system as an inner product between the state defined by the physical boundary and the state defined by the reference boundary.  A generic state of the $G$-gauge theory can be written in the field configuration basis as 
\begin{align}
    |\Psi\>=\sum_{A\in H^1[M,G]} \Psi_{A}|A\>,
\end{align}
where $A$ is the gauge field on $M$.\footnote{For finite, potentially non-abelian group $G$, the notation of non-abelian cohomology $H^1[M,G]$ (which is just a set) simply denotes the set of all gauge bundles on $M$. A gauge bundle and its associated gauge field $A$ on $M$ are completely determined by the holonomy on the noncontractible cycles, up to conjugation action on the holonomoy, i.e., a map $f\colon\pi_1(M)\rightarrow G$ with $f$ identified with $f'$ if their images are related by conjugation with an element $g\in G$. When $G$ is an abelian group, the notation coincides with the usual cohomology of $M$ with coefficient $G$, which is also denoted as $H^1[M,G]$.}
The all-charge condensed reference boundary corresponds to the Dirichlet boundary 
\begin{align}
    |\sf{Ref},A_0\>=\sum_{A\in H^1[M,G]} \delta_{A,A_0}|A\>,
\end{align}
that fixes the value of the gauge field to $A_0$ on the reference boundary. $A_0$ also sets the expectation value of the Wilson loops of condensed gauge charges on different non-contractible cycles.
If the physical boundary is in the state 
\begin{align}
    |\sf{Phys}\>=\sum_{A\in H^1[M,G]}\Z_{\sf{T}}[A]|A\>,
\end{align}
then the sandwich has partition function 
\begin{align}
    \<\sf{Ref},A_0|\sf{Phys}\>=\Z_{\sf{T}}[A_0],
\end{align}
which is the partition of a \oned{} system $\sf{T}$ coupled to a background $G$-gauge field $A_0$. 

The bulk $G$-gauge theory also has a standard Neumann boundary condition when the twist $\omega$ is trivial. This is a topological boundary described by the state 
\begin{align}
    |N, \widehat{A}_0\>:=\sum_{A\in H^1[M,G]} e^{i \int_{M} \widehat{A}_0\cup A}|A\>, ~\widehat{A}_0\in H^1[M,\widehat{G}]
\end{align}
where $\widehat{G}:=\hom(G,U(1))$ is the Pontryagin dual of $G$. In the condensed matter language, this boundary is the all-flux condensed boundary, and $\widehat{A}_0$ sets the expectation values of the 't Hooft loops of the condensed fluxes on different non-contractible cycles.

Gauging a global symmetry $G$ in \oned{} corresponds in the \SymTO{} to changing the reference boundary from Dirichlet to Neumann. It is direct to verify that taking the inner product with the Neumann boundary state gives the partition function of the gauged theory $\sf{T}/G$. Recall that the reference boundary defines the symmetry represented by the \SymTO{}. Formally, the Dirichlet boundary defines the symmetry of the effective \oned{} system to be $\text{Vec}_G$, while the Neumann boundary defines the symmetry to be $\rep(G)$. The fact that these two symmetries are described by the same \SymTO{} with different boundary conditions is the direct consequence of the fact that there is a topological manipulation relating the two symmetries. In this case the topological manipulation is gauging $G$. The \SymTO{} provides a general framework for analyzing such topological manipulation. Namely any topological manipulation, such as gauging subgroups and stacking with invertible phases, does not change the \SymTO{}, and corresponds to only changing the reference boundary in the sandwich construction.

This shows the advantage (and limitation) of the \SymTO{} -- it isolates \emph{topological} aspects of the symmetry from dynamical properties (which are potentially non-topological in gapless systems). For gapped symmetric states, all universal long-wavelength aspects of symmetries are topological and the \SymTO{} fully captures the universal properties of the physical system. For gapless states, such as conformal field theories (CFTs), the \SymTO{} does not fully constrain the operator content and scaling dimensions of the physical system, but does capture topological consistency conditions, and may enable a characterization or classification of symmetry-enriched gapless phases of matter~\cite{Kong_2018,Kong_2020_2,Kong_2021_2,Chatterjee_2023_2,wen2023classification,huang2023topological,bhardwaj2023club,bhardwaj2024hasse}.

A major result of this work is extending this advantage of \SymTO{} to fermionic symmetries. In Section~\ref{sec: bz}, we show that bosonization, which is also a topological manipulation but maps fermionic symmetries into bosonic symmetries, also corresponds to a change of reference boundary of the \SymTO{}.

\subsection{Physical boundaries and anyon condensation}
In the \SymTO{} framework different $\EA$-phases are represented by different boundary conditions on the physical boundary, and it is known that boundaries of \twod{} topological orders have intimate relation with anyon condensation.~\cite{Kong_2017,kong2014anyon,Kong_2018,Kong_2020_2,Kong_2021_2,Chatterjee_2023_2} Namely, for every boundary condition\footnote{More precisely we assume the boundary is either gapped or CFT-like. } of a \twod{} topological order $\EC$, there is an associated anyon condensation. An anyon is condensed on a boundary if it can be moved from the bulk to this boundary and absorbed. Or equivalently if it can be dragged out of the boundary without costing any energy. Anyons that are condensed simultaneously need to satisfy a set of consistency conditions. First, all condensed anyons must be bosons. Moreover, 
if $a$ and $b$ are both condensed, then at least one fusion outcome of $a\times b$ is necessarily also condensed, 
These conditions make the condensate have the structure of a separable, connected, commutative algebra. See reference~\cite{kong2014anyon} for a formal definition. Due to the physical origin of this structure, we will call it a condensable algebra instead. For abelian topological orders, a condensable algebra is equivalent to a subgroup of the fusion group, such that all elements of this subgroup have trivial self and mutual statistics.

Anyon condensation in a bulk \twod{} topological order occurs as a phase transition that results in a new phase. For Abelian topological orders, the new topological order can be viewed as formed by the remaining deconfined anyons, i.e. anyons that braid trivially with the condensate.
In a non-Abelian theory, an anyon of the initial phase may become partially confined/deconfined, and it is less straightforward to identify the new topological order resulting from condensation. In general a composite anyon of the original phase becomes a simple anyon of the new phase. The composite anyons needs to be invariant under fusing with the condensate, since the condensate is the vacuum of the new phase. This gives the anyons in the new phase the structure of a module over the condensable algebra $\A$. That is to say, an excitation of the new phase (possibly confined) is a non-simple anyon of the old phase$\M=\oplus_i c_i a_i$, and there is a product $\nu_M: \A \otimes \M\to \M$. The product needs to satisfy some natural consistency conditions~\cite{kong2014anyon}. Deconfined excitations are the modules such that braiding with the condensate commute the product, these are formally called local modules.

A useful formula regarding anyon condensation is the dimension formula. If we denote the topological order obtained by condensing $\A$ in $\EC$ by $\EC/\A$, then
\begin{align}
    \dim(\EC/\A)=\dim(\EC)/\dim(\A)
\end{align}
where $\dim(\EC)=\sqrt{\sum_{a\in \EC} d_a^2}$ is the total quantum dimension of the topological order, and if $\A=\sum c_a a$, then $\dim(\A)=\sum_a c_a d_a$.
If a condensation results in a trivial topological order, the condensable algebra is called Lagrangian.  We see from the dimension formula that a condensable algebra is Lagrangian if and only if its dimension equals the total quantum dimension of the topological order: $\dim(\A)=\dim(\EC)$.

The relation between anyon condensation and boundary condition of a topological order is as follows. A gapped boundary has a Lagrangian condensation while a gapless boundary necessarily has a non-Lagrangian condensation. In the \SymTO{} setup the reference boundary is always gapped, and different $\EA$-phases are realized by choosing different boundary conditions on the physical boundary. Therefore  gapped $\EA$-phases correspond to Lagrangian condensations of $\sf{Z}[\EA]$, while gapless $\EA$-phases corrspond to non-Lagrangian condensations of $\sf{Z}[\EA]$. In the gapped boundary case the boundary is fully determined by the condensation, in the sense that all excitations on the boundary can be calculated from the Lagrangian algebra. Physically they are  the anyons confined on the boundary. When the boundary is gapless, the condensation on the boundary determines only the gapped excitations, and the gapless excitations on the boundary can not be fully determined by the condensation alone. However it is known that if the gapless boundary is described by some CFT, the \SymTO{}constrains the partition function of the CFT via its modular transformation properties~\cite{Chatterjee_2023_2}.

In the bosonic \SymTO{}, the reference boundary condenses all symmetry charges, the corresponding algebra is called the electric algebra\cite{zhang2023anomalies}. The electric algebra in $D(G,\omega)$ is
\begin{align}
    \A_{\sf{ref}}^b:=\oplus_{R\in \rep(G)}d_R e_R.\label{eq:bref}
\end{align}
Here we use $e_R$ to denote a gauge charge carrying the representation $R$, and $d_R$ is the dimension of the representation. The subscript $b$ stands for bosonic, stressing that the reference boundary is bosonic, as oppose to the fermionic reference boundary we will introduce latter. $\A_{\sf{ref}}^b$ has a compact description as the group algebra $\bC[G]$. Here $\bC[G]$ is the linear space of complex functions on $G$, and the product of the algebra is given by the product of functions (point-wise multiplication). A natural basis for $\bC[G]$ is the delta functions $|g\>, g\in G$. $\bC[G]$ carries a natural $G$-representation $g\cdot |h\>:=|gh\>$, which constitutes a highly reducible representation. When decomposed into irreducible representations (irreps), $\bC[G]\cong \oplus_R d_R R$. The quantum dimension of $\A_{\sf{ref}}^b$ is $\dim(\A_{\sf{ref}}^b)=\sum_R d_R^2=|G|=\dim(D(G,\omega))$, therefore $\A_{\sf{ref}}^b$ is Lagrangian.

If a gauge charge is also condensed on $\sf{B_{phys}}$, then one can form a Wilson line of this charge that connects the two boundaries without creating any excitation. Since vertical charge lines in the \SymTO{} correspond to local symmetry charge of the represented \oned{} system, this means condensing gauge charges on the reference boundary results in a non-zero order parameter for the represented \oned{} system, signaling spontaneous breaking of the symmetry. Therefore to describe symmetric $\EA$-phases, the physical boundary must not condense any gauge charges. Such a condensation/ condensable algebra is called magnetic~\cite{zhang2023anomalies}.

\subsection{Condensable algebras and gapless SPTs\label{sec:algebra} }
In the \SymTO{} sandwich construction different $G$-phases correspond to different anyon condensations on the physical boundary. Gapped and gapless SPTs~\cite{Scaffidi_2017,Verresen_2021,Thorngren_2021,li2023decorated,Wen_2023,li2023intrinsicallypurely,yu2024quantum,PhysRevB.108.245135,su2024gapless,yu2024universal,zhang2024quantum,zhong2024topological,PhysRevB.108.245135, PhysRevLett.129.210601,florescalderón2023topological,ando2024gauging}, being phases of matter enriched by symmetry, should have correspondence in the \SymTO{} framework. Indeed, it has been shown that \SymTO{} gives a complete characterization for gapped and gapless bosonic SPTs. More concretely, there is a one-to-one correspondence between condensable algebras of $D(G)$ and gapped or gapless SPTs protected by $G$. We briefly review this correspondence here, the full theory can be found in~\cite{wen2023classification}. Some details of quantum double models are also reviewed in Appendix ~\ref{app:DG algbera}.

A bosonic $\Gamma$-(g)SPT has a symmetry extension structure summarized by the sequence~\cite{Thorngren_2021,Wen_2023,li2023intrinsicallypurely}
\begin{align}
    1\to N\to \Gamma\to G\to 1.
\end{align}
The physical meaning of $N,G$ is as follows. Excitations that carry non-zero charges of the subgroup $N$ are gapped. Below the gap to these gapped excitations, the symmetry is effectively the quotient $G=\Gamma/N$. $N$ is also called the gapped symmetry, and $G$ the gapless symmetry. A gapped SPT is the case where all excitations are gapped, i.e. $N=\Gamma$. The low energy gapless sector of a gSPT can carry an anomaly of the effective symmetry $G$. This is called the emergent anomaly of the gSPT. When the emergent anomaly is non-trivial, the gSPT is called an intrinsically gapless SPT(igSPT)~\cite{thorngren2019fusion,Wen_2023}, since the resulting edge modes of this SPT cannot be realized in any gapped SPT with the same symmetry.  The topological aspects of the gSPT are determined by two functions $(\eta,\epsilon)$, where $[\eta]\in H^2[N,U(1)]$ determines the SPT-class of the gapped degrees of freedom, and $\epsilon: N\times \Gamma\to U(1)$ determines the charges of edge $N$-actions under $\Gamma$. The two functions are subject to a set of natural consistency conditions and equivalence relations. The edge modes of a gSPT is fully determined by the data $(\eta,\epsilon)$. The emergent anomaly of the low energy effective symmetry $G$ can also be computed from $(\eta,\epsilon)$~\cite{wen2023classification}. 

On the \SymTO{} side, the condensable algebras of $D(\Gamma)$ have been classified in~\cite{davydov2009modular,davydov2016lagrangian}. A condensable algebra of $D(\Gamma)$ is determined by the data $(H, N, \eta, \epsilon)$, and is denoted by $\A[H,N,\eta,\epsilon]$. Here $H$ is a subgroup of $\Gamma$, $N$ is a normal subgroup of $H$. $\eta\in \Z^2[N,U(1)]$ is a 2-cocycle of $N$, and $\epsilon$ is a $U(1)$-valued function: $N\times H\to U(1)$. The functions $\eta, \epsilon$ need to satisfy a set of consistency equations.  When $H\neq G$, the condensable algebra $\A[H,N,\eta,\epsilon]$ necessarily contains gauge charges, therefore condensing a $\A[H,N,\eta,\epsilon]$ at $\sf{B_{phys}}$ with $H$ strictly smaller than $\Gamma$ describes an SSB phase of $\Gamma$. Thus for describing symmetric phases we consider $H=\Gamma$. The algebra is then Lagrangian if and only if $N=\Gamma$. In general, in the \SymTO{} sandwich construction, condensing $\A[\Gamma,N,\eta,\epsilon]$ on the physical boundary corresponds to a \oned{} phase whose gapped symmetry is $N$ and gapless symmetry is $G=\Gamma/N$. Therefore gapped phases are represented by algebras with $N=\Gamma$. In this case the function $\epsilon$ is fully determined by $\eta$. Thus a magnetic and Lagrangian algebra of $D(\Gamma)$ is fully determined by a 2-cocycle of $\Gamma$. This agrees with the classification of \oned{} $\Gamma$-SPTs.\footnote{In the classification of SPTs two cocycles differ by a coboundary describe the same phase, this is matched in \SymTO{} by the fact that two condensable algebras are isomorphic if the 2-cocycles differ by a coboundary.} When $N\neq \Gamma$, the condensation is non-Lagrangian and corresponds to a gapless SPT(gSPT) of $\Gamma$ with gapped symmetry $N$. In this case the data $(\eta,\epsilon)$ defining a condensable algebra have the same structure as those defining a $\Gamma$-gSPT. Namely they satisfy the same set of consistency conditions and equivalence relations. The emergent anomaly of the corresponding gSPT is matched in \SymTO{} by the post-condensation twist: condensing $A[\Gamma,N,\eta,\epsilon]$ changes the \SymTO{} to a twisted gauge theory of the quotient group: $D(G)_\omega$, and the twist $\omega\in \Z^3[G,U(1)]$ is exactly the emergent anomaly of the corresponding gSPT. The edge modes of the gapped/gapless SPTs can also be recovered in \SymTO{} by the method we described before.  

The \SymTO{} provides a unified theory for gapped and gapless SPTs: gapped SPTs correspond to Lagrangian magnetic algebras and gapless SPTs correspond to non-Lagrangian magnetic algebras. We will explore generalization of this correspondence to fermionic gSPTs in Section~\ref{sec:gSPT}, and a more complete treatment will appear in a separate work~\cite{wenFgSPT}.

\begin{figure*}[t]
    \includegraphics[width=1\columnwidth]{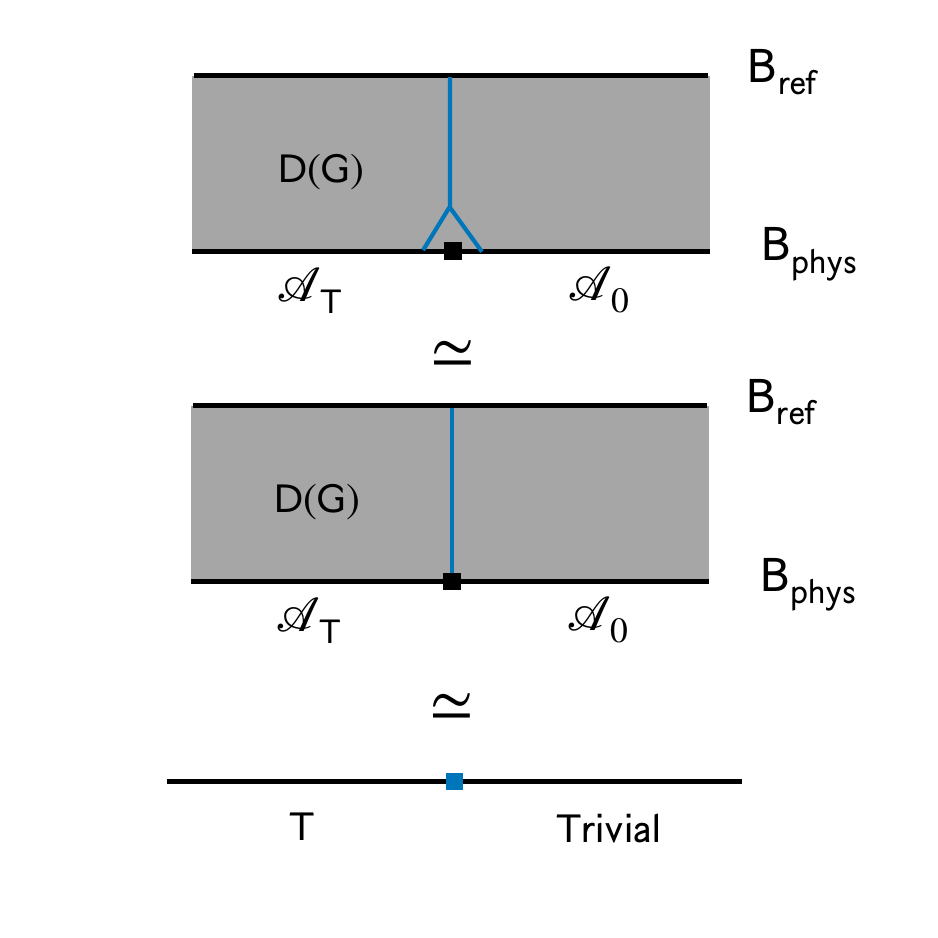}
    \vspace{-0.2 in}
    \caption{\textbf{Edge modes in \SymTO{}.} The physical boundary has two regions of different condensations separated by a point defect, represented by a black box. The all-flux condensation $\A^0$ corresponds to the trivial symmetric phase. $\A_{\sf{T}}$ is the condensation that corresponds to a non-trivial \oned{} phase $\sf{T}$. If there is an anyon string that can end on the point defect as well as the reference boundary, then this anyon line becomes a local zero mode supported on the edge of $\sf{T}$ after dimensional reduction.}
    \label{fig:SymTOedgemodes}
\end{figure*}
\begin{figure*}
    \centering
    \includegraphics[width=1.8\columnwidth]{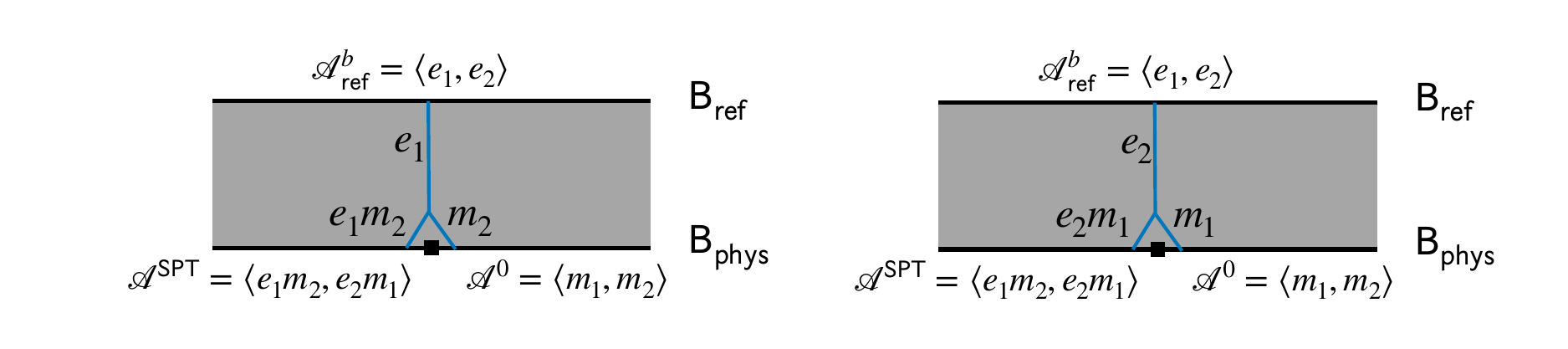}
    \caption{\textbf{Edge modes of the cluster chain from \SymTO{}.} The \SymTO{} bulk is $D(\bZ_2\times \bZ_2)$-two copies of toric codes. The trivial paramagnet phase of $\bZ_2\times \bZ_2$ is represented by the canonical magnetic condensation $\A^0=\<m_1,m_2\>$. The non-trivial SPT phase, i.e. the cluster chain, is represented by the magnetic condensation $\A^1=\<e_1m_2,e_2m_1\>$. The two panels show two linearly independent edge modes in the \SymTO{}. Notice these two edge modes anti-commute, giving rise to a two-fold GSD for every edge.}
    \label{fig:cluster-chain-edge-modes}
\end{figure*}

\subsection{Edge modes from \SymTO{}\label{sec:SymTOedgemodes}}
One characteristic of gapped or gapless SPT phases is their edge modes. This is a set of zero modes acting on the edge of the system whose algebra is protected by the symmetry. They are responsible for the GSD with open boundaries and characterize the topological properties of the phases.  To study edge modes in \SymTO{} we consider the following setup. For a symmetry group $G$ with no anomaly there is always a canonical magnetic algebra that corresponds to condensing all gauge fluxes: $\A^{0}:=\oplus_{C} m_C$, the sum is over conjugacy classes of $G$ and $d_{m_C}=|C|$. Since $\dim(\A^0)=\sum_C |C|=|G|$ this algebra is Lagrangian. Condensing $\A^{0}$ on $\sf{B_{phys}}$ produces a gapped symmetric phase which is identified as the trivial $G$-phase, i.e. the vacuum. Now to study edge modes of a phase $\sf{T}$, we consider putting the corresponding boundary condition $\A^{\sf{T}}$ on a finite region of $\sf{B_{phys}}$, while the rest of $\sf{B_{phys}}$ has the canonical magnetic condensation $\A^0$, see Fig.~\ref{fig:SymTOedgemodes}. Then edge modes correspond in the \SymTO{} to anyon line operators that are localized at the interface between $\A^{\sf{T}}$ and $\A^0$, such that no excitations are created. As an illustration, we show the \SymTO{} representation of edge modes of the nontrivial $\bZ_2\times \bZ_2$-SPT (a.k.a the cluster chain) in Fig.~\ref{fig:cluster-chain-edge-modes}.

\section{\SymTO{} for fermions: Structure and Examples\label{sec:Z2FSymTO}}
We now seek a generalization of the \SymTO{} framework for fermionic systems. 
The key features of the \SymTO{} that we wish to preserve are that:
\begin{enumerate}
    \item It reduces to the original system by dimensional reduction/sandwich construction.
    \item Non-trivial defect lines on the reference boundary have the same structure as symmetry defects of the original system.
    \item Anyons condensed on the reference boundary have the same structure as local charges of the original system.
\end{enumerate}

The sandwich construction of the fermionic \SymTO{} proceeds similarly to that for bosons with two key differences: first, the gauge group of the \SymTO{} contains the fermion parity subgroup $\bZ^F_2$ to account for the conservation of fermion number parity, $(-1)^F$, and second, to describe a system with fermion excitations we must explicitly introduce local (ungauged) fermion excitations into the otherwise bosonic \SymTO{} sandwich.\footnote{In our construction, these local fermion excitations will be located on the reference boundary of a bosonic \SymTO{}. We mention that there are other proposals that introduce local fermions into the bulk of the \SymTO{} construction \cite{Gaiotto2020,freed2023topological,KantaroLecture}. See Section \ref{sec: summary} for more discussion. }

In this section, we illustrate the basic ideas of this approach through a series of examples of how various familiar gapped topological and SPT phases, and gapless critical points arise in the fermion \SymTO{} framework. In Section~\ref{sec:gSPT} below, we also construct a new fermionic intrinsically gapless SPT phase , characterized by a low-energy emergent symmetry with a fermionic anomaly.
In Section~\ref{sec:general}, we formalize this construction, and address various technical details related to spin structures.

\subsection{Construction of the fermionic \SymTO{}}
The symmetry group, $G^F$, of a fermion system necessarily includes fermion parity $\bZ_2^F$ generated by $(-1)^F$ where $F$ is the number of fermions (modulo two), as a normal subgroup. $G^F$ can be regarded as an extension of the bosonic symmetry, $G^B=G^F/\bZ_2^F$, by fermion parity $\bZ_2^F$.
Following the bosonic construction, we choose the \SymTO{} as the \twod{} $G^F$ gauge theory. For example, without any microscopic (bosonic) symmetries, $G^B=1$, $G^F=\bZ_2^F$ remains non-trivial, enabling a description of distinct fermion gapped phases such as the trivial and Kitaev chain topological superconductor phases which have no symmetry-distinction.
We consider for now the case where the symmetry splits into the product of the fermion parity subgroup and the bosonic symmetry group: $G^F=\bZ_2^F\times G^B$. In this case the $D(\bZ_2^F)$ sector of the \SymTO{} $D(G^F)$ contains an abelian bosonic flux $m^F$ and an abelian charge $e^F$,\footnote{We caution that, unlike in the SPT literature where $m_f$ is sometimes used to denote a fermionic statistics, here we use a superscript $m^F$ to denote that this is the flux associated with the fermion parity sub-group of the gauge-group $G^F$.} and a fermionic dyon $f=e^F\times m^F$.

The reference boundary, $\sf{B_{ref}}$, must be chosen so that the condensed anyons have the right structure of local symmetry charges of $G^F$. Since a $\bZ_2^F$-charge is necessarily fermionic, it looks as if this requires condensing particles with fermion statistics, which is not directly possible. To proceed, we introduce an auxiliary local fermion, $c$, that lives only at $\sf{B_{ref}}$. 
We then create a gapped reference boundary, $\sf{B_{ref}}$, by condensing the \emph{bosonic} pair $f\times c$. 
In addition, following the bosonic \SymTO{} we also condense all the $G^B$ gauge charges $e^R, R\in \rep(G^B)$  at $\sf{B_{ref}}$.

We note that this approach is analogous to that for obtaining fermion models from bosonic string-net models for quantum doubles~\cite{}. Notice that the physical boundary $\sf{B_{phys}}$ is bosonic. We have chosen to locate the fermionic degrees of freedom on the reference boundary so that all topological aspects of the symmetry, including the statistics of symmetry charges, are encoded by the reference boundary.
\subsubsection{Local fermion excitations\label{sec:local fermion}}
This choice of reference boundary condition automatically produces a fermion excitation at the physical boundary, $\sf{B_{phys}}$, which can be thought of as the charge of $\bZ_2^F$. Namely, consider a short, vertical $f$ line segment that spans from $\sf{B_{phys}}$ to $\sf{B_{ref}}$, where its end is decorated by a local fermion operator $c$ such that it can be absorbed $cf$ condensate at $\sf{B_{ref}}$. However, the other end at $\sf{B_{phys}}$ cannot be absorbed, and instead leaves behind an excitation. 
Under dimensional reduction, this line operator becomes a local point-like operator with Fermi exchange statistics inherited from the $c$ decoration. Also, notice that with the fermionic reference boundary, a single $f$-anyon can now by created by a local operator. Thus, the fermionic reference boundary turns the emergent fermions, $f$, into \textit{local} physical fermions of the effective \oned{} system.
\begin{figure*}[t]
    \includegraphics[width=1.5\columnwidth]{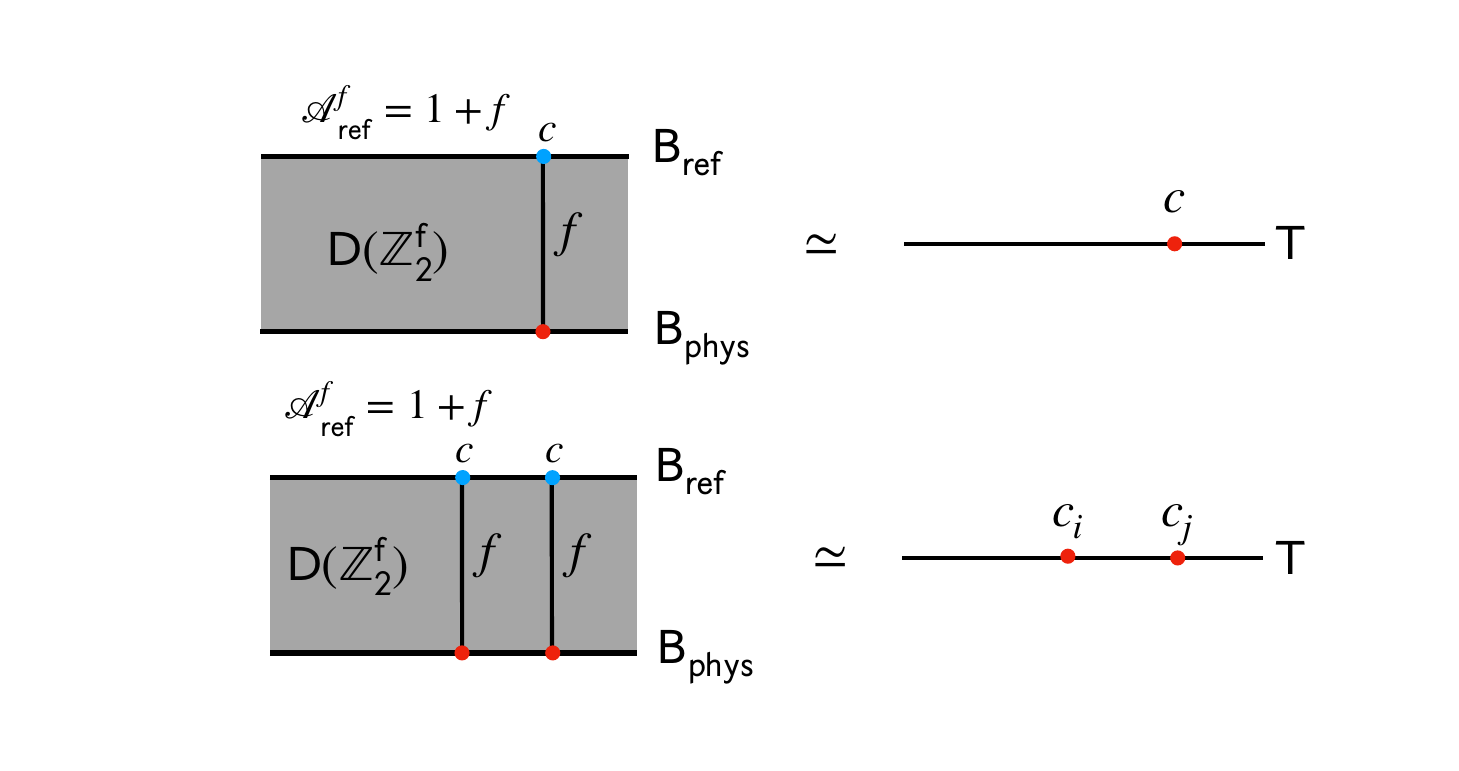}
    \caption{\textbf{$\bZ_2^F$-\SymTO{} setup.}  Top: An $f$-line can end on $\sf{B_{ref}}$, with the endpoint dressed by a local fermion operator $c$, represented by a blue dot. Notice there is no excitation created on $\sf{B_{ref}}$. The excitation is created at the other end of the $f$-line, represented by a red dot. After dimensional reduction this vertical line becomes a local fermionic operator. Bottom: The local fermion operators $c$ attached to $f$-lines give anti-commutation relation between two vertical $f$-lines. After dimensional reduction, this becomes the standard anti-commutation relation between local fermionic operators.}
    \label{fig:FSymTOsetup1}
\end{figure*}

\subsubsection{Induced spin structure\label{sec:spin1}}
\begin{figure*}[t]
    \includegraphics[width=2\columnwidth]{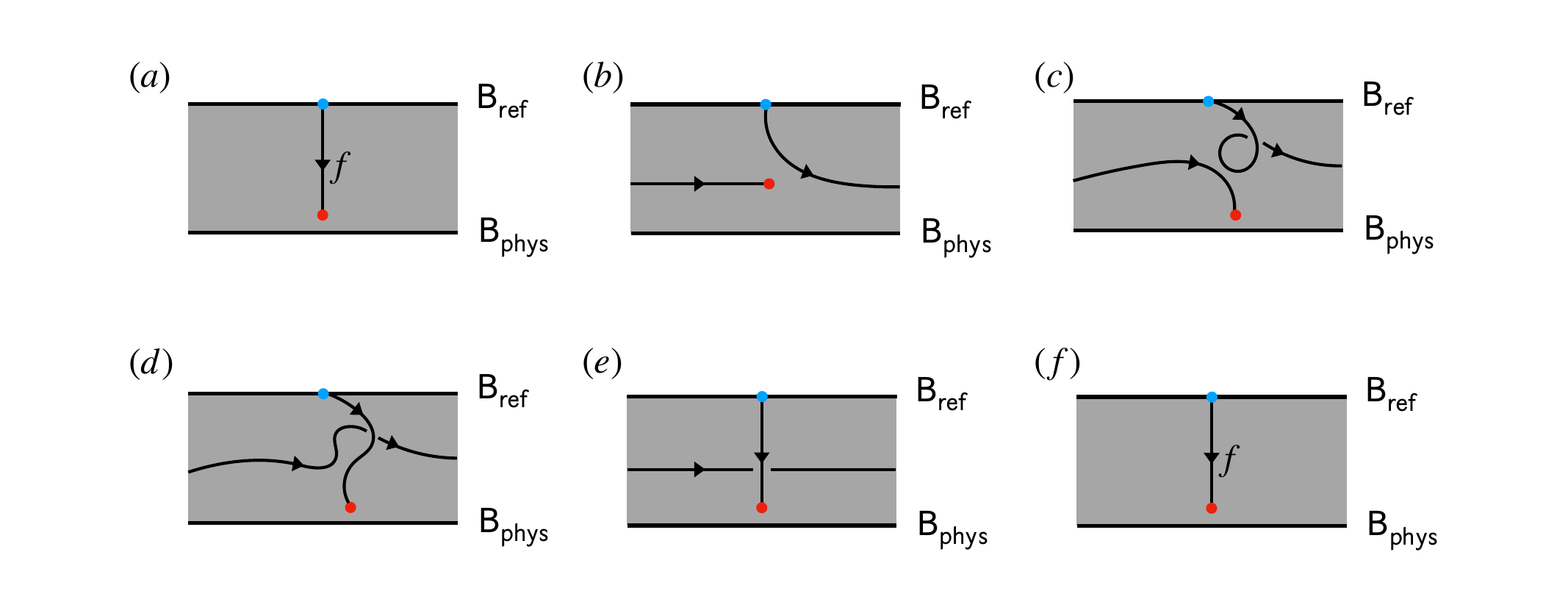}
    \vspace{-0.2 in}
    \caption{\textbf{Diagrammatic argument for spin-structure dependence of the $f$-condensed boundary.} (a). An $f$-string is used to create an $f$ anyon (red dot) in the bulk. The $f$-string ends on the fermionic reference boundary and has a fermionic endpoint (blue dot). The created anyon $f$ should be viewed as a local physical fermion of the effective \oned{} system represented by the sandwich. (b). The $f$ anyon (red dot) moves around the horizontal direction by extending the $f$-string and returns to its initial location. (c). We perform a braiding for the $f$-string, which results in a -1 sign. (d) We perform an $F$-move to re-arrange the string into two separate strings: a vertical short string and a horizontal noncontractible loop. The $F$-move does not produce any phases. (e) Deform the strings into straight lines. (f). The horizontal $f$-loop is absorbed by the $f$-condensate at $\sf{B_{ref}}$, producing a phase $\<W_f(\gamma)\>_{\sf{B_{ref}}}$. Thus the overall phase accumulated in this process is $-\<W_f(\gamma)\>_{\sf{B_{ref}}}$. }
    \label{fig:spinstructure}
\end{figure*}

A crucial difference between fermions and bosons, is that fermions can only be defined on manifolds that permit a spin structure. 
Roughly speaking, a spin structure gives a consistent set of $\pm 1$ phases obtained when transporting a fermion around a closed loop in spacetime.
In the following, we will mainly consider the case where $\sf{B_{ref}}$ has the structure of a two-torus, $T^2$, in which case the spin structure may be specified by periodic (P) or anti-periodic (AP) boundary condition along the space-like and time-like noncontractible loops. We will therefore denote the spin structure as (P/AP,P/AP) where the first (second) argument indicates the boundary conditions in space (time) respectively. To define the $fc$-condensed reference boundary will require choosing a spin structure on $\sf{B_{ref}}$, and this spin structure is exactly the spin structure of the effective \oned{} system represented by the \SymTO{} sandwich.
Specifically, the spin structure of the physical fermions is associated with the phase obtained when absorbing an $f$ loop into the reference boundary. To see this, consider a short $f$-segment that creates an $f$-anyon in the \SymTO{} bulk (see Fig.~\ref{fig:FSymTOsetup1}). According to our discussion in~\ref{sec:local fermion}, this $f$-anyon corresponds to the local physical fermion excitation of the effective \oned{} system. Now imagine transporting this local fermion by dragging the $\sf{B_{phys}}$ end of the $f$-segment around a spatial cycle, $\gamma$. As shown in Fig.~\ref{fig:spinstructure}, with a single fermion exchange (resulting in a $(-1)$ phase) this configuration can be deformed into the original fermion excitation (short, open $f$-segment), and a closed $f$-string loop corresponding to a Wilson loop operator $W_f(\gamma)$. The $f$ loop can then be absorbed into $\sf{B_{ref}}$ producing a phase $\<W_f(\gamma)\>_{\sf{B_{ref}}} = \pm 1$. The resulting overall phase is $(-1)\<W_f(\gamma)\>_{\sf{B_{ref}}}$. Therefore the boundary condition on $\gamma$ is related to the value of $f$-loop by the following relation:
\begin{align}
    P/AP\Leftrightarrow \<W_f(\gamma)\>_{\sf{B_{ref}}}=-1/+1.
\end{align}
Throughout the remainder of this section, we will choose the convention that there is a unit eigenphase for absorbing an $f$-loop into both the space- or time- cycle $\gamma$ of $\sf{B_{ref}}=T^2$: $\<W_f(\gamma)\>_{\sf{B_{ref}}}=+1$. This corresponds to AP boundary conditions for the physical fermions along the two cycles of $\sf{B_{phys}}$. We note, however, that the boundary condition for the physical fermions can be toggled by inserting  a fermion parity flux ($m^F$) defect line as desired. Below we will use this freedom to relate the partition functions of fermion phases for different boundary conditions, through the \SymTO{} framework.

\subsubsection{Symmetry defects}
In addition to the standard symmetry defects of the bosonic part of the symmetry, $G^B$, there are additional line defects associated with the $\bZ_2^F$ subgroup of the \SymTO{}: the horizontal $m^F$-line and $e^F$-line. Since both defects anticommute with the $\bZ_2^F$ charge (short vertical decorated $f$ strings described above), and neither can be trivially absorbed into $\sf{B_{ref}}$ (unlike for the bosonic $\sf{B_{ref}}$), we have the freedom of choosing either as the fermion parity symmetry. 
These two choices are essentially equivalent, and the resulting \SymTO{} dictionaries differ only by a relabelling of $e^F\leftrightarrow m^F$ (which is an automorphism of the \SymTO{}). We fix this ambiguity by choosing $m$-line as the fermion parity symmetry generator. 
Below, we will see that this ambiguity precisely stems from the physical ambiguity between labeling the trivial and topological superconducting phases (Kitaev chain) of a Majorana chain with periodic boundary conditions.

We emphasize that this ambiguity is physical and not a deficiency of the \SymTO{} framework. Consider a \oned{} lattice of Majorana fermions. The trivial and topological superconducting phases differ only by a convention for which pairs of Majorana modes constitute atomic ``sites", and which ones lie along bonds of the chain. With periodic boundary conditions, we can transform from one phase to the other through translation by one site. This translation symmetry exactly corresponds to the exchange symmetry $e^F\rightarrow m^F$ in $D(\bZ_2^F)$.

\begin{figure*}[t]
\vspace{-0.2 in}
    \includegraphics[width=1.5\columnwidth]{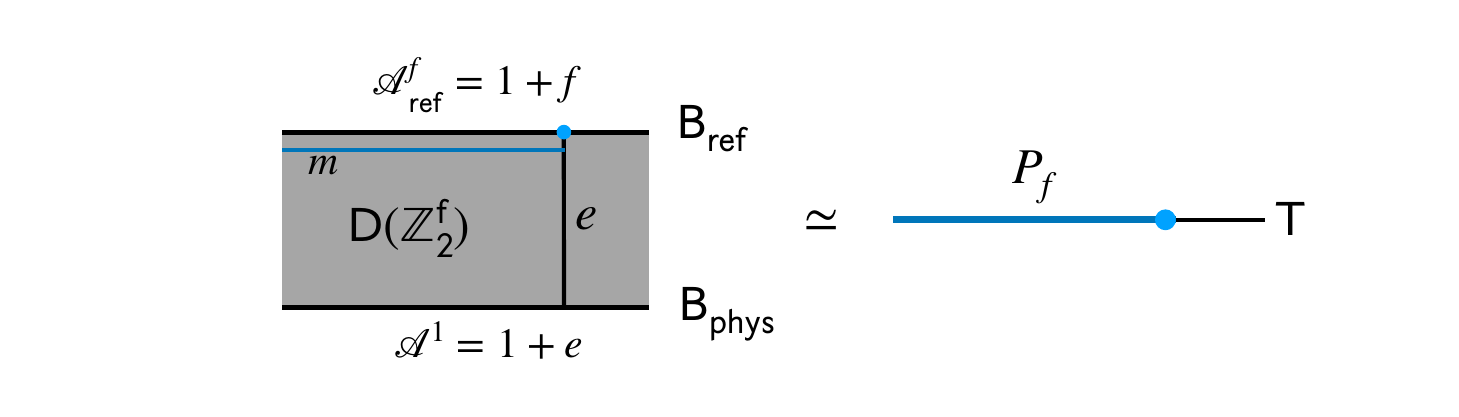}
    \vspace{-0.2 in}
    \caption{\textbf{Kitaev chain order parameter in \SymTO{}.} In the \SymTO{} we consider an $e$-line that emanate from $\sf{B_{phys}}$, it approaches the reference boundary and becomes an $m$-line (blue line) near $\sf{B_{ref}}$. After dimensional reduction this becomes a charge of $\bZ_2^F$ (blue dot) that lives at the end of a $\bZ_2^F$-defect line. Since in the \SymTO{} there is no excitation created anywhere, this non-local operator now has non-zero vacuum expectation value. }
    \label{fig:kitaevop}
\end{figure*}

\subsection{Example: Fermionic gapped phases with $G^F=\bZ_2^F$}
We begin by studying gapped phases of one-dimensional systems without any additional symmetries other than conservation of fermion parity, $(-1)^{F}$ where $F$ is the number of fermions (modulo two), which has group structure $G^F=\bZ_2^F$ (where the $F$ superscript is just a label reminding that this $\bZ_2$ factor arises from fermion parity). 
There are two gapped phases of such a \oned{} fermion system without any symmetries: a trivial superconductor and topological superconductor (Kitaev chain) with unpaired Majorana edge modes. In a Majorana chain with periodic boundary conditions, there is a fundamental ambiguity between these phases, which differ only by a (Majorana) translation: the labeling of ``trivial" and ``topological" amounts to declaring which pairs of Majorana fermions are paired into atomic sites, and which pairs forms inter-site bonds. An unambiguous property is that the interface between these two phases carries an unpaired Majorana zero mode with radical quantum dimension $\sqrt{2}$.

\begin{figure*}[t]
    \includegraphics[width=1.3\columnwidth]{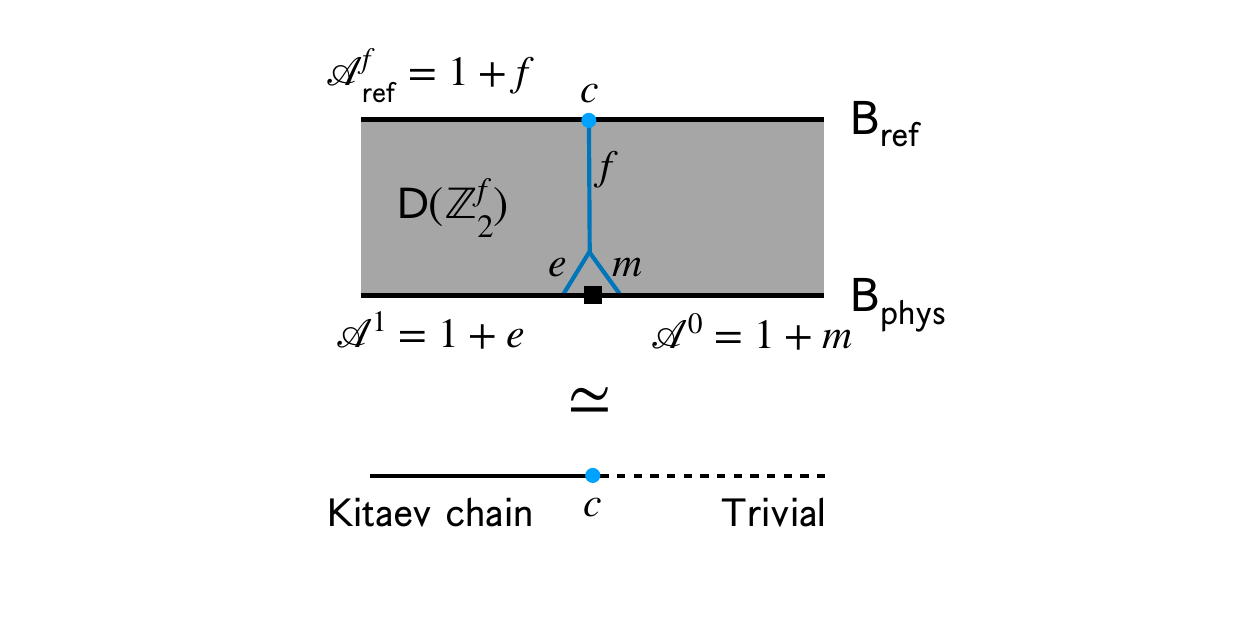}
    \vspace{-0.3in}
    \caption{\textbf{Majorana edge mode in \SymTO{}.} On $\sf{B_{phys}}$ there is an interface between $\A^e$-boundary condition and $\A^m$ boundary condition (black box). An $f$ line can now connect $\sf{B_{ref}}$ and this interface without costing any energy. There is a local fermion operator attached to the endpoint of this line on $\sf{B_{ref}}$ (blue dot). After dimensional reduction this line becomes an edge mode. This edge mode is fermionic and squares to 1. The point defect on the physical boundary has quantum dimension $\sqrt{2}$. These properties indicate the edge mode is a Majorana mode.  }
    \label{fig:kitaevedge}
\end{figure*}

Parallely, there are two distinct gapped boundaries of the \SymTO{} corresponding respectively to condensing $m$ or $e$ at the physical boundary $\sf{B_{phys}}$ in the sandwich construction. 
The interface between these two gapped boundaries hosts a twist defect with quantum dimension $\sqrt{2}$. Further, this twist defect is associated a short vertical line operator that becomes a local Majorana fermion zero mode under dimensional reduction (See Fig.~\ref{fig:kitaevedge}). This line operator consists of a short vertical $f$ line that terminates on the reference boundary decorated by a local fermion, $c$, and on the physical boundary by splitting $f\rightarrow e\times m$, and absorbing the $e$ and $m$ anyons in the appropriate domains.

We resolve this ambiguity by fixing the convention that the microscopic  fermion parity symmetry is generated by the $m$-line (as described above). As we now show, this choice selects the $m$ condensed boundary as the trivial phase and the $e$ condensed boundary as the topological phase.
For this purpose, we note that the local action of the fermion parity symmetry $m$ on a finite boundary interval corresponds to acting with an $m$ line that spans the interval, and terminates on the edges of the interval.
Consider first the $m$-condensed physical boundary (with condensable algebra $\A_m^{phys}:=1+m$). Here, the fermion parity defect line $m$ can be freely absorbed by the physical boundary, so the local action of fermion parity is trivial -- i.e. the system is a trivial insulator.
By contrast, for the $e$-condensed physical boundary (with condensable algebra $\A_e^{phys}:=1+e$, the fermion parity cannot be trivially localized to a finite boundary region, \emph{unless} its end points are ``decorated" by pulling an $f$ string from the reference boundary (See Fig.~\ref{fig:kitaevop}). 
This decorated $m$ string, with fermionic ends, corresponds to a non-local string order parameter of the topological superconducting phase~\cite{Kitaev_2001,Fidkowski_2011,Bultinck_2017,PhysRevB.90.115141}.

We note that, for a bosonic system with $\bZ_2$ symmetry, the $e$-condensed physical boundary condition would correspond to a spontaneous symmetry breaking (SSB) state. There, this arose because $e$ was also condensed at the reference boundary. 
Unlike for ordinary bosonic symmetries, $\bZ_2^F$ cannot be spontaneously broken, since its charges (the order parameter of the symmetry) are fermionic and can never have long-range SSB order. In the \SymTO{} description, SSB arises when the symmetry defect line cannot be absorbed into either boundary, such that the sandwich, and the sandwich with a symmetry-defect ``condiment" layer correspond to distinct, degenerate ground states. Importantly, this cannot happen in the fermionic \SymTO{} construction, since there is no way to condense the fermion end of the $f$ string on the physical boundary $\sf{B_{phys}}$(recall the physical boundary in our construction stays bosonic). Moreover, for both the $e$ and $m$ condensed $\sf{B_{phys}}$ phases an $m$ string condiment layer that wraps through the bulk of the sandwich can either by directly absorbed into the physical boundary, or dressed by an $f$ string from the reference boundary and then absorbed into the $e$-string boundary, indicating that the system has a unique symmetric ground-state.

In addition to matching the structure of edge states, we can also directly use the \SymTO{} to reproduce the partition function of gapped phases, as we now illustrate.

\subsubsection{Spin structure and partition function of gapped phases \label{sec:spin-example}}
In the continuum a fermionic system is defined on a manifold with a spin structure (or a Pin-structure for un-orientable manifolds), which can loosely be thought of as a ``background gauge field of fermion parity'' in the sense that the spin structure defines whether a fermion obtains a minus sign when dragged around a non-contractible cycle of the manifold. A spin structure on the torus can specified by periodic or anti-periodic boundary condition along two non-contractible loops (generators of $H_1[T^2]$). Therefore we will denote a spin structure on the torus by $(P/AP,P/AP)$.

\paragraph{Review: Partition function of the Kitaev chain.}
 The partition function of a fermionic system depends on the spin structure, $\eta$. For example, the partition function of the Kitaev chain is given by $\arf(\eta)$ in the IR limit, where $\arf(\eta)=\pm 1$ is the Arf invariant. 
 The values of the Arf invariant on the torus are
\begin{align}
\arf(\eta) = 
    \begin{cases}
      1 & \eta=(P,AP), (AP,P), (AP,AP)\\
      -1 & \eta=(P,P)
    \end{cases}       
\end{align}
Physically this comes from that fact that the ground state of the topological phase of the Kitaev chain has opposite fermion parities for periodic and anti-periodic boundary conditions, whereas the trivial phase is adiabatically connected to a product state that is completely insensitive to boundary conditions.
The partition function of the topological phase of the Kitaev chain can then be computed as follows. We denote the ground-states with $P/AP$ boundary conditions as $|GS_{P/AP}\>$ respectively, and define the normalized partition function on a torus ($T^2$) with $X^1,X^0=P/AP$ boundary conditions in the space and time cycles as $\Z[T^2,(X^1,X^0)]$. Then the ordinary zero-temperature thermal partition function has $AP$ boundary conditions in time, from which we identify: 
\begin{align}
\Z[T^2,(P,AP)] & = \tr |GS_P\>\<GS_P|=1 \nonumber\\
\Z[T^2,(AP,AP)] & = \tr |GS_{AP}\>\<GS_{AP}|=1
\end{align}
Partition functions with periodic boundary condition in time can be obtained from these by inserting a $(-1)^{P_f}$ into the trace which acts like a spacelike defect of fermion parity in a path integral evaluation of the trace: 
\begin{align}
    \Z[T^2,(X,P)] = \<GS_X|(-1)^{P_F}|GS_X\>.
\end{align}
The key defining property of the topological superconductor is that the ground-state fermion parity of $P$ and $AP$ boundary conditions are opposite. The fermion parity of either alone is non-universal, and can be toggled by adding an additional occupied fermion orbital, only the relative fermion parity for $P$ vs $AP$ boundary conditions is well defined. For definiteness, we will fix the convention that the ground-state with periodic boundary conditions has odd fermion parity, in which case:
\begin{align}
\Z[T^2,(P,P)]
& = \tr (-1)^{P_f} |GS_P\>\<GS_P|
\nonumber\\
&=\<GS_P|
(-1)^{P_F}|GS_P\> 
\nonumber\\
&= -1 \nonumber\\
\Z[T^2,(AP,P)]
&= \tr (-1)^{P_f} |GS_{AP}\>\<GS_{AP}|\nonumber\\
&=\<GS_{AP}|(-1)^{P_F}|GS_{AP}\> 
\nonumber\\
&= 1 
\end{align}

\paragraph{Partition function of the Kitaev chain from \SymTO{}:}
We will now show how this partition function can be obtained from  \SymTO{} methods, deferring some technical details to Section~\ref{sec:general}. 
Partition function of the \SymTO{} sandwich is given by an inner product between states defined on $\sf{B_{ref}}$ and $\sf{B_{phys}}$. We have given the boundary states corresponding to the Dirchelet and Neumann boundary condition in~\ref{sec:bg} in the field configuration basis. Here we will find it useful to introduce another basis for boundary state, suitable for a torus boundary geometry. This is the so-called anyon label basis, $\{|a\>\}$, where the label $a$ goes through all simple anyons of the bulk \SymTO{}. This basis is defined by the property that time-like anyon loops act on them as $W_b^t|a\>=\sum_c N^c_{ba}|c\>$, and space-like loops act on them as $W^x_b|a\>=\frac{S_{a,b}}{S_{a,0}}|a\>$, where $N^{c}_{ba}$ is the fusion matrix of the \SymTO{}, and $S_{a,b}$ is the $S$-matrix of the \SymTO{}.  The state $|a\>$ can be thought of as being prepared by a path integral in the interior of the torus, with an anyon loop of $a$ inserted in the center. 

The anyon basis is particularly useful when a boundary is specified by an anyon condensation. If the anyon condensation is given by a condensable algebra $\A=\sum_a c_a a$, then it is straightforward to derive that the corresponding boundary state can be written in the anyon basis as $|\A\>:=\sum_a c_a|a\>$. Notice this state has the property that $W^{t/x}_a$ both have eigenvalue 1, for any $a\in \A$.

\begin{figure*}[t]
    \includegraphics[width=2\columnwidth]{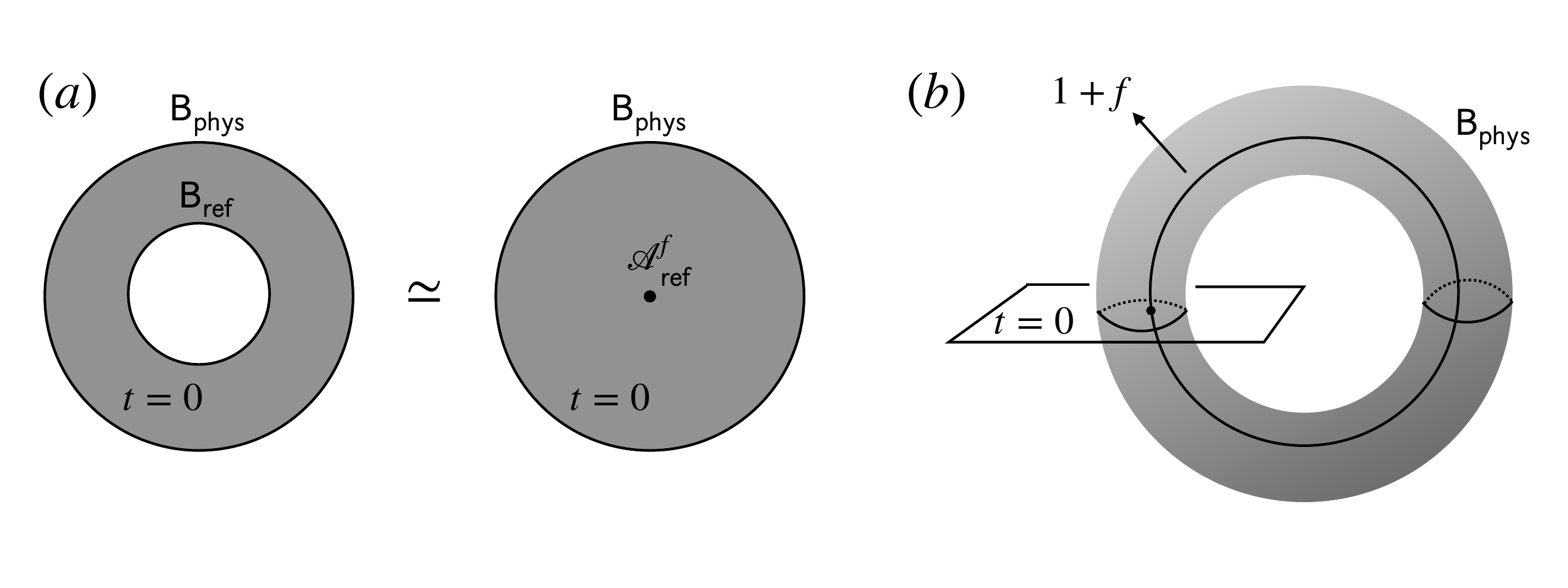}
    \vspace{-0.2 in}
    \caption{1. The reference boundary is shrunk to a point and becomes an insertion of the algebra $\A_{\sf{ref}}^f$. 2. The spacetime picture, the reference boundary becomes a world line of $1+f$ inserted in the center of the solid torus. }
    \label{fig:f-bdry2}
\end{figure*}

For the $f$-condensed reference boundary of the $\bZ_2^F$-\SymTO{}, it is natural to expect the reference boundary to be given by the state $|1\>+|f\>$. Notice that the state $|1\>+|f\>$ has the property that $f$-loops in the space- and time-cycles both have eigenvalue 1. To see this, we can shrink the reference boundary to a point, then in the spacetime picture the reference boundary is now reduced to an insertion of an $1+f$ anyon string in the center of a solid torus. See Fig.~\ref{fig:f-bdry2}. Then an $f$-loop in the time direction can be fused with the $1+f$ insertion: $W^t_f(|1\>+|f\>)=|f\>+|1\>$. Thus $f$-loops in the time direction has eigenvalue 1. An $f$-loop in the space direction on the other hand is performing a braid with the $1+f$ insertion. Since this braiding is trivial, the space-like $f$-loops also have eigenvalue $1$ when acting on the reference boundary. Then according to the discussion on the relation between $f$-loop eigenvalues and spin structures in~\ref{sec:spin1}, we see that the state $|1\>+|f\>$ describes the $f$-condensed boundary with $(AP,AP)$ spin structure.

%

With this established, we can start to compute the partition function of the Kitaev chain via \SymTO{}. The Kitaev chain is obtained by imposing an $e$ condensation ($\A^{e}=1+e$)  at $\sf{B_{phys}}$. The partition function of the \SymTO{} sandwich is given by the inner product between the state on the inner circle and the state on the outer circle: $\Z[T^2,(AP,AP)]=\<\A_{\sf{ref}}|\A_{\sf{phys}}\>$, with $|\A_{\sf{ref}}\>=|1\>+|f\>$ and $|\A_{\sf{phys}}\>=|1\>+|e\>$, resulting in: $\Z[T^2,(AP,AP)]=1$.

The spin structure (boundary conditions) for the physical fermion can be adjusted by inserting fermion parity flux lines through space or time-like cycles of the \SymTO{} bulk.
Consider changing the spatial boundary condition by inserting an $m$-loop along the time cycle.
Again, if we shrink the reference boundary to a point, the condensate $\A_{\sf{ref}}=1+f$ becomes an insertion of a $1+f$ world-line in the solid torus (see Fig.~\ref{fig:f-bdry2}).
Fusing the time-like $m$ loop  with the $1+f$ condensed $\sf{B_{ref}}$ line changes it into an $m+e$ line,
 resulting in: $$\Z[T^2,(P,AP)]=(\<e|+\<m|)(|\A_{\sf{phys}}\>)= 1.$$

 Similarly, inserting an $m$-loop encircling the spatial cycle produces $(AP,P)$ boundary conditions. The $m$-loop has a $(-1)$ braiding phase with the $f$-insertion at $\sf{B_{ref}}$, resulting in:
\begin{align}
    \Z[T^2,(AP,P)]&=(\<1|+\<f|)W^{x}_m|\A_{\sf{phys}}\>\nonumber\\
&=\<1|\A_{\sf{phys}}\>-\underset{0}{\underbrace{\<f|\A_{\sf{phys}}\>}}=1
\end{align}
Here $W^x_m$ is the $m$-loop in the spatial direction. 

Lastly, (P,P) boundary conditions are obtained by inserting $\bZ_2^F$ symmetry defects ($m$ loops) in both space and time directions. Fusing the time-like defect with the $1+f$ insertion at $\sf{B_{ref}}$ and taking into account the braiding between the space-like defect and the $f$-line, yields
$$\Z[T^2,(P,P)]=\underset{0}{\underbrace{\<m|\A^{e}\>}}-\<e|\A^{e}\>=-1.$$

Combining these results, we confirm that the proposed fermionic \SymTO{}  procedure correctly reproduces the known partition function for the Kitaev chain. 

In comparison, to deduce the trivial gapped phase, corresponding to an $m$-condensate ($\A_{\sf{phys}}= \A^{m}=1+m$), one can use precisely the same calculation with $e\leftrightarrow m$. The sole change is that: 
$$\Z^{\rm trivial}[T^2,(P,P)]=\<m|\A^{m}\>-\underset{0}{\underbrace{\<e|\A^{m}\>}}-=+1$$
 so that $\Z^{\rm trivial}=1$ for all boundary conditions as required for a topological trivial phase.

\subsection{Example: Fermionic SPTs with $G^F=\bZ_2\times \bZ_2^F$}
The \SymTO{} framework also captures gapped symmetry-protected topological (SPT) states. In this section, we illustrate the main principles through the simplest example: $G^F =\bZ_2\times \bZ_2^F$.

It is known in the literature (see e.g. \cite{Tang2012}) that different \oned{} topological phases for $\bZ_2\times \bZ_2^F$ are labeled by a pair of $\bZ_2$-valued topological invariants, $(\nu_+,\nu_-)$ that respectively label the number of Majorana edge states with even (+) or odd (-) $\bZ_2$ symmetry charge. The stacking rule for these phases is $(\nu_+,\nu_-)\boxtimes(\nu'_+,\nu'_-) = (\nu_++\nu'_+,\nu_-+\nu'_-)$, with addition taken modulo $2$. The four phases have a group structure $\bZ_2^2$ under stacking.

\begin{figure*}\centering
    \includegraphics[width=2.2\columnwidth]{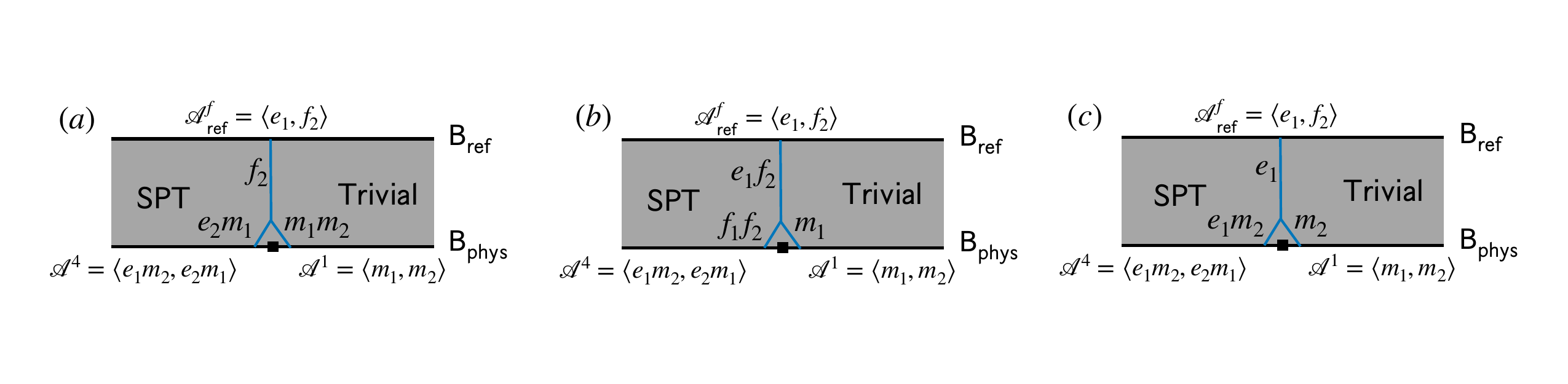}
    \vspace{-0.4in}
    \caption{\textbf{Edge modes of the $(\nu_+=1,\nu_-=1)$ SPT.} The left half of the sandwich represents the non-trivial SPT phase of $\bZ_2\times \bZ_2^F$, while the right half of the sandwich represents the trivial phase. (a). A $\bZ_2$-neutral fermionic edge mode. (b). A $\bZ_2$-charged fermionic edge mode. (c). A $\bZ_2$-charged bosonic edge mode, obtained from the product of the fermion modes shown in (a),(b). }
    \label{fig:Z2edgemode}
\end{figure*}

To construct a \SymTO{} description, we consider a $\bZ_2\times \bZ_2^F$ gauge theory, $D(\bZ_2)\times D(\bZ_2^F)$. The anyons of this theory can be labeled by two copies of those for a $\bZ_2$ gauge theory: $\{1,e,m,f=e\times f\}$.
 
We label anyons from $D(\bZ_2)$ with a subscript 1 and those from $D(\bZ_2^F)$  with a subscript 2.
According to the general recipe outlined above, the reference boundary according to the general recipe~Eq.\eqref{eq:fref} is given by condensing the symmetry gauge-charge $e_1$, and a bound state $f_2\times c$ of the $\bZ_2^F$ emergent fermion. This corresponds to the condensation algebra: $\A^f_{\sf{ref}}=1+e_1+f_1+e_2+f_1e_2$.

Distinct gapped phases correspond to different Lagrangian (fully confining) condensations on $\sf{B_{phys}}$. We list all Lagrangian condensations of $D(\bZ_2\times \bZ_2^F)=D(\bZ_2)\boxtimes D(\bZ_2^F)$, and the corresponding SPT phases in Table~\ref{tabel:spt}.
%
\begin{table*}
\begin{center}
\def\arraystretch{1.2}
\begin{tabular}{|l |c|c| }
\hline
Condensation Algebra & Description & SPT Invariant \\
\hline
 $\A^1=1+m_1+m_2+m_1m_2$ & Trivial  & $(\nu_+=0,\nu_-=0)$ \\
 $\A^2=1+m_1+e_2+m_1e_2$ & neutral Kitaev chain & $(\nu_+=1,\nu_-=0)$\\
 $\A^3=1+e_1e_2+m_1m_2+f_1f_2$ & $\bZ_2$-charged Kitaev chain & $(\nu_+=0,\nu_-=1)$ \\
 $\A^4=1+e_1m_2+e_2m_1+m_1m_2$ & SPT & $(\nu_+=1,\nu_-=1)$\\
$\A^5=1+e_1+m_2+e_1m_2$& SSB & $\times$ \\
$\A^6=1+e_1+e_2+e_1e_2$& Kitaev chain + SSB & $\times$ \\
\hline
\end{tabular}
\end{center}
\caption{{\bf Gapped $G^F = \bZ_2\times \bZ_2^F$ phases from \SymTO{}. } The correspondence between condensable algebras of $D(\bZ_2^2)$ and \oned{} $\bZ_2\times \bZ_2^F$-phases. SPT indicates an invertible topological phase that becomes trivial without symmetry protection, and SSB denotes spontaneous symmetry breaking of the $\bZ_2$ symmetry.
}
\label{tabel:spt}
\end{table*}
The first four phases are symmetric, since the symmetry charge, $e_1$, is not directly condensed on $\sf{B_{phys}}$. The last two have spontaneous symmetry broken (SSB) $\bZ_2$ symmetry, since the $e_1$ condensate on both boundaries defines a local order parameter given by a short $e_1$ segment spanning from $\sf{B_{ref}}$ to $\sf{B_{phys}}$. 
For the symmetric phases, the SPT invariants can be directly confirmed by constructing edge states operators for the interface between the SPT and trivial phase (or by checking their bulk string-order parameter). As an example, we show the edge modes of the $(\nu_+=1,\nu_-=1)$ phase, represented by the $\A^4$ condensation, in Fig.~\ref{fig:Z2edgemode}.
Comparing this set of phases to the known classification, we find that the \SymTO{} provides a complete description.
A natural question is how the SPT stacking rule is manifested in \SymTO{}, this question was addressed for bosonic symmetries in~\cite{turzillo2023duality}, we will address the fermionic case in Section~\ref{sec: stacking}.

\subsection{Example of a critical point: The Majorana CFT}
Having confirmed that the proposed fermionic \SymTO{} formulation reproduces various known gapped phases, we now examine phase transitions between these gapped phases, focusing on the critical point separating the Kitaev and trivial gapped phases with no additional symmetries ($G^F=\bZ_2^F$).
It is well known that one possible phase transition between these phases is a massless Majorana CFT with with fixed-point Lagrangian:
\begin{align}
L = i\int dtdx \left[\psi_R(\d_t-\d_x)\psi_R +  \psi_L(\d_t+\d_x)\psi_L\right]
\end{align}
where $\psi_{L,R}$ are left- and right- moving Majorana (real fermion) fields.
This Majorana CFT is related to the Ising CFT by bosonization (gauging fermion parity by summing over the different boundary conditions for the fermion operators).

In the following, we show that the (ungapped) ``nothing condensed" physical boundary where neither $e$ nor $m$ are condensed is compatible with this CFT, and follow the method developed in~\cite{huang2023topological} to compute the partition function of the Majorana CFT from that for the bosonic Ising CFT. We note in passing, that, the ``nothing condensed" boundary does not fully determine the CFT in the gapless phase, but only topological aspects such as how symmetries are implemented on scaling operators. For example, there are other CFTs (e.g. the tricritical Ising CFT, or non-minimal models) that are also compatible with the $\bZ_2^F$ \SymTO{}, but describe more exotic multi-critical points. We do not consider these more exotic options here.

\subsubsection{Review of (bosonic) Ising CFT}
To set the stage, we briefly review some important facts about the Ising CFT and its \SymTO{} description~\cite{huang2023topological}.
The Ising CFT belongs to the family of rational CFTs (RCFT). An RCFT has finite number of primary operators and the Hilbert space on a circle decomposes into a finite direct sum
\begin{align}
    \H=\bigoplus_{a,b}M_{ab}\H_a\otimes \overline{\H}_b
\end{align}
where $\H_a$ ($\overline{H}_b$) is an irreducible representation of the left (right) chiral algebra and $M_{ab}$ are positive integers. The representations of a chiral algebra form a modular tensor category $\EC$, therefore the labels $a,b$ can be thought of as taking values in anyons of a 2+1D chiral topological order $\EC$ and its time reversal $\overline{\EC}$. This relation between 1+1D RCFT and 2+1D chiral topological orders can also be understood from a sandwich construction. Consider the non-chiral topological order $\Z[\EC]=\EC\boxtimes \overline{\EC}$ and a sandwich construction with physical boundary being the state 
\begin{align}
    |\Psi\>=\sum_{a,b\in \EC}\chi_a\overline{\chi}_b|a,b\>.
\end{align}
Here $\chi_a, \overline{\chi}_b$ are the characters of the chiral,anti-chiral algebra representations respectively. 
The reference boundary can be expanded in the anyon basis as:
\begin{align}
    |\A\>=\sum_{a,b\in \EC}M_{a,b}|a,b\>.
\end{align}
Then the sandwich has partition function 
\begin{align}
    \<\A|\Psi\>=\sum_{a,b}M_{a,b}\chi_a\overline{\chi}_b,
\end{align}
which is the partition function of the RCFT. This sandwich construction can be viewed as the \SymTO{} for the non-invertible categorical symmetry $\EC$.  

As an example, the Ising CFT is a RCFT with three primaries $1,\sigma,\psi$ which form the Ising category. By the above construction it can live on the boundary of the double Ising theory and corresponds to the state
\begin{align}
    |\Psi_{\sf{Ising}}\>=|\chi_1|^2|1,1\>+|\chi_\sigma|^2|\sigma,\overline{\sigma}\>+|\chi_\psi|^2|\psi,\overline{\psi}\>.\label{eq: Ising state}
\end{align}

\subsubsection{Majorana CFT from \SymTO{}}
\begin{figure*}[t]
    \includegraphics[width=1.8\columnwidth]{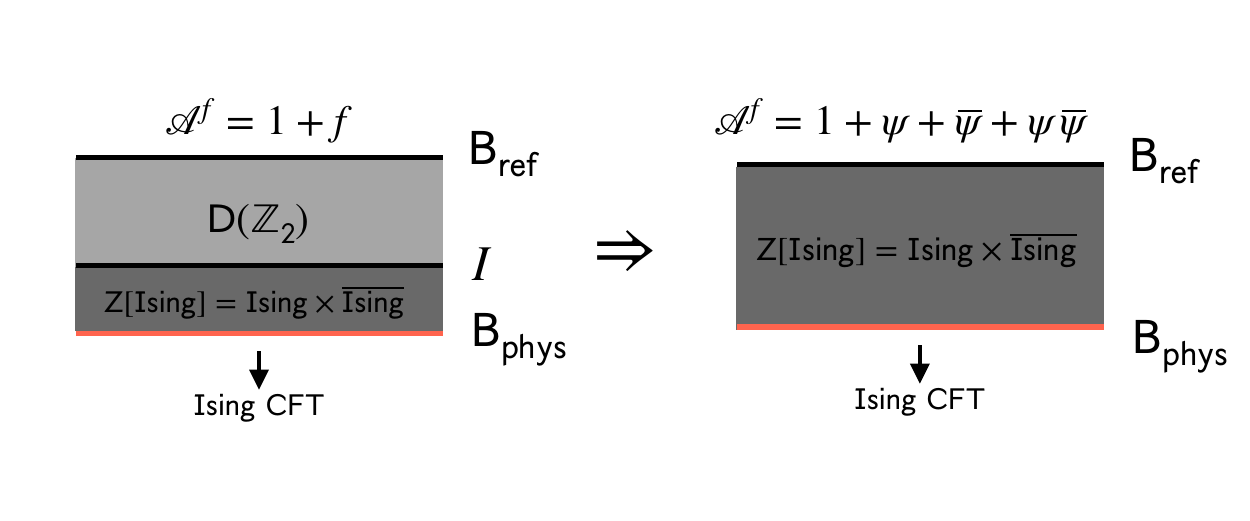}
    \vspace{-0.4in}
    \caption{\textbf{The ``club sandwich'' construction.} Left: Below the domain wall $\mathcal{I}$ is the double Ising theory with Ising CFT living on the physical boundary. Above the domain wall $\mathcal{I}$ is the $\bZ^F_2$-\SymTO{}with fermionic reference boundary $1+f$. When the double Ising layer is thin, this setup is a $\bZ^F_2$-\SymTO{} sandwich describing a gapless $\bZ^F_2$-symmetric(fermionic) system. Right: the domain wall is moved up and fused with the reference boundary. The entire bulk becomes the double Ising theory. The reference boundary has the condensation $1+\psi+\overline{\psi}+\psi\overline{\psi}$.}
    \label{fig:IsingCFT}
\end{figure*}

To make the connection between the Ising CFT and the $\bZ_2^F$-\SymTO{}, let us replace the $\sf{B_{phys}}$ boundary of the \SymTO{} sandwich by a thin slab of Ising CFT. See Fig.~\ref{fig:IsingCFT}
Crucially, there exists an invertible, gapped domain wall, $\mathcal{I}$, between the double Ising theory and the toric code, across which Toric code anyons can  ``tunnel" to those of the Ising quantum double topological order according to:
\begin{align}
    &1\leftrightarrow (1,\overline{1})+(\psi,\overline{\psi})\nonumber\\
    &e\leftrightarrow (\sigma,\overline{\sigma})\nonumber\\
    &m\leftrightarrow (\sigma,\overline{\sigma})\nonumber\\
    &f\leftrightarrow (1,\overline{\psi})+(\psi,\overline{1}).\label{eq: tctods}
\end{align}
without creating excitations at the boundary.
The resulting three layer, ``club sandwich"~\cite{bhardwaj2023club}, with a $\bZ_2^F$ gauge theory interfaced by $\mathcal{I}$ to a slab of Ising quantum double $\sf{Z[Ising]}$, sandwiched between the fermion condensed reference boundary $\sf{B_{ref}}$, and the physical boundary state \eqref{eq: Ising state}, now describes a gapless fermionic \oned{} system. As we will see, instead of the Ising CFT the SymTO sandwich is now equivalent to the Majorana CFT. Crucially, the Majorana CFT is a fermionic CFT whose partition function depends on the spin structure. 

We now compute the partition function of the sandwich to reveal this spin structure dependence. 
First, move the invertible domain wall $\mathcal{I}$ towards $\sf{B_{ref}}$ and fuse it with $\sf{B_{ref}}$. After this the bulk of the sandwich becomes entirely the double Ising theory. According to the rules $1\to 1+\psi\overline{\psi}, f\to \psi+\overline{\psi}$, the fermionic reference boundary condensation $1+f$ becomes $1+\psi\overline{\psi}+\psi+\overline{\psi}$. Therefore our initial club sandwich is topologically equivalent to a double Ising topological order sandwiched between the $1+\psi\overline{\psi}+\psi+\overline{\psi}$ condensed reference boundary and the Ising CFT physical boundary~\eqref{eq: Ising state}. 

The partition function of the sandwich can now be computed as the inner product (recalling that the reference state $|1\>+|f\>$ corresponds to $(AP,AP)$ boundary conditions): 
\begin{align}
    &\Z[T^2,(AP,AP)]\nonumber\\
    &=(\<1,\overline{1}|+\<\psi, \overline{\psi}|+\<\psi,\overline{1}|+\<1,\overline{\psi}|)(|\Psi_{\sf{Ising}}\>)\nonumber\\
    &=|\chi_1|^2+|\chi_\psi|^2+\chi_\psi\chi_1^*+\chi_1\chi_\psi^*=|\chi_1+\chi_\psi|^2.
\end{align}
We obtain other boundary conditions by inserting fermion parity defect lines as we did for the Kitaev chain in~\ref{sec:spin-example}. Consider changing the spatial boundary condition by inserting a symmetry defect in the time direction.  The partition function of the sandwich with $(P,AP)$ boundary condition for the physical fermions is then given by $(\<e|+\<m|)\mathcal{I}(|\Psi_{\sf{Ising}}\>)$. Applying the transformation rule~\eqref{eq: tctods}, this is equal to
\begin{align}
    \Z[T^2,(P,AP)]=(\<\sigma,\overline{\sigma}|+\<\sigma,\overline{\sigma}|)|\Psi_{\sf{Ising}}\>=2|\chi_\sigma|^2.
\end{align}

To change the boundary condition along the time direction we insert a symmetry defect along the spatial direction. The partition function with $(AP,P)$ boundary condition is then
\begin{align}
    &\Z[T^2,(AP,P)]\nonumber\\
    &=(\<1|+\<f|)W^{x}_m \mathcal{I}|\Psi_{\sf{Ising}}\>\nonumber\\
    &=\<1|\mathcal{I}|\Psi_{\sf{Ising}}\>-\<f|\mathcal{I}|\Psi_{\sf{Ising}}\>\nonumber\\
    &=|\chi_1|^2+|\chi_\psi|^2-\chi_\psi\chi_1^*-\chi_1\chi_\psi^*=|\chi_1-\chi_\psi|^2.
\end{align}
Here $W^x_m$ is an $m$-loop in the spatial direction.
Lastly, we obtain the $(P,P)$ boundary condition by inserting symmetry defects in both space and time directions. We have the partition function: $\<e|\mathcal{I}|\Psi_{\sf{Ising}}\>-\<m|\mathcal{I}|\Psi_{\sf{Ising}}\>=|\chi_\sigma|^2-|\chi_\sigma|^2=0$.

In summary, the \SymTO{} computation gives
\begin{align}
    &\Z[T^2,(AP,AP)]=|\chi_1+\chi_\psi|^2,\nonumber\\
    &\Z[T^2,(P,AP)]=2|\chi_\sigma|^2,\nonumber\\
    &\Z[T^2,(AP,P)]=|\chi_1-\chi_\psi|^2,\nonumber\\
    &\Z[T^2,(P,P)]=0
\end{align}
This matches perfectly with the torus partition functions of the Majorana CFT~\cite{thorngren2019anomalies}.

\section{Intrinsically fermionic and gapless SPTs from \SymTO{}\label{sec:gSPT}}
As we reviewed in Section~\ref{sec:bg}, \SymTO{} provides a full characterization for bosonic gapless SPTs, including their classification, edge modes, group extension structures, and emergent anomalies.

Given such success, it is natural to ask whether  there is also a correspondence between \SymTO{} and fermionic gSPT,  whether the \SymTO{} could assist us in constructing new models of fermionic gSPTs and understand topological aspects of them, etc. We provide evidence by studying a \SymTO{} sandwich construction, which turns out to describe a fermionic gapless SPT with symmetry $\bZ_8\times \bZ_2^F$ and an emergent $k=2$ $\bZ_2\times \bZ_2^F$ fermionic anomaly. We show that this igSPT has the intriguing property of having a half-charged Majorana edge mode. 

Previously, models of fermionic igSPTs~\cite{Thorngren_2021} have only realized bosonic anomalies: i.e. are essentially bosonic igSPTs realized from the bosonic (spin- or Cooper pair) degrees of freedom in a Mott insulator of fermions.
While this may give a physical mechanism for their experimental realization, the resulting low energy properties (edge states and gapless modes) are indistinguishable from those of a purely bosonic system.
Therefore it is an interesting question whether there exists an ``intrinsically-fermionic" igSPTs whose topological features can not arise in a purely bosonic system, but involve gapless fermion degrees of freedom in the IR.  Below we construct precisely such an intrinsically fermionic igSPT via the \SymTO{} methods developed above. 
While lattice models and field theories of this state could in principle be constructed without reference to \SymTO{}, we found that the \SymTO{} actually serves as a useful algebraic, and graphical tool to quickly prototype these phases. In this way, the \SymTO{} can also be a practical technique for constructing new physics, as well as re-interpreting known phases.

We consider the gapless sector of the igSPT to have the fermionic symmetry $\bZ_2\times \bZ_2^F$ and an emergent anomaly. Anomalies of this symmetry in \oned{}~ have an $\bZ_8$-classification. The $\nu\in \bZ_8$ anomaly can be realized by $\nu$ ``flavors" of Majorana CFTs: 
For the case of interests to us here,  anomalies of $\bZ_2\times \bZ_2^F$ correspond to \twod~ $\bZ_2$-symmetric topological superconductors, which can be viewed as $\nu$ stacks of $\bZ_2$-neutral $p+ip$ superconductor and $\bZ_2$-charged $p-ip$ superconductor pairs.  
The symmetry-preserving gapless edge of these 2+1D SPTs are described by the field theory:
\begin{align}
L = \sum_{a=1}^{\nu} i\int dtdx& [\psi_{R,a}(\d_t-\d_x)\psi_{R,a}\nonumber\\
 &+\psi_{L,a}(\d_t+\d_x)\psi_{L,a}]
\label{eq:flavored_majoranas}
\end{align}
where $a$ is a flavor index, $\psi_{L/R,a}$ are left/right (L/R) moving Majorana fermions, and the anomalous $\bZ_2\times \bZ_2^F$ symmetry acts by: 
\begin{align}
\begin{pmatrix}
\psi_{L,a} \\ \psi_{R,a}
\end{pmatrix} 
\overset{\bZ_2}\longrightarrow 
\begin{pmatrix}
\psi_{L,a} \\ -\psi_{R,a}
\end{pmatrix}, 
~~~~
\begin{pmatrix}
\psi_{L,a} \\ \psi_{R,a}
\end{pmatrix} 
\overset{\bZ_2^F}\longrightarrow 
\begin{pmatrix}
-\psi_{L,a} \\ -\psi_{R,a}
\end{pmatrix}. 
\end{align}
With interactions, the anomalies have a $\bZ_8$ group structure (topologically $\nu\simeq \nu+8$)~\cite{Cheng_2018,Kapustin_2015,Wang_2018}.

Previously we have established the \SymTO{} for non-anomalous $G^F$ symmetry. In order to describe fermionic igSPTs via \SymTO{}we need to discuss the description of anomalous fermionic symmetry in \SymTO.

\subsection{Anomalous fermionic symmetries in \SymTO{}\label{sec:anomalous SymTFT}}
The \SymTO{} of a system with an anomalous bosonic symmetry is generally given by the twisted gauge theory obtained from gauging the corresponding higher-dimensional SPT whose boundary realizes the anomalous symmetry.  This can be understood from the fact that any \oned~ anomalous system can be realized on the boundary of a \twod~ SPT, and the \SymTO{} is simply obtained by gauging this \twod~ SPT. Following this principle we may formulate \SymTO{} for an anomalous fermionic symmetry(with no gravitational anomaly) as follows. An anomaly of $G^F$ in \oned{}~ corresponds to an SPT of $G^F$ in \twod, i.e. a (non-chiral)topological superconductor with symmetry $G^F$. Then the \SymTO{} is given by gauging this topological superconductor, which is a non-chiral topological order that we will call a twisted $G^F$-gauge theory, although the twist here no longer has a cohomology classification.  Independent of the twist, the gauge charges of the twisted $G^F$-gauge theory have the same structure as those of an untwisted one, the super-Tannakian category $\rep(G,P_f)$. Namely the gauge charges are labelled by representations of the group $G^F$, and a gauge charge in the representation $R$ is bosonic/fermionic if fermion parity is represented as $+1/-1$. Therefore we can still define the reference boundary of the \SymTO{} sandwich as the one where all gauge charges are condensed(with the understanding that fermionic charges are condensed together with a local fermion introduced on the reference boundary). This completes our construction for anomalous fermionic symmetry.

Recall that the \twod{} SPT corresponding to the level $\nu$ anomaly of $\bZ_2\times \bZ_2^F$ is $\nu$ stacks of $\bZ_2$-neutral $p+ip$ superconductor and $\bZ_2$-charged $p-ip$ superconductor pairs. The theory obtained by gauging $G^F=\bZ_2\times \bZ_2^F$ in the 2+1D bulk of this SPT can be deduced from the stacking structure; this topological order will become the \SymTO{} for the anomalous $\bZ_2\times \bZ_2^F$ symmetry. 
Denote the symmetry ($\bZ_2$) flux by $\Phi$, and fermion-parity ($\bZ_2^F$) flux by $\Phi_F$. Whereas $\Phi_F$ effects both the $p\pm ip$ layers (all fermions are charged under $\bZ_2^F$), the symmetry flux, $\Phi$, only affects the $p-ip$ layers.
For a single $p\pm ip$ superconductor, the flux (superconducting vortex), carries an unpaired Majorana zero mode with topological spin $e^{\pm i\pi/4}$.
Thus, for $\nu=1$, the $\Phi$ flux will be such a non-Abelian particle, and the corresponding anomalous edge will also have a non-Abelian character that prevents it from having a tensor-product Hilbert space structure.
We anticipate that this obstacle will prevent the realization of the $\nu=1$ edge anomaly as an igSPT -- since this anomaly has a ``gravitational" character that cannot be alleviated merely by extending the symmetry group.
Hence, we instead focus on even $\nu$. 

The $\nu=4$ ($n=2$) anomaly is a self-anomaly of the bosonic, $\bZ_2$, part of the symmetry group and does not involve $\bZ_2^F$. Equivalently, the 2+1D $\nu=4$ SPT is topologically-equivalent to a bosonic SPT, and the fermions are merely spectators.
Therefore, the only interesting candidate for an intrinsically-fermionic igSPT is $\nu=2$ ($\nu=6\simeq -2$ will be similar), on which we now focus.

We refer to the 2+1D topological order obtained by gauging the symmetry of this $\nu=2$ SPT: $D^F[\bZ_2\times \bZ_2^F,\nu=2]$. 
The simple anyons are generated by fluxes $\phi,\phi_F$, and their corresponding charges, denoted by $b$ and $f$. The $\bZ_2$ symmetry flux is a $\pi$-flux in two $p+ip$ layers. Since a stack of two $p+ip$ superconductors is equivalent to an integer quantum Hall state with unit Hall conductance, this $\pi$ flux therefore binds $1/2$ of a symmetry-charge and acquiring quantum spin $e^{i\pi /4}$. 
The $\bZ_2^F$-flux is a $\pi$ flux in all $p+ip$ and $p-ip$ layers. It therefore induces equal and opposite contributions to spins from the $+$ and $-$ layers, and is a boson with fractional $\bZ_2$ charge (from the $+$ layers).
The $\Phi^2$ is a $2\pi$-flux in two $p+ip$ layers, therefore it binds a charged electron and is identified with $f\times b$. $\phi_F^2$ is a boson, but it binds a charged electron and 7 uncharged electrons, therefore it is charged under $\bZ_2$ and is identified with $b$. Thus the gauged $\nu=2$ SPT is completely generated by $\Phi$ and $\Phi_F$. They both have order 4. The braiding phase between $\Phi_F$ and $\Phi$ is the same as the braiding phase between two $\Phi$s, which is square of the exchange phase: $\theta_{\phi,\phi_F}=e^{i\pi /2}$. This completes the construction of the gauged $\nu=2~\bZ_2\times \bZ_2^F$ SPT, which is the \SymTO{} for the $\nu=2$ anomalous $\bZ_2\times \bZ_2^F$. 

\subsection{Fermionic igSPT with $\bZ_8$ symmetry}
The topological properties of igSPTs stem from an emergent anomalous symmetry, $G^F$. Instead of arising at the boundary of a higher-dimensional gapped $G^F$ SPT, the anomaly is ``cured" by gapped degrees of freedom that transform under a larger group $\Gamma^F$, which is an extension of $G^F$. In the \SymTO{} description, this requires that an untwisted $\Gamma^F$ gauge theory can be reduced to a twisted $G^F$ gauge theory by partially-confining (non-Lagrangian) anyon condensation, $\A$.

For the intrinsically-fermionic $\nu=2$ $G^F=\bZ_2\times \bZ_2^F$ anomaly above, one solution is to extend the $\bZ_2$ symmetry by $\bZ_4$ to $\Gamma^F = \bZ_8\times \bZ_2^F$. In other words, a fermionic system with $\bZ_8$ symmetry can exhibit an intrinisically-fermionic igSPT with the $\nu=2$ $\bZ_2\times \bZ_2^F$ anomaly.

\paragraph{Lifting the anomaly}
Denote the anyons of $D(\bZ_8 \times \bZ_2^F)$ as $a_{1,2}$ with $a\in {1,e,m}$ and composites thereof, and where $1,2$ subscripts refer to the $\bZ_8,\bZ_2^F$ gauge subgroups respectively. As above, denote the canonical fermion charge as $f_2=e_2m_2$. Then, a suitable partially-confining condensation that yields the desired twisted  $D(\bZ_2\times \bZ_2^F,\nu=2)$ topological order is obtained by condensing the anyon $a =e_1^6m_1^2e_2m_2$. The remaining deconfined anyons are generated by $\phi = e_1m_1$, and $\phi_F = e_1^2m_2$, and are topologically identical to those of the anomalous \SymTO{} $D(\bZ_2\times \bZ_2^F,\nu=2)$.

\paragraph{Sandwich construction}
A sandwich construction of the igSPT is obtained by choosing a $\Gamma^F=\bZ_8\times \bZ_2$ 
gauge theory for the bulk \SymTO{}, the canonical reference boundary $\A_{\sf{ref}}^f = \<e_1,e_2m_2c\>$, and the partially-confining (non-Lagrangian) physical boundary, $\A_{\sf{phys}} = \<e_1^6m_1^2e_2m_2\>$.
The physical boundary is necessarily gapless (or symmetry broken). A candidate field theory that matches the anomaly is given by two copies of the Majorana CFT (Eq.~\ref{eq:flavored_majoranas} with $\nu=2$), or equivalently a massless Dirac fermion with a chiral symmetry anomaly.

The $\bZ_8$ symmetry is generated by inserting a horizontal $m_1$ line ``condiment" in the sandwich. For the igSPT sandwich, a pair of symmetry generators, $m_1^2$ line, can be freely absorbed into the boundaries, by splitting $m_1^2 = a \times (e_1^6f_2)$ and absorbing the $a$ into the physical boundary and the $e_1^6f_2$ into the reference boundary. Thus, the $\bZ_8$ symmetry is effectively reduced to an IR $\bZ_2$ symmetry. Denoting elements of $\Gamma^F=\bZ_8\times \bZ_2^F$ as $(n,m)$, the group extension structure is given by the short exact sequence: 
\begin{align}
   1\to \bZ_4\xrightarrow{i} \bZ_8\times \bZ_2^F\to\bZ_2^{\text{IR}}\times \bZ_2^F\to 1.
\end{align}
Here the embedding map $i$ is $\bZ_4\ni 1\to (2,0)\in \bZ_8\times \bZ_2^F$.

\begin{figure*}[t]
    \includegraphics[width=1\columnwidth]{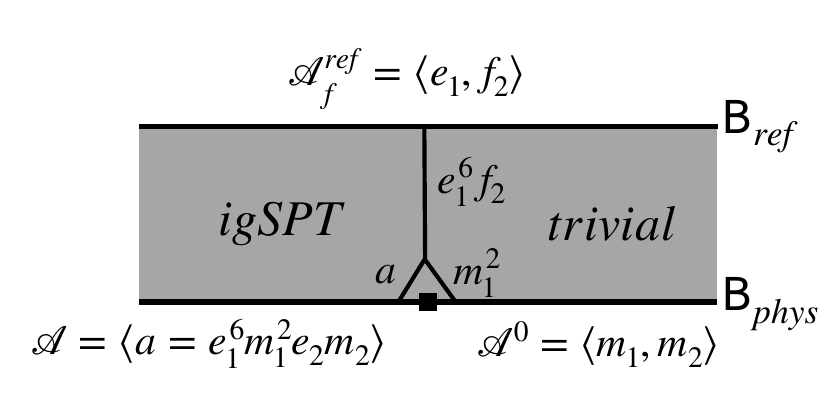}
    \caption{\textbf{Edge mode of the $\bZ_8\times \bZ_2^F$ igSPT.} The edge mode is a verticle $e_1^6f_2$ line at the interface between igSPT and trivial phase. Ths operator has charge $1/2$ under the $\bZ_2^{\text{IR}}$ symmetry(a horizontal $m_1$-line), and charge $1$ under fermion parity(a horizontal $m_2$-line).}
    \label{fig:fermion_igspt_edge}
\end{figure*}

\paragraph{Edge modes} A defining property of igSPTs is the presence of symmetry-protected topological edge states, that, despite the gapless bulk, are confined to the edge. Moreover, these edge states cannot be realized in any gapped topological phase with the same symmetry.
In the \SymTO{} sandwich construction, edge modes are revealed by consider a spatial interface between two different gapped physical boundaries : a trivial phase described by condensing symmetry fluxes: $\A^0 = \<m_1,m_2\>$, and the igSPT boundary, $\A = \<a=e_1^6m_a^2f_2\>$~ (Fig~\ref{fig:fermion_igspt_edge}).
The short string segment of $e_1^{6}f_2$ localized near the trivial/igSPT interface represents a zero-energy edge state. Decorating the $\sf{B_{ref}}$ end of this segment by a local (Majorana) fermion, $\gamma = \frac12(c+ic^\dagger)$, allows it to be absorbed into the fermionic condensate. At $\sf{B_{phys}}$, the segment can split into an $a$ anyon and an $m_1^2$ excitation that can respectively be absorbed into the igSPT ($\A$) and trivial $(\A^0)$ condensates. Since the ends of the segment can be absorbed, the resulting operator does not produce an energetic excitation. Moreover, the segment operator has a non-trivial symmetry action on the ground-space of the system. The symmetry generators are horizontal $m_1$ or $m_2$ lines. The zero mode operator anticommutes with the $m_2$ line, as required since the mode is a fermion and is charged under the physical fermion parity. Moreover, the zero mode operator has exchange phase $ (-i)=e^{-2\times 2\pi i/8}$ with $m_1$, indicating that it carries ($-2$) units of the $\bZ_8$ symmetry charge (in terms of the IR $\bZ_2$ symmetry, the fermionic zero mode appears to carry a ``half" symmetry charge). 

\paragraph{Ground-state degeneracy with open boundaries}
For an igSPT of length $L$, with open boundary conditions (surrounded by trivial phases), the igSPT edge zero-modes span a four-fold ground-space degeneracy. Defining this ground-space degeneracy for a gapless system takes some care. In a system of length for the multi-component Majorana CFT the bulk gap scales $\sim 1/L$, whereas the topological ground-space degeneracy is exponentially small $\sim e^{-L/\xi}$ where $\xi$ is a microscopic correlation length related to the $a$-condensation energy scale. 
To see this, note that the edge mode operator shown in Fig.~\ref{fig:fermion_igspt_edge} fails to commute up to an $e^{2i\pi/4}$ phase with a horizontal $m_1$ line (which measures the symmetry charge). This algebra cannot be represented on a unique ground-state, but rather requires a minimum of four degenerate ground-states.
Importantly, unlike for gapped SPTs, this ground-space degeneracy does not factorize into a pair of local Hilbert spaces for each edge. Rather, there is a single, non-local dimension-four ground-space shared by the entire chain. 

\subsection{Field theory description}
Despite its purely formal appearance, the \SymTO{} sandwich construction actually directly hints at a potential microscopic mechanism for realizing the igSPT phase. 
The insights from \SymTO{} can also guide the construction of lattice-models and field-theory descriptions, as we now illustrate.

Consider a \oned{} fermion chain with $\bZ_8$ symmetry. The $e_1^6m_1^2e_2m_2$ condensation can then be interpreted via the \SymTO{} dictionary between anyons and generalized charges. The $m_1^2$ anyon represents a strength-two domain wall of the $\bZ_8$ symmetry.   Thus the $e_1^6m_2^2$ part of the condensate corresponds to decorating strength two domain walls with $6=-2$ symmetry charges of the $\bZ_8$. 
This decorated domain wall has fermionic statistics, which can be screened by binding them to a local fermion (recall that in the fermion \SymTO{} construction, the $e_2m_2$ anyon is the ``shadow" of the local fermion excitation on the physical boundary), and then condensing this composite object.

The \SymTO{} picture suggests a simple bosonized field theory description of the igSPT phase.
We start from a pair of Luttinger liquids, consisting of a $\bZ_8$-charged boson field $e^{i\varphi_1}$, and its dual vortex $e^{i\varphi_2}$, and a left and right moving fermion field $\psi_{L,R} = \eta_{L,R}e^{i\phi_{L,R}}$ where $\eta$ are anticommuting Klein factors. The (Euclidean/imaginary-time) action density for this theory reads:
\begin{align}
    \mathcal{L}_0 =& \frac{i}{4\pi} \sum_{I,J=1,2}\d_\tau \varphi_I \sigma^x_{IJ} \d_x\varphi_J + 
    \nonumber\\ 
    &+\frac{i}{4\pi} \sum_{I,J =L,R} \d_\tau\phi_{I}\sigma^z_{I,J}\d_x\phi_J +H_v
    ,
\end{align}
where $H_v$ is the Hamiltonian containing (non-universal) terms that describe the velocity of excitations such as $v_{IJ}\d_x\varphi_I\d_x\varphi)_J$ where $v_{IJ}$ is a symmetric, positive matrix (and similarly for the $\Phi$ fields), and $\sigma^{x,z}$ are Pauli matrices in the $I,J$ flavor space(s).
Quantizing the theory yields commutation relations: $e^{i\alpha\d_x\varphi_{I}(x)/2\pi}\varphi_{J}(y)e^{-i\alpha\d_x\varphi_{I}(x)/2\pi} = \varphi_{J}(y)+\alpha\sigma^x_{IJ}~\delta_{x,y}$, and (and similarly for $\phi_I$ with $\sigma^x_{IJ}\leftrightarrow \sigma^z_{IJ}$), where $\alpha$ is a constant.

The $\bZ_8$ symmetry generator, $g$, and $\bZ_2^F$ generator, $(-1)^F$ acts on the fields as:
\begin{align}
    g&:\varphi_1 \rightarrow \varphi_1+2\pi/8,
    \nonumber\\
    (-1)^F&:\phi_{L/R} \rightarrow \phi_{L,R}+\pi
    \label{eq:LLsym}
\end{align}
(other, unlisted fields transform trivially under the symmetry in question).
Crucially this symmetry action is non-anomalous, and $\mathcal{L}_0$ may emerge as the continuum description of a lattice model with on-site symmetry action.

\paragraph{Domain wall operators}
According to the commutation relations above, in the quantized theory, these global symmetry transformations restricted to the interval $I=[x_1,x_2]$ are implemented by operators $g|_I = e^{i\int_{a}^bdx \d_x \varphi_2/8}$ and $(-1)^F|_I = e^{i\int_{x_1}^{x_2}(\phi_R-\phi_L)/2}$. Correspondingly, the operators $m_1(x) = e^{i\varphi_2(x)/8}$ and $m_2(x) = e^{i(\phi_R(x)-\phi_L(x))/2}$, create symmetry domain walls at position $x$.

If the symmetry were gauged, these domain wall operators would correspond to boundary-termination of a a gauge-flux ($m_{1,2}$) line operator.
From this identification, one can immediately recognize a local operator that plays the role of the decorated symmetry domain wall $a=m_1^2e_1^{-2}f_2$ whose condensation in the \SymTO{} produces the igSPT boundary with anomalous $\bZ_2\times\bZ_2^F$ symmetry action as:
$$ 
a(x) = e^{i\left(
2\varphi_2(x)/8-2\varphi_1(x)+\phi_R(x)
\right)}.
$$
This operator has trivial self-statistics, $[a(x),a(y)]=0$. 
The operator is non-local due to the fractional coefficient in front of the $\varphi_2$ field, and cannot be directly added to the Hamiltonian, however its fourth power can. 
Specifically, the $a$-anyon condensation can be implemented by adding a Hamiltonian term:
\begin{align}
H_{\rm igSPT} = -\Delta\int dx \cos\left[
\varphi_2(x)-8\varphi_1(x)+4\phi_R(x)
\right]
\end{align}
where $\Delta$ is an energy scale (depending on the Luttinger liquid parameters, this term may or may not be perturbatively relevant, however, it can always be made relevant by cranking up the coefficient $\Delta$).
$H_{\rm igSPT}$ is clearly invariant under the terms in Eq.~\ref{eq:LLsym}, and when relevant, gaps out the linear combination of fields appearing in the cosine term, and also fields that are conjugate to this linear combination.

\paragraph{Emergent anomaly}
A basis for the resulting low energy fields that commute with $H_{\rm igSPT}$ is:
\begin{align}
    \tilde{\phi}_R &= \phi_R-4\varphi_1
    \nonumber\\
    \tilde{\phi}_L &= \phi_L
    ,
\end{align}
and an effective low-energy (IR) action-density for the null-space of $H_{\rm igSPT}$ takes the form of a (bosonized) massless Dirac fermion: $$\mathcal{L}_{\rm IR} = \frac{i}{4\pi}\sum_{IJ}\d_\tau \tilde{\phi}_I\sigma^z_{IJ} \tilde{\phi}_J.
$$
The operators $\eta_{L,R}e^{\pm\tilde{\phi}_{L,R}}$ are fermionic fields, and their arguments transform under the symmetry as:
\begin{align}
    g&:\begin{pmatrix}
    \tilde{\phi}_R \\ 
    \tilde{\phi}_L
    \end{pmatrix}
    \rightarrow 
    \begin{pmatrix}
    \tilde{\phi}_R -\pi\\ 
    \tilde{\phi}_L
    \end{pmatrix}
    \nonumber\\
    (-1)^F&:\begin{pmatrix}
    \tilde{\phi}_R \\ 
    \tilde{\phi}_L
    \end{pmatrix}
    \rightarrow 
    \begin{pmatrix}
    \tilde{\phi}_R +\pi\\ 
    \tilde{\phi}_L +\pi
    \end{pmatrix}
\end{align}
which is indeed the anomalous $\bZ_2\times \bZ_2^F$ symmetry action for the edge of the $\nu=2$ fermion SPT~\cite{Else_2014}.

\paragraph{Edge modes} The igSPT edge modes can be identified by considering an interface with $H_{\rm igSPT}$ added for $x>0$, and a trivial mass term for $x<0$:
\begin{widetext}
\begin{align}
H_{\rm interface} = -\Delta_{\rm igSPT} \int_0^L dx\cos\left(
\varphi_2-8\varphi_1+4\phi_R
\right) - 
\Delta_{\rm trivial} \int_{x\notin [0,L]} dx \left[
\cos\varphi_2 + \cos(\phi_L+\phi_R)
\right]
\end{align}
\end{widetext}

To  simplify the analysis, we take the extreme limit $\Delta\rightarrow \infty$ (or equivalently consider the RG fixed point limit where these terms are relevant).
In this limit, the field configurations are pinned locally to the minima of the cosine. For the igSPT region the ground-space is given by the four distinct minima of the corresponding cosine:
\begin{align}
    \phi_R = 2\varphi_1-\varphi_2/4 + \frac{n\pi}{2}
\end{align}
The operator $e^{2i\varphi_1(x)}$ creates a kink between these minima whenever $0<x<L$, which costs a finite energy due to the gradient terms in $H_v$, except precisely at the edges: $x=0$ or $x=L$.
In the trivial regions, there is a unique local ground-state given by $\phi_L=-\phi_R$ and $\varphi_2=0$.
The operators $e^{\varphi_1(x)}$ and $e^{i\phi_{L,R}(x)}$ create local Kink excitations in these ground-state configurations, except when inserted exactly at the interface.

In the \SymTO{}, the zero-mode operator is implemented by an $e_1^{-2}f_2$ line segment
 that splits into an $a=m_1^2e_2^{-2}f_2$ anyon that terminates on the igSPT side, and an $m_1^{-2}$ anyon that terminates on the trivial side, and which is decorated by a local fermion on the physical boundary
(Fig.~\ref{fig:fermion_igspt_edge}), decorated by a local fermion on the physical boundary. 
Translating these operators into the continuum theory gives a zero-mode operator:
\begin{align}
    \Psi^\dagger(0,L) = \eta_R\exp\left\{i\left[
    \phi_R(0,L)-2\varphi_1(0,L)
\right]\right\}
\end{align}
$\Psi^\dagger$
commutes with the cosine terms in both $H_{\rm igSPT}$ and $H_{\rm trivial}$, i.e. it is indeed a zero mode.
Crucially, this operator is only a zero mode when inserted at the igSPT/trivial interface ($x=0$ or $L$). Otherwise, as described above, the term creates (massive) kinks in the local ground-state configuration of the fields. Moreover, by inspection, $\Psi^\dagger$ has odd fermion parity and adds $\bZ_8$ two units of $\bZ_8$ symmetry charge ($g^\dagger\Psi^\dagger g = e^{i\pi/2}\Psi^\dagger$).
As a technical comment, the Klein factor $\eta_R$ is required to ensure that $\Psi$ anticommutes with all local fermion excitations such as $\eta_L e^{i\phi_L(x)}$ (as physically required for any odd-fermion parity excitation).
\drew{}

\section{\SymTO{} for fermions: Systematics and Technical Details\label{sec:general}}
Having illustrated the fermionic \SymTO{} construction for a variety of specific examples, we now generalize this framework to general finite, unitary fermionic symmetries.

Conceptually, the construction proceeds much as the case without symmetry above, with some technical modifications. The bulk \SymTO{} is a $G^F$ gauge theory, whose symmetry charges divide into bosonic and fermionic ones. At the fermionic reference boundary, in the sandwich constructions, one condenses all symmetry gauge charges by binding the fermionic ones to a local fermion $c$ to convert them into condensable bosons. This reference boundary is again fermionic and requires choosing a $G^F$-spin structure.
The remainder of this section is devoted to making the qualitative description above mathematically explicit.

\subsection{General symmetries\label{sec: FSymTO general}}
Having worked out the topological holography for fermions without a general symmetry, we now consider the \SymTO{} for a general fermionic symmetry $G^F$. 
The fermion parity generates a central subgroup $\bZ_2^F\lhd G^F$, and the quotient $G^B = G^F/\bZ_2^F$ is referred to as the ``bosonic part'' of the symmetry. 
Consider a \oned{} fermionic system $\sf{T}$ with symmetry $G^F$ living on $M$. To start, we assume the symmetry is not anomalous. Following the construction of \SymTO{} for bosonic symmetries, we may take the \twod{} \SymTO{} for $G^F$ to be the $G^F$-gauge theory. To complete the construction we once again consider the sandwich construction. We must choose an appropriate reference boundary condition so that the sandwich is equivalent to the original system $\sf{T}$.

\subsubsection{Condensation on the reference boundary\label{sec:general condensation}}

Symmetry ``charges" of $G^F$ corresponded to representations of of $G^F$, which are labeled by a $\bZ_2^F$ fermion parity $|R| = 0/1$ when the representation $R$ has even/odd fermion parity respectively.
Fermi-statistics requires that braiding two symmetry charge excitations with representations $R,R'$ results in the exchange phase $(-1)^{|R|\cdot|R|'}$, and similarly charges should have quantum spin (self-statistics) $(-1)^{|R|}$.
Formally, the local charges of $G^F$ form the super-Tannakian category $\rep(G^F,P_f)$~\cite{Bruillard_2017}.

Recall anyons in $D(G^F)$ are labeled by pairs $(\sigma, r)$, where $\sigma$ is a conjugacy class of $G^F$ and $r\in\rep(\C_G(\sigma))$ is a representation of the centralizer of $\sigma$. The simple anyons correspond to the cases where the conjugacy class is minimal (not a union of two smaller conjugacy classes) and the representation $r$ is irreducible. When the conjugacy class is the trivial one $\sigma=\{1\}$, the anyons $(1,R\in \rep(G))$ are called charges and we denote them by $e_R$. When the representation $r$ is the trivial one, the anyons $(\sigma,1)$ are called fluxes and we denote them by $m_\sigma$. Anyons that are neither charges nor fluxes are called dyons. For details of quantum double models, see Appendix~\ref{app:DG algbera}.

Since all the gauge charges in $D(G^F)$ are bosonic, the set of all gauge charges $e_R, R\in \rep(G^F)$ does not have the correct structure of $\rep(G^F,P_f)$.
Then, to transmute the statistics of an odd fermion parity gauge-charge charge into a fermion we can bind a fermion parity flux to it to form a fermionic dyon. Thus the condensation having the structure of $\rep(G^F,P_f)$ is given by the algebra
\begin{align}
    \A_{\sf{ref}}^f=\left(\bigoplus_{|R|=0}d_R e_R\right)\bigoplus\left(\bigoplus_{|R|=1}d_R e_Rm^F\right).
    \label{eq:fref}
\end{align}
where $d_a$ is the quantum dimension of anyon, $a$.
The $e_Rm^F$ anyons are generalization of the $f$-particle of toric code.  Similar to the $\bZ_2^F$-\SymTO{} case, the fermionic anyons $e_Rm^F$ must be bound to a local fermion $c$ in order to obtain a condensable object with bosonic statistics.

\begin{figure*}[t]
    \includegraphics[width=1.5\columnwidth]{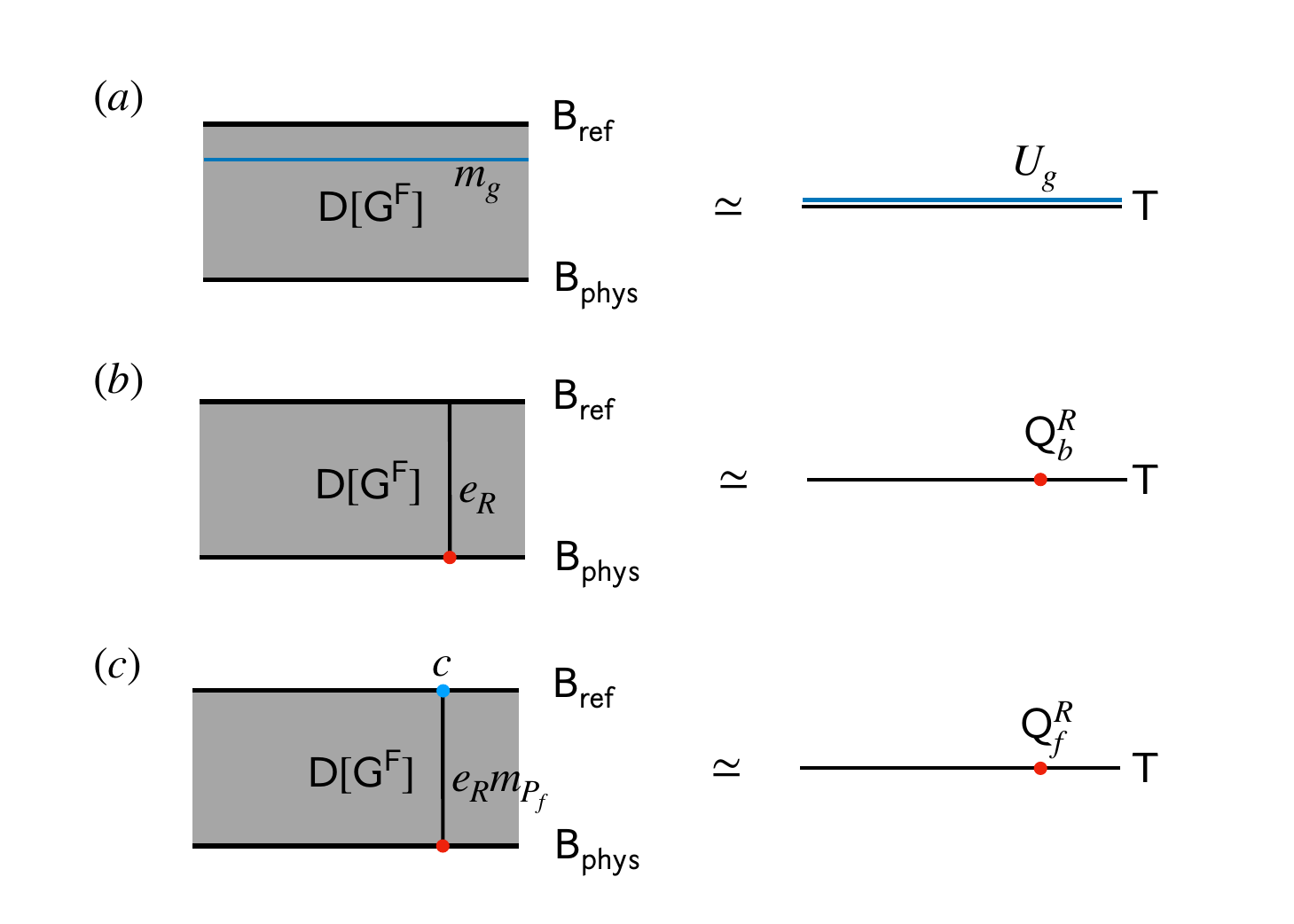}
    \caption{\textbf{\SymTO{} for $G^F$.} (a). A flux line near the reference boundary becomes the symmetry defect of $\sf{T}$ after dimensional reduction. (b). A vertical charge line carrying even representation, $R(P_f)=1$ becomes a local bosonic charge of $\sf{T}$ after dimensional reduction. (c). An odd charge line, $R(P_f)=-1$, is bound with the fermion parity flux line, and its endpoint on $\sf{B_{ref}}$ is dressed by a local fermion operator(blue dot). After dimensional reduction this line becomes a  charged local fermion. The statistics of these local fermions is guaranteed by the local fermion operator on $\sf{B_{ref}}$(blue dot).  The red dots stand for excitations.}
    \label{fig:fsetup4}
\end{figure*}

Since the condensation involves fermions, the algebra $\A_{\sf{ref}}^f$ is not a usual commutative Frobenius algebra. Instead it has the structure of a super-commutative (i.e. with $\bZ_2^F$ graded commutation relations) Frobenius super-algebra~\cite{Aasen_2019,Lou_2021}. Specifically, we can define even and odd sectors $\A_{\sf{ref}}^{f,0}, \A_{\sf{ref}}^{f,1}$ that contain the bosonic and fermionic charges respectively.
Let us rewrite $\A_{\sf{ref}}^f$ to reveal its super-structure. The even sector is isomorphic to the group algebra $\bC[G^F]$ tensored with the trivial $\bZ_2^F$-representation: 
\begin{align}
    \A_{\sf{ref}}^{f,0}=\left(\bigoplus_{R,|R|=0}d_R e_R\right)=\bC[G]\otimes_{\bZ_2^F}\bC_+.
\end{align}
Here $\bC_+$ is the trivial 1d representation of $\bZ_2^F$: $P_f\cdot z=z, \forall z\in \bC_+$. $\bC[G]\otimes_{\bZ_2^F}\bC_+$ is spanned by $|g\>\otimes 1$ modulo the relation $|P_f g\>\otimes 1\cong |g\>\otimes P_f\cdot 1=|g\>\otimes 1$. If we decompose the first factor $\bC[G]$ into direct sum of irreps of $G^F$, then the sector $V^R\otimes_{\bZ_2^F}\bC_+$ is projected out if $R$ is odd. Similarly the odd sector $\A_{\sf{ref}}^{f,1}$ is the tensor product between $\bC[G]$ and the non-trivial representation of $\bZ_2^F$:
\begin{align}
    \A_{\sf{ref}}^{f,1}=\left(\bigoplus_{R,|R|=1}d_R e_Rm^F\right)=\bC[G]\otimes_{\bZ_2^F}\bC_-.
\end{align}
$\bC[G]\otimes_{\bZ_2^F}\bC_-$ is spanned by $|g\>\otimes 1$ modulo the relation $|P_f g\>\otimes 1\cong |g\>\otimes P_f\cdot 1=-|g\>\otimes 1$. The even representations in $\bC[G]$ are projected out by the $\bC_-$ factor. Put these two sectors together we arrive at a compact expression
\begin{align}
    \A_{\sf{ref}}^f&=\bC[G]\otimes_{\bZ_2^F} (\bC_+\oplus \bC_-)\nonumber\\
    &=\bC[G]\otimes_{\bZ_2^F} \bC^{1|1}.
\end{align}
We used $\bC^{1|1}$ to denote the super vector space with a 1d even sector and a 1d odd sector. $\bC^{1|1}$ carries a grading-preserving algebra structure, generated by a single Majorana fermion operator $\gamma$ in the odd sector. The grading of $\A_{\sf{ref}}^f$ is then inherited from that of $\bC^{1|1}$. The algebra structure of $\A_{\sf{ref}}$ is defined by
\begin{align}
    &(|g\>\otimes_{\bZ_2^F} \gamma^i )\cdot(|h\>\otimes_{\bZ_2^F} \gamma^j )=
    \nonumber\\
    &~~~~\delta_{g,h}(-1)^{j\cdot gh^{-1}}|g\>\otimes_{\bZ_2^F} \gamma^{i+j},\label{eq:product}
\end{align} 
where $\delta_{g,h}$ is 1 if $gh^{-1}$ is in $\bZ_2^F$ and 0 otherwise. It is not hard to verify that this algebra is connected and separable.
In order for $\A_{\sf{ref}}^f$ to define a consistent condensation the algebra also needs to be super-commutative, this means the following diagram commutes,
\[\begin{tikzcd}[ampersand replacement=\&]
	{\A_{\sf{ref}}^f\otimes \A_{\sf{ref}}^f} \& {\A_{\sf{ref}}^f} \\
	{\A_{\sf{ref}}^f\otimes \A_{\sf{ref}}^f}
	\arrow["\mu", from=1-1, to=1-2]
	\arrow["{R (-1)^{|\cdot|\times|\cdot|}}"', from=1-1, to=2-1]
	\arrow["\mu"', from=2-1, to=1-2]
\end{tikzcd}\]
Here, the map $\mu$ taking the product~Eq.\eqref{eq:product},
and $R(-1)^{|\cdot|\times |\cdot|}$ is the ``super-braiding'', defined as 
\begin{align}
    R(-1)^{|\cdot|\times |\cdot|}(a\otimes b)=(-1)^{|a|\cdot |b|}R(a\otimes b).
\end{align}
Indeed the diagram commutes, 
\begin{align}
    &\mu\circ R(-1)^{|\cdot|\times |\cdot|}((|g\>\otimes_{\bZ_2^F} \gamma^i)\otimes (|h\>\otimes_{\bZ_2^F} \gamma^j))
    \nonumber\\
    &=(-1)^{ij}\mu((P_f^i\cdot|h\>\otimes_{\bZ_2^F} \gamma^j)\otimes (|g\>\otimes_{\bZ_2^F} \gamma^i))
    \nonumber\\
    &=(-1)^{ij}(-1)^{ij}\mu((\cdot|h\>\otimes_{\bZ_2^F} \gamma^j)\otimes (|g\>\otimes_{\bZ_2^F} \gamma^i))
    \nonumber\\
    &=\delta_{g,h}(-1)^{hg^{-1}\cdot i}|g\>\otimes_{\bZ_2^F}\gamma^{i+j}
    \nonumber\\
    &=\delta_{g,h}(-1)^{hg^{-1}\cdot j}|h\>\otimes_{\bZ_2^F}\gamma^{i+j}
    \nonumber\\
    &=\mu((|g\>\otimes_{\bZ_2^F} \gamma^i)\otimes (|h\>\otimes_{\bZ_2^F} \gamma^j)).
\end{align}
Since $\dim(\A_{\sf{ref}}^f)=\sum_{|R|=0}d_R^2+\sum_{|R|=1}d_R^2=|G^F|$, the algebra $\A_{\sf{ref}}^f$ is Lagrangian and defines a gapped boundary. We therefore define our \SymTO{} sandwich construction by choosing $\A_{\sf{ref}}^f$-condensation as the reference boundary. The \SymTO{} construction is shown in Fig.~\ref{fig:fsetup4}.

We also note that the fermionic anyons $e_Rm^F,~|R|=1$ are still deconfined on $\sf{B_{ref}}$: they have trivial braiding with all anyons condensed on $\sf{B_{ref}}$. They are all equivalent to the local fermion $c$, due to the condensate $e_Rm^Fc$. Therefore, up to the condensate the only non-trivial deconfined excitation on the reference boundary is the local fermion $c$. This means the reference boundary can be viewed as a trivial fermionic system, whose topological excitations are described by the super-fusion category $\sf{sVec}$. This structure matches with the construction in~\cite{KantaroLecture}.
\subsection{Spin and twisted spin structures of reference boundary \label{sec:spin}}
As we have demonstrated in Section~\ref{sec:spin}, fermionic reference boundary with the condensation $1+f$ is not unique. In order to fully specify a reference boundary condition, a spin structure on $\sf{B_{ref}}$ must be specified in addition to the condensation $1+f$. In Section~\ref{sec:spin} we showed through a diagrammatic arguement that this spin structure is tied to the expectation value of $f$-loops on the reference boundary. For instance, when $\sf{B_{ref}}\cong T^2$, the boundary state $|1\>+|f\>$, for which the $f$-loops in the time- or space-direction both have value $1$, corresponds to the spin structure (AP,AP). This dependence of the reference boundary on spin structure is crucial to the theory of \SymTO{} for fermionic symmetries, since a fermionic system is defined on a manifold with spin structure. More generally, the symmetry of a fermionic system can be viewed as a group $G^F$ that contains fermion parity $\bZ_2^F$ as a subgroup. $G^B=G^F/\bZ_2^F$ is referred to as the bosonic symmetry of the system.  Such a $G^F$-symmetric system is defined on a manifold with a twisted version of the usual spin structure, called a $G^F$-spin structure. A $G^F$-spin structure is a pair $(\eta,a_b)$, where $\eta$ is a usual spin structure, and $a_b$ is a $G^B$-background gauge field satisfying certain constraint. Therefore, in order to describe $G^F$-symmetric fermionic systems in the \SymTO{} framework, we must analyze how the fermionic reference boundary with the condensation Eq.~\eqref{eq:fref} depends on $G^F$-spin structures. 

In this section we discuss this dependence in detail. We take the approach that the reference boundary can be viewed as a boundary state, expanded in field configuration basis as
\begin{align}
    |\Psi_{\sf{ref}}\>=\sum_{A\in H^1[M, G^F]}\Psi_{\sf{ref}}(A)|A\>.
\end{align}
We will see that there are multiple allowed boundary states with the anyon condensation $\A_{\sf{ref}}^f$, each labelled by a  $G^F$-spin structure $(\eta,a_b)$. We denote these reference states by $|\EF(\eta,a_b)\>$, with $\EF$ standing for ``fermionic". For a given physical boundary state $|\Psi_{phs}\>$, the partition function of the sandwich is given by the inner product 
\begin{align}
    \Z_{\Psi}(\eta,a_b):=\< \EF(\eta,a_b)|\Psi_{\sf{phys}}\>
\end{align}
which is the partition function of a \oned{} system in the presence of a $G^F$-spin structure $(\eta,a_b)$. 

We will begin by reviewing definition and properties of spin structure. We then discuss the reference boundary state of the $\bZ_2^F$-\SymTO{} for arbitrary boundary manifold and spin structure. Finally we generalize to $G^F$-\SymTO{}, and derive the boundary states $|\EF(\eta,a_b)\>$ for arbitrary $G^F$-spin structure $(\eta,a_b)$. 

\subsubsection{Review of  spin structure\label{sec:spin review}}
Recall that the spin structure on a torus amounts to choosing periodic(P) or anti-periodic(AP) boundary conditions along a basis for the noncontractible loops of the torus. This gives us all four spin structures on the torus. \footnote{ Mathematically, AP/P boundary condition corresponds to bounding/non-bounding spin structure of the noncontractible loop, and is also known as NS(Neveu-Schwarz)/R(Ramond) sector in the context of string theory.} On a general two dimensional (spacetime) manifold $M$, the spin structure can be similarly defined as a choice of P or AP boundary condition for every loop on the manifold. We can then define an indicator $q(\gamma)$, where $\gamma$ is a loop on $M$, and $q(\gamma)=1,-1$ when the boundary condition on $\gamma$ is AP,P respectively.
The value of this indicator function should be invariant under deformation of the loop, and can be extended to any $\bZ_2$-coefficient formal sum of loops by deforming the formal sum into a single loop. Therefore $q$ is a function on the homology group $H_1[M,\bZ_2]$.  A fermion loop around the composite cycle $\gamma_1+\gamma_2$ can be deformed into separate loops encircling $\gamma_1$ and $\gamma_2$, where the deformation involves a number of fermion exchanges equal to $\gamma_1\cdot\gamma_2$: Here $\gamma_1\cdot \gamma_2$ is the number of intersection points of $\gamma_1$ and $\gamma_2$, which is homologically invariant $\mod 2$. This requires that $q$ satisfy the consistency condition:
\begin{align}
    q([\gamma_1]+[\gamma_2])=q([\gamma_1])q([\gamma_2])(-1)^{\gamma_1\cdot \gamma_2}.
\end{align}
Functions satisfying this identity are called quadratic forms on $H_1[M,\bZ_2]$. 

Alternatively, the spin structure can be equivalently defined as a $\bZ_2$-valued $1$-cochain, $\eta$ that trivializes the 2nd Stiefel-Whitney class: $d\eta = w_2$.
There is a one-to-one correspondence between quadratic forms, $q_\eta$ and the associated cochain, $\eta$.
Any two spin structures, $\eta,\eta'$ satisfy $d(\eta-\eta')=0$, i.e. any two spin structures are related by the addition of a flat, $\bZ_2$-valued gauge field.  
Formally, the set of spin structures is referred to as a torsor of $H^1[M,\bZ_2]$.

The Arf-invariant, which appears in the partition function of the Kitaev chain, is defined by summing over the loops in the quadratic form:
\begin{align}
    \arf(\eta):=\frac{1}{\sqrt{H_1[M,\bZ_2]}}\sum_{\gamma\in H_1[M,\bZ_2]} q_\eta(\gamma),
\end{align}
which takes values in $\{1,-1\}$ and only depends on the spin structure of the manifold.

\subsubsection{Spin structure in the $\bZ_2^F$-\SymTO{}}
We now generalize our discussion of spin structure dependence in section~\ref{sec:spin1} for the $\bZ_2^F$-\SymTO{} to general boundary manifolds. We can simply apply the argument of moving an $f$-anyon in
Fig.~\ref{fig:spinstructure} to any loop $\gamma$ on the boundary $\sf{B_{ref}}$. Following the same procedure as in Fig.~\ref{fig:spinstructure}, we can see that the boundary condition along $\gamma$ is still related to the expectation value of an $f$-loop on $\gamma$ via
\begin{align}
    P/AP\Leftrightarrow \<W_f(\gamma)\>_{\sf{B_{ref}}}=-1/+1.
\end{align}
Comparing with the definition of the quadratic form $q_\eta$, we see that the expectation value of the $f$-loops on $\sf{B_{ref}}$ is simply $q_\eta$:
\begin{align}
    \<W_f(\gamma)\>_{\sf{B_{ref}}}=q_\eta(\gamma).\label{eq: floop-spin}
\end{align}
This relation tells us that the collection of values of $f$-loops on the reference boundary is in one-to-one correspondence with the spin structure of the reference boundary. Below we will utilize this relation to determine the boundary state with given a spin structure.\footnote{We note that this continuum argument can also be explicitly verified in microscopic toric-code-like lattice models with $cf$-condensed boundaries. We provide this lattice construction in Appendix~\ref{app: f-bdry lattice}.}

We now compute the state of the reference boundary with a given spin structure $\eta$. We write an $f$-condensed boundary state in the field configuration basis as 
\begin{align}
    |\EF\>=\sum_a \F(a)|a\>,
\end{align}
here $a\in H^1[\sf{B_{ref}},\bZ_2]$ is the value of $\bZ_2$-gauge field on the boundary, $\EF$ stands for ``fermionic" reference boundary state.
Our goal is now to determine the coefficients $\F(a)$ such that the state $|\EF\>$ corresponds to an $f$-condensed boundary with spin structure $\eta$. An $f$-condensed boundary should be able to absorb an $f$-loop on it. We can write an $f$-loop on $\sf{B_{ref}}$ as a product of an $e$-loop and an $m$-loop,
\begin{align}
    W_f[\gamma]:=W_e[\gamma]W_m[\gamma], \gamma\in H_1[\sf{B_{ref}},\bZ_2],
\end{align}
where we choose an arbitrary framing for $\gamma$ so that $W_f[\gamma]$ is unambiguously defined. 
The $e$-loop $W_e[\gamma]$ is the Wilson loop of the $\bZ_2$-gauge field $a$:
\begin{align}
    W_e[\gamma]=(-1)^{\oint_\gamma a}=(-1)^{\int_{\sf{B_{ref}}}a\cup\tilde{\gamma}}
\end{align}
where we used the Poincare dual of $\gamma$, denoted by $\tilde{\gamma}\in H^1[\sf{B_{ref}},\bZ_2]$, to write the Wilson loop as an integration over the entire boundary $\sf{B_{ref}}$. On the other hand the $m$-loop is the Wilson loop of the dual gauge field $\hat{a}$,
\begin{align}
    W_m[\gamma]=(-1)^{\oint_\gamma \hat{a}}=(-1)^{\int_{\sf{B_{ref}}}\hat{a}\cup \tilde{\gamma}}
\end{align}
The $e/m$-loops act on the basis $|a_0\>$ as follows,
\begin{align}
    W_e[\gamma]|a_0\>&=(-1)^{\int_{\sf{B_{ref}}}a_0\cup\tilde{ \gamma}}|a_0\>,\\
    W_m[\gamma]|a_0\>&=(-1)^{\int_{\sf{B_{ref}}}\hat{a}\cup \tilde{\gamma}}|a_0\>=|a_0+\tilde{\gamma}\>.
\end{align}
Therefore the $f$-loop acts on the boundary state as
\begin{align}
    W_f[\gamma]|\EF\>
    &=\sum_{a}(-1)^{\int_M a\cup\tilde{\gamma}}\F(a)|a+\tilde{\gamma}\>
    \nonumber\\
    &=\sum_{a} (-1)^{\int_M a\cup\tilde{\gamma}}\F(a+\tilde{\gamma})|a\>.
\end{align}

\begin{figure*}
      \includegraphics[width=1.8\columnwidth]{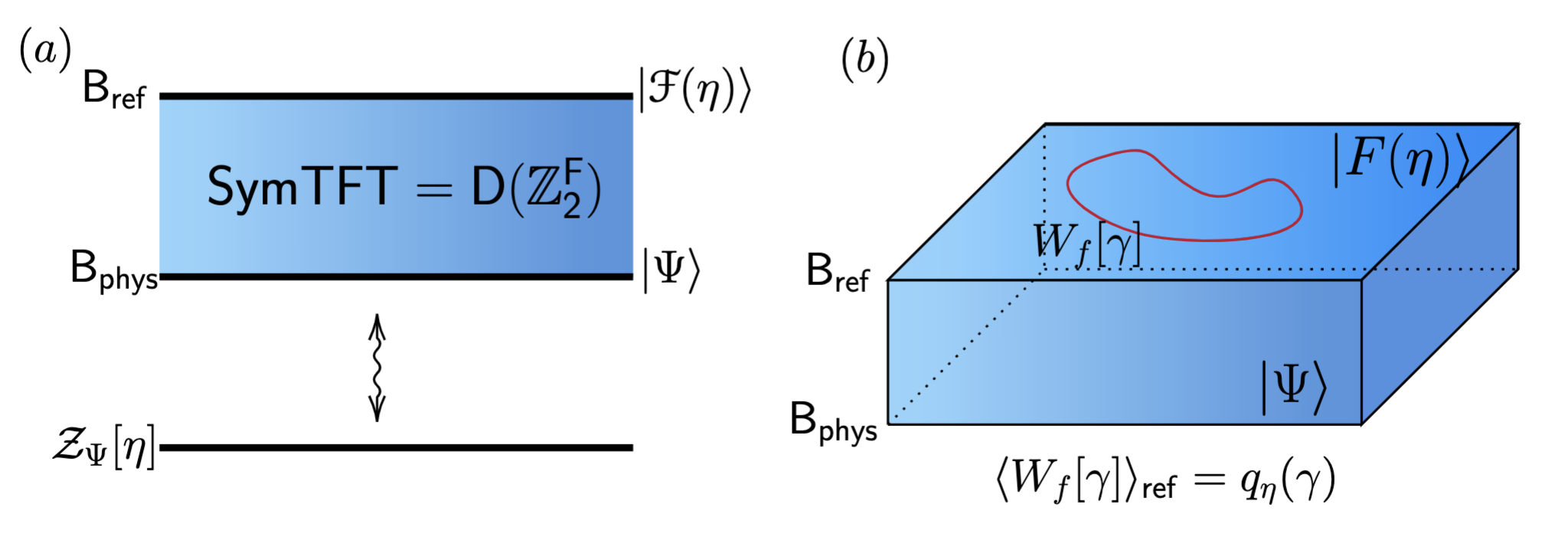}
    \caption{Schematic picture of the \SymTO{} for $\bZ_2^F$.  (a). The \SymTO{} is the $\bZ_2$ quantum double topological order(toric code).  The reference boundary depends on a spin structure $\eta\in C^1[\sf{B_{ref}},\bZ_2]~d\eta=w_2$, the corresponding state is given by Eq.\eqref{eq:fbdry}. The sandwich with $|\EF(\eta)\>$ reference boundary describes a \oned{} fermionic system with spin structure $\eta$.  (b) For a boundary spin structure $\eta$, an $f$-loop $W_f(\gamma)$ on the reference boundary has eigenvalue given by the quadratic form of the spin structure, $q_\eta(\gamma)$. }
    \label{fig:Z2F schem}
\end{figure*}

Requiring $|\EF\>$ to be an eigenstate of $W_f[\gamma]$ leads to the condition (we will write $\tilde{\gamma}$ as $\gamma$ from now on)
\begin{align}
    \sum_{a} (-1)^{\int_M a\cup\gamma}\F(a+\gamma)|a\>=
    f(\gamma)\sum_{a} \F(a)|a\>,
\end{align}
for some eigenvalues $f(\gamma)$. By comparing coefficients on both sides we have
\begin{align}
    (-1)^{\int_M a\cup\gamma}\F(a+\gamma)=f(\gamma)\F(a), 
\end{align}
for all $a,\gamma\in H^1[\sf{B_{ref}},\bZ_2]$. Setting $a=0$ we see $\F(\gamma)=f(\gamma)\F(0)$. Therefore $\F$ satisfies the condition
\begin{align}
    \frac{\F(a+\gamma)}{\F(0)}=(-1)^{\int_M a\cup\gamma}\frac{\F(\gamma)}{\F(0)}\frac{\F(a)}{\F(0)},
\end{align}
which states that $\F(a)/\F(0)$ is a quadratic form on $H^1[\sf{B_{ref}},\bZ_2]$.  
We may then write $\F(a)=\F(0)q_\eta(a)$ for some spin structure $\eta$. Up to normalization an $f$-condensed boundary is then determined by a spin structure on $\sf{B_{ref}}$, and given by the state
\begin{align}
    |\EF(\eta)\>\propto\sum_{a\in H^1[\sf{B_{ref}},\bZ_2]}q_\eta(a)|a\>.    
\end{align}
The state $|\EF(\eta)\>$ is an eigenstate of $f$-loops on $\sf{B_{ref}}$ with eigenvalue $q_\eta(\gamma)$. Therefore according to the relation Eq.\eqref{eq: floop-spin}, the state $|\EF(\eta)\>$ is exactly the state corresponding to the spin structure $\eta$ on $\sf{B_{ref}}$.  

However, there is still a phase ambiguity for every state $|\EF(\eta)\>$. These phases for different $\eta$s are not independent of each other. To see this, notice that in the \SymTO{} fermion parity is represented by an $m$-line on $\sf{B_{ref}}$, therefore insertion of an $m$-line should change the spin structure according to $W_m[\gamma]|\EF(\eta)\>=|\EF(\eta+\gamma)\>$. 
Therefore once a state $|\EF(\eta)\>$ with a given spin structure $\eta$ is chosen, the states with other spin structures are uniquely determined by the relation $W_m[\gamma]|\EF(\eta)\>=|\EF(\eta+\gamma)\>$. This will fix the phase ambiguity of the states $|\EF(\eta)\>$ up to an overall common phase, which is inconsequential. To see this, assume the state $|\EF(\eta)\>$ has an arbitrary phase factor $\T(\eta)$:
\begin{align}
    |\EF(\eta)\>=\T(\eta)\sum_a q_\eta(a)|a\>.
\end{align}
Then the action of an $m$-loop on the state is
\begin{align}
    W_m[\gamma]|\EF(\eta)\>&=\T(\eta)\sum_a q_\eta(a)|a+\gamma\>\nonumber\\
    &=\T(\eta)\sum_a q_\eta(a+\gamma)|a\>.
\end{align}
The state $|\EF(\eta+\gamma)\>$ on the other hand is given by
\begin{align}
    |\EF(\eta+\gamma)\>=\mathcal{T}(\eta+\gamma)\sum_a q_{\eta+\gamma}(a)|a\>.
\end{align}
By comparing coefficients we have the relation
\begin{align}
 \mathcal{T}(\eta)q_\eta(a+\gamma)=\mathcal{T}(\eta+\gamma)q_{\eta+\gamma}(a).
\end{align}
Summing over $a\in H^1[M,\bZ_2]$ on both sides, we have
\begin{align}
    \mathcal{T}(\eta)\arf(\eta)=\mathcal{T}(\eta+\gamma)\arf(\eta+\gamma).
\end{align}
We see that the solution for $\mathcal{T}$ is $\mathcal{T}(\eta)=c\cdot \arf(\eta)$, where $c\in \bC^\times$ is arbitrary. Take $c=1$, we have the phase-fixed states 
\begin{align}
    |\EF(\eta)\>&=\arf(\eta)\sum_{a\in H^1[M,\bZ_2]}q_\eta(a)|a\>
    \nonumber\\
    &=\sum_{a\in H^1[M,\bZ_2]}\arf(\eta+a)|a\>,\label{eq:fbdry}
\end{align}
where we used relation $\arf(\eta+a)=\arf(\eta)q_\eta(a)$.  

Eq.\eqref{eq:fbdry} completes our construction for the $\bZ_2^F$-\SymTO{}. The schematic of the \SymTO{} construction is shown in Fig.~\ref{fig:Z2F schem}.

As an application of this result, we can obtain the partition function of the Kitaev chain on any orientable manifold as follows: the physical boundary is the $e$-condensed Dirichlet boundary $|a_0=0\>$, therefore the partition function with spin structure $\eta$ is
\begin{align}
    \Z_\eta^{\text{Kitaev}}=\<\EF(\eta)|a_0=0\>=\arf(\eta).
\end{align}

\subsubsection{Review of $G^F$-spin structure}\label{sec:GFspin}

A general fermionic symmetry is described by the short exact sequence
\begin{align}
1\to \bZ_2^F\xrightarrow{i} G^F\xrightarrow{\pi} G^B\to 1.\label{eq:GF2}
\end{align}
For the convenience of later discussion, let us also specify \textit{set-theoretic} section maps $s: G^B\to G^F, t: G^F\to \bZ_2^F$ that satisfy $\pi\circ s=\text{id}_{G^B},~t\circ i=\text{id}_{\bZ_2^F}$. The extension class of this sequence is $\rho:=\delta s\in Z^2[G^b,\bZ_2^F]$. The section maps allow us to write elements of $G^F$ as pairs $(j,g^b),~j\in \bZ_2^F,~g^b\in G^B$. Then $s:g^b\mapsto (0,g^b)$ and $t:(j,g^b)\mapsto j$. The section maps are not unique and not group homomorphisms in general, and in the context of the fermionic \SymTO{} we expect different choices of the section maps will give the same \SymTO{}.\footnote{Still, given a microscopic fermionic system which is a tensor product of local fermionic Hilbert spaces, it is most natural to identify $G^B$ as local symmetry actions, as discussed in e.g. \cite{Bulmash2022}. Hence, there is a canonical choice of the section map $s$, and different choices $s$ may be considered different physical systems. We do not work with a microscopic lattice system in this paper and hence will ignore this subtlety.}

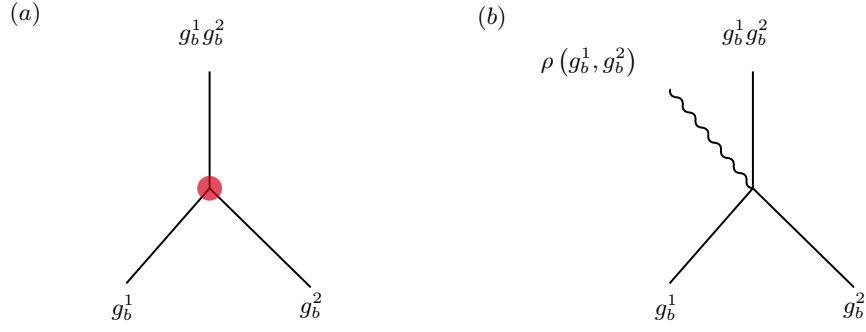
\begin{figure*}
    \centering   
\tikzset{every picture/.style={line width=0.75pt}} 

\begin{tikzpicture}[x=0.75pt,y=0.75pt,yscale=-1,xscale=1]

\draw    (126.5,56) -- (126.5,115) ;
\draw    (126.5,115) -- (177.5,165) ;
\draw    (126.5,115) -- (84.5,163) ;
\draw  [draw opacity=0][fill={rgb, 255:red, 208; green, 2; blue, 27 }  ,fill opacity=0.7 ] (120.25,115) .. controls (120.25,111.55) and (123.05,108.75) .. (126.5,108.75) .. controls (129.95,108.75) and (132.75,111.55) .. (132.75,115) .. controls (132.75,118.45) and (129.95,121.25) .. (126.5,121.25) .. controls (123.05,121.25) and (120.25,118.45) .. (120.25,115) -- cycle ;
\draw    (400.5,56) -- (400.5,115) ;
\draw    (400.5,115) -- (451.5,165) ;
\draw    (400.5,115) -- (358.5,163) ;
\draw    (400.5,115) .. controls (398.15,114.79) and (397.08,113.52) .. (397.28,111.17) .. controls (397.49,108.82) and (396.42,107.55) .. (394.07,107.34) .. controls (391.72,107.13) and (390.65,105.86) .. (390.85,103.51) .. controls (391.06,101.16) and (389.99,99.89) .. (387.64,99.69) .. controls (385.29,99.48) and (384.22,98.21) .. (384.42,95.86) .. controls (384.62,93.51) and (383.55,92.24) .. (381.2,92.03) .. controls (378.85,91.82) and (377.78,90.55) .. (377.99,88.2) .. controls (378.19,85.85) and (377.12,84.58) .. (374.77,84.37) .. controls (372.42,84.16) and (371.35,82.89) .. (371.56,80.54) .. controls (371.76,78.19) and (370.69,76.92) .. (368.34,76.71) .. controls (365.99,76.51) and (364.92,75.24) .. (365.12,72.89) .. controls (365.33,70.54) and (364.26,69.27) .. (361.91,69.06) .. controls (359.56,68.85) and (358.49,67.58) .. (358.69,65.23) -- (358.5,65) -- (358.5,65) ;

\draw (75.5,167.4) node [anchor=north west][inner sep=0.75pt]    {$g^{1}_{b}$};
\draw (170.5,166.4) node [anchor=north west][inner sep=0.75pt]    {$g^{2}_{b}$};
\draw (109.5,28.4) node [anchor=north west][inner sep=0.75pt]    {$g^{1}_{b} g^{2}_{b}$};
\draw (24,19.4) node [anchor=north west][inner sep=0.75pt]    {$( a)$};
\draw (349.5,167.4) node [anchor=north west][inner sep=0.75pt]    {$g^{1}_{b}$};
\draw (444.5,166.4) node [anchor=north west][inner sep=0.75pt]    {$g^{2}_{b}$};
\draw (383.5,28.4) node [anchor=north west][inner sep=0.75pt]    {$g^{1}_{b} g^{2}_{b}$};
\draw (292,42.4) node [anchor=north west][inner sep=0.75pt]    {$\rho \left( g^{1}_{b} ,g^{2}_{b}\right)$};
\draw (260,21.4) node [anchor=north west][inner sep=0.75pt]    {$( b)$};

\end{tikzpicture}

    \caption{(a).The $G^B$-defect network has non-zero curvature when viewed as a $G^F$-defect network, represented by the red dot. (b) Adding a $\rho(g_b^1,g_b^2)$ defect line to the junction resolves the curvature. The collection of the $\bZ_2^F$ defect lines constructed in this way form a $\bZ_2^F$-gauge field $\tau$ that has the same curvature as the $G^B$-gauge field: $d\tau=d(a_b^*(s))=a_b^*(\rho)$.}
    \label{fig:G^F defect}
\end{figure*}

A fermionic system with symmetry $G^F$ couples to $G^F$-spin structures. A $G^F$-spin structure can be thought of as a pair $(\eta,a_b)$, where $\eta$ is a usual spin structure that satisfies $d\eta=w_2$, and $a_b$ is a background $G^B$-gauge field, $da_b=0$, such that $[a_b^*(\rho)]=0\in H^2[M,\bZ_2^F]$. We can take a trivialization of $a^*_b(\rho)$: $\tau\in C^1[M,\bZ_2^F],~ d\tau=a^*_b(\rho)$. Physically this corresponds to the following picture. The gauge field configuration of $a_b$ can also be viewed as a $G^B$-defect network. When a $g_b^1$ defect line and a $g_b^2$ defect line meet, they become the $g_b^1g_b^2$ defect line. This satisfies the fusion rule of defects when they are viewed as $G^B$-defects, but violates fusion rule when viewed as $G^F$-defects. In fact, the violation is given by $\rho(g_b^1,g_b^2)$, which can be viewed as a curvature(or a flux) at the junction of the three defect lines. This curvature can also be viewed as associated with a $\bZ_2^F$-gauge field $\tau$, whose defect lines look like the wiggly line in Fig.~\ref{fig:G^F defect}. This relation is exactly the equation $d\tau=a^*_b(\rho)$.

Now consider the fermions on a 2 dimensional manifold with a $G^F$-spin structure $(\eta,a_b)$. The fermions carry charges of $G^F$, labelled by representations $R\in \rep(G^F)$, such that $R(P_f)=-1$. There are also bosonic charges living on the manifold(e.g. pairs of charged/uncharged fermions), that carry representations such that $R(P_f)=1$. The spin structure $\eta$ specifies AP/P boundary condition along every loop on the manifold. When a fermion is moved along a loop $\gamma$, it will experience the spin structure $\eta$, therefore obtains a $-1$ sign if $q_\eta(\gamma)=1$. It will also experience the $G^B$ gauge field, as well as the $\tau$-field that comes from the fusion of $G^B$-defects. Therefore the action on the fermion when it is moved around $\gamma$ is 
\begin{align}
   \Theta(\eta,a_b)[\gamma]&= -q_\eta(\gamma) R\left(\oint_\gamma a_b^*(s)+\tau\right)\nonumber\\
   &=(-1)^{1+\oint_\gamma \tau}q_\eta(\gamma)R\left(\oint_\gamma a_b^*(s)\right).
\end{align}
This action can be divided into three parts. The first part is $-q_\eta(\gamma)$, which comes from the AP/P condition along $\gamma$. The second part is the holonomy of the $G^B$-gauge field, $\oint_\gamma a_b^*(s)$, which acts on the fermion via the representation $R$. The last part is $(-1)^{\oint_\gamma \tau}$, which can be thought of as the fermion parity action accumulated along $\gamma$.
\subsubsection{$G^F$-spin structure in $G^F$-\SymTO{}}
We now consider the fermionic reference boundary in the \SymTO{} for a general $G^F$-symmetry. The anyon condensation is given by the algebra Eq.$\eqref{eq:fbdry}$. However, there are multiple boundaries with this anyon condensation, distinguished by  $G^F$-spin structures on the boundary. We can repeat the moving $f$-anyon argument in Fig.~\ref{fig:spinstructure} to a general fermionic charge and any loop $\gamma$. Following the same procedure as in Fig.~\ref{fig:spinstructure}, we see that the expectation value of the fermionic charge loop is negative the phase accumulated from moving the fermionic charge around, 
\begin{align}
    \<W_{e_Rm^F}(\gamma)\>&=-\Theta(\eta,a_b)=q_\eta(\gamma) R\left(\oint_\gamma a_b^*(s)+\tau\right)\label{eq:floopgeneral}
\end{align}
This is the generalized form of the relation Eq.~\eqref{eq: floop-spin}. Similarly, for bosonic charges we have the relation
\begin{align}
    \<W_{e_R}(\gamma)\>=R\left(\oint_\gamma a_b^*(s)\right)
\end{align}
These two relations tells us that the $G^F$-spin structure on the boundary can be distinguished by expectation values of the loop operators of the condensed anyons. Next we utilize these relations to determine the boundary state with a given $G^F$-spin structure. 

The $G^F$-\SymTO{} is a $G^F$-gauge theory. The $G^F$-gauge field, $A_f\in C^1[M,G^F]$, can be divided into a $\bZ_2^F$-gauge field $A_{\bZ_2}\in C^1[M,\bZ_2^F]$, and a $G^B$-gauge field $A_{b}\in C^1[M,G^B]$, with relation $dA_{\bZ_2}=A_{b}^*(\rho)$ and $A_f=A_{\bZ_2}+A_b^*(s)$. 
Here, the bulk gauge field in the \SymTO{} is denoted by upper case $A$, in contrast to the $a_b$ that appears in the $G^F$-spin structure.
Any boundary state can then be written in the field configuration basis as
\begin{align}
|\Psi\>=\sum_{A_{\bZ_2},A_b}\Psi(A_{\bZ_2},A_b)|A_{\bZ_2},A_b\>,
\end{align}
where the sum is taken over fields satisfying $dA_{\bZ_2}=A_b^*(s)$.

\begin{widetext}

We claim that the fermionic reference boundary with $G^F$-spin structure $(\eta,a_b)$ is given by the state
\begin{align}
    |\EF(\eta,a_b)\>:=\sum_{A_{\bZ_2},dA_{\bZ_2}=a_b^*(\rho)}\frac{\arf(\eta+A_{\bZ_2}+\tau)}{\sqrt{H^1[M,\bZ_2]}}|A_{\bZ_2},A_b=a_b\>.\label{eq:f-bdry-2}
\end{align}

For this to be well-defined, notice $d(\eta+A_{\bZ_2}+\tau)=w_2+2a_b^*(\rho)=w_2$, thus $\eta+A_{\bZ_2}+\tau$ is a valid spin structure. To show $|\EF(\eta,a_b)\>$ is a valid $\A_{\sf{ref}}^f$-condensed boundary, consider a fermionic Wilson loop on $\sf{B_{phys}}$, $W_{e_R}[\gamma]W_{m^F}[\gamma]$. Here $W_{e_R}[\gamma]$ is a bosonic Wilson loop of a charge carrying an irrep $R$ with $R(P_f)=-1$. It acts on the field configuration basis as 
\begin{align}
    W_{e_R}[\gamma]|A_f\>=R(\oint_\gamma A_f)|A_f\>.
\end{align}
$W_{m^F}[\gamma]$ is a fermion parity flux loop that acts as $W_{m^F}[\gamma]|A_f\>=|A_f+\gamma\>$, where we again used Poincare duality to represent the loop $\gamma$ as an element of $H^1[M,\bZ_2^F]$.  Therefore the effect of the fermionic Wilson loop on the state $|\EF(\eta,a_b)\>$ is 

\begin{align}
    W_{e_R}[\gamma]W_{m^F}[\gamma]|\EF(\eta,a_b)\>=\sum_{A_{\bZ_2},dA_{\bZ_2}=a_b^*(\rho)}\frac{\arf(\eta+A_{\bZ_2}+\tau)}{\sqrt{H^1[M,\bZ_2]}}R\left(\oint_\gamma A_f\right)|A_{\bZ_2}+\gamma,A_b=a_b\>.
\end{align}

\begin{figure*}
\includegraphics[width=0.9\columnwidth]{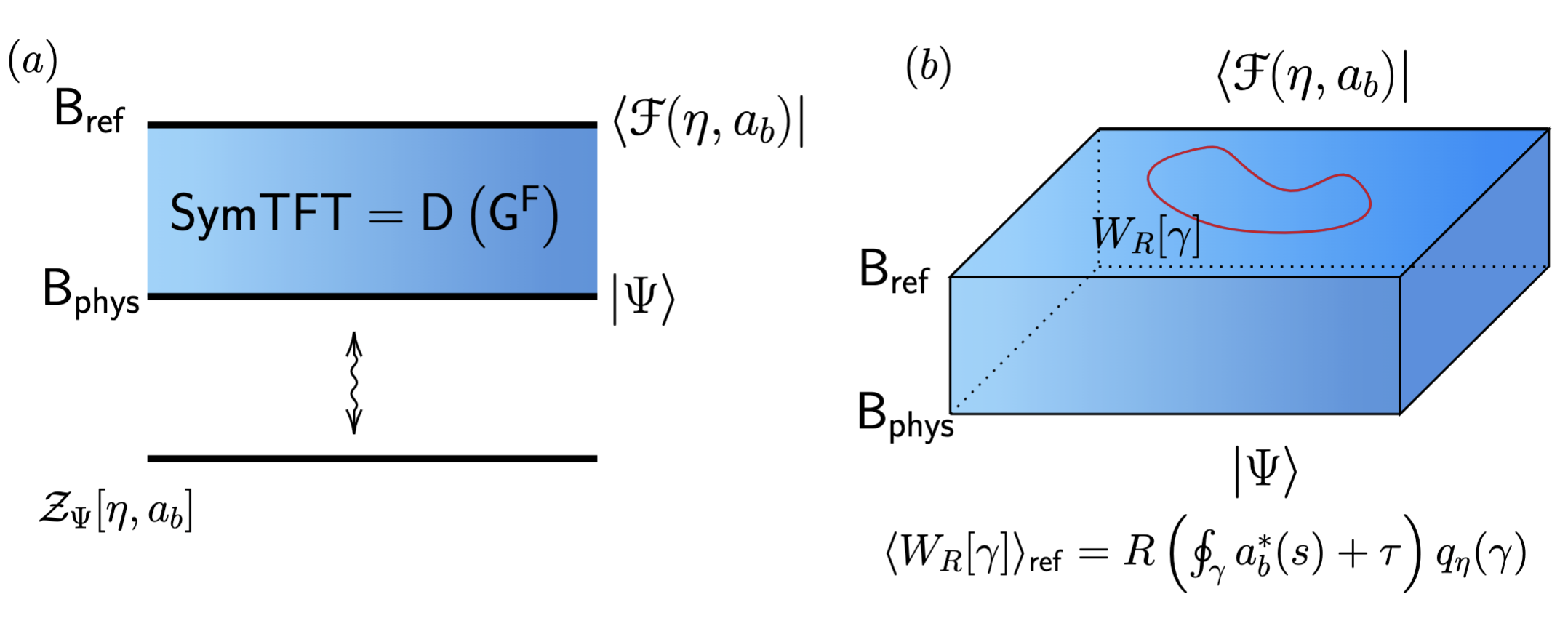}

    \caption{Schematic picture of the $G^F$-\SymTO{} construction. (a). The \SymTO{} is the $G^F$-quantum double theory. The reference boundary with $G^F$-spin structure $(\eta,a_b)$ is given by the state Eq.\eqref{eq:f-bdry-2}. The \SymTO{} sandwich with $|\EF(\eta,a_b)\>$ reference boundary describes a \oned{} $G^F$-symmetric system coupled to $G^F$-spin structure $(\eta,a_b)$. (b). A fermionic Wilson loop carrying representation $R$ has non-zero expectation value on the reference boundary. The expectation value depends on the $G^F$-spin structure.}
    \label{fig:GF scheml}
\end{figure*}
We can write $A_f$ as
\begin{align}
    A_f=(A_{\bZ_2}+\tau)+(A_b^*(s)+\tau).
\end{align}
Then $d(A_{\bZ_2}+\tau)=0$ and $A_{\bZ_2}+\tau$ is a flat $\bZ_2$-gauge field. Therefore we have
\begin{align}
    &W_{e_R}[\gamma]W_{m^F}[\gamma]|\EF(\eta,a_b)\>\nonumber\\
    &=\sum_{A_{\bZ_2},dA_{\bZ_2}=a_b^*(\rho)}\frac{\arf(\eta+A_{\bZ_2}+\tau)}{\sqrt{H^1[M,\bZ_2]}}
    R\left(
    \oint_\gamma (A_{\bZ_2}+\tau)+(A^*_b(s)+\tau)\right)
    |A_{\bZ_2}+\gamma,A_b=a_b\>
    \nonumber \\ 
    &=\sum_{A_{\bZ_2},dA_{\bZ_2}=a_b^*(\rho)}\frac{\arf(\eta+A_{\bZ_2}+\tau+\gamma)}{\sqrt{H^1[M,\bZ_2]}}R\left(\oint_\gamma a_b^*(s)+\tau\right)(-1)^{\int_M (A_{\bZ_2}+\tau)\cup \gamma}|A_{\bZ_2},A_b=a_b\>
\end{align}
where we used the condition that $R$ is an odd representation, thus $R|_{\bZ^F_2}(g)=(-1)^g\mathbb{I}_{d_R}$. Now using the relation
\begin{align}
    \arf(\eta+A_{\bZ_2}+\tau+\gamma)=\arf(\eta+A_{\bZ_2}+\tau)q_{\eta+A_{\bZ_2}+\tau}(\gamma)=\arf(\eta+A_{\bZ_2}+\tau)q_{\eta}(\gamma)(-1)^{\int_M (A_{\bZ_2}+\tau)\cup \gamma},
\end{align}
we arrive at the expression
\begin{align}
    W_{e_R}[\gamma]W_{m^F}[\gamma]|\EF(\eta,a_b)\>=R\left(\oint_\gamma a^*_b(s)+\tau\right)q_{\eta}(\gamma) |\EF(\eta,a_b)\>.
\end{align}
Comparing with relation Eq.~\eqref{eq:floopgeneral}. we see that  $|\EF(\eta,a_b)\>$ indeed corresponds to the fermionic boundary with spin structure $(\eta,a_b)$. 
\end{widetext}
The reference boundary state $|\EF(\eta,a_b)\>$ completes our \SymTO{} for a general fermionic symmetry $G^F$. The structure is summarized in Fig.~\ref{fig:GF scheml}.

In the following we use this framework to study fermionic SPT phases, reproducing their partition function and stacking rules from \SymTO{}. We will also discuss duality in \SymTO{}, generalizing the \SymTO{} construction for bosonic dualities to include bosonization.

\section{Fermionic SPT phases from \SymTO{}\label{sec: SPT}}
We now discuss fermionic SPTs in \SymTO{}.  Let us first review the cohomology and super-cohomology characterizations of \oned{} fermionic SPTs~\cite{Kapustin_2018,turzillo2023duality}. 

The cohomology characterization of a $G^F$-SPT is a cocycle $\omega\in Z^2[G^F,U(1)]$ and its slant product with fermion parity $\nu(g):=\omega(g,F)-\omega(F,g)$.\footnote{We use the convention that $U(1)=\bR/\bZ=[0,1]/(0\simeq 1)$ and write multiplication in $U(1)$ as addition mod 1.} Physically the phase $\exp(2\pi i \omega)$ is the projective phase of edge symmetry actions, and $\nu$ keeps track of the fermion parity of edge symmetry operators. Since $\nu$ is determined by $\omega$ it would seem redundant to include $\nu$ in the characterization. But $\nu$ participates in the stacking rules of fermionic SPTs in a non-trivial way. More concretely, the stacking rule of $G^F$-SPTs  is not the same as the group structure of the cohomology group $H^2[G^F,U(1)]$, and $\nu$ will encode this difference. Therefore we will still label fermionic SPTs by the pairs $(\omega,\nu)$. 

The super-cohomology characterization  originates in the decorated domain wall structure of fermionic SPTs. It consists of two functions $\alpha\in C^2[G^B,U(1)],\beta\in C^1[G^B,\bZ_2]$. Physically $\alpha$ is the projective phase of edge $G^B$ symmetry actions, and $\beta$ keeps track of the fermion parity of these edge operators. But since $G^B$ symmetry actions are not closed under composition if fermion parity extension is non-trivial, $\alpha$ is in general not a cocycle. $\alpha,\beta$ can be written as pull-backs of $\omega,\nu$ via the section map $s$: $\alpha=s^*(\omega), \beta=s^*(\nu)$. They are subject to relation 
\begin{align}
    \delta \alpha=\frac{1}{2} \beta\cup \rho,~\delta \beta=0
\end{align}
The cohomology invariants can be recovered by the super-cohomology invariants via the relation 
\begin{align}
    \omega=p^*(\alpha)+\frac{1}{2}p^*(\beta)\cup t,~\nu=p^*(\beta).
\end{align}

\subsection{Partition function of fermionic SPTs from \SymTO{}}
Now we return to the \SymTO{} and derive the partition function of fermionic SPTs. The physical boundary corresponding to the $G^F$-SPT with cohomology invariants $\omega,\nu$ is given by the boundary state 
\begin{align}
    |\omega\>:=\sum_{A_f} e^{2\pi i\int_{M}A_f^*(\omega)}|A_f\>.
\end{align}
Write $A_f=A_{\bZ_2}+A_b^*(s)$, and use the relation $\omega=p^*(\alpha)+\frac{1}{2}p^*(\beta)\cup t$, we can write $A^*_f(\omega)$ as 
\begin{align}
    A^*_f(\omega)&=(A_{\bZ_2}+A_b^*(s))^*(p^*(\alpha)+\frac{1}{2}p^*(\beta)\cup t)\\
    &=  A_b^*(\alpha)+\frac{1}{2}A_b^*(\beta)\cup A_{\bZ_2}.
\end{align}
\begin{widetext}
The reference boundary is given by~\eqref{eq:f-bdry-2}. Take the inner product between the reference boundary state and the physical boundary state, we obtain:
\begin{align}
    &\<\EF(\eta,a_b)|\omega\>=\sum_{A_{\bZ_2},dA_{\bZ_2}=a^*_b(\rho)}\frac{\arf(\eta+A_{\bZ_2}+\tau)}{\sqrt{|H^1[M,\bZ_2]|}}\exp\left(2\pi i \int_{M} a_b^*(\alpha)+\frac{1}{2}a_b^*(\beta)\cup A_{\bZ_2}\right)\nonumber\\
    &=\sum_{A_{\bZ_2},dA_{\bZ_2}=a^*_b(\rho)}\frac{\arf(\eta)}{\sqrt{|H^1[M,\bZ_2]|}}q_{\eta}(A_{\bZ_2}+\tau)\exp\left(2\pi i \int_{M} a_b^*(\alpha)+\frac{1}{2}a_b^*(\beta)\cup (A_{\bZ_2}+\tau)+\frac{1}{2}a_b^*(\beta)\cup \tau\right)\nonumber\\
    &=\sum_{A_{\bZ_2},dA_{\bZ_2}=a^*_b(\rho)}\frac{\arf(\eta)}{\sqrt{|H^1[M,\bZ_2]|}}q_{\eta+a_b^*(\beta)}(A_{\bZ_2}+\tau)\exp\left(2\pi i \int_{M} a_b^*(\alpha)+\frac{1}{2}a_b^*(\beta)\cup \tau\right)\nonumber\\
    &=\arf(\eta)\arf(\eta+a_b^*(\beta))\exp\left(2\pi i \int_{M} a_b^*(\alpha)+\frac{1}{2}a_b^*(\beta)\cup \tau\right)\nonumber\\
    &=q_\eta(a_b^*(\beta))\exp\left(2\pi i \int_{M} a_b^*(\alpha)+\frac{1}{2}a_b^*(\beta)\cup \tau\right)
\end{align}
In summary, we have 
\begin{align}
    \<\EF(\eta,a_b)|\omega\>&=q_{\eta}(a_b^*(\beta))\exp\left(2\pi i \int_{M}a^*_b(\alpha)+\frac{1}{2}a^*_b(\beta)\cup \tau\right).
\end{align}
This is exactly the partition function of the $(\omega, \nu)$ SPT in the presence of a $G^F$-spin structure $(\eta,a_b)$.
\end{widetext}

\subsection{Stacking rules of fermionic SPTs from \SymTO{}\label{sec: stacking}}
We have shown that the \SymTO{} is capable of describing fermionic invertible phases. In particular, we have shown how to reveal edge modes of SPT phases via \SymTO{}, and how to compute the partition function of fermionic SPTs via \SymTO{}. 

Another crucial ingredient to the classification of invertible phases is their group structure under stacking.
Stacking is a physical operation that outputs a new phase from two input phases. Take two $G$-symmetric invertible phases $\EF^{1,2}$. Their stacking $\EF^1\boxtimes \EF^2$ is obtained by two steps. In the first step we form stack the two phases on top of one another, mathematically: defining a new system as the tensor product $\EF^2$  of $\EF^1$. This results in a system with an enlarged symmetry $G\times G$. Then in the second step we turn on generic interactions between the two phases that preserve the diagonal symmetry $\{(g,g)|g\in G\}< G\times G$ while preserving the gap of the combined system. The resulting system can the be viewed as another $G$-symmetric invertible phase, and this process defines a group structure on the set of invertible phases.

For example, in the $G^F=\bZ_2\times \bZ_2^F$ case we obtained four invertible phases from \SymTO{}. But there are two possible group structure on this set of invertible phases, either $\bZ_2^2$ or $\bZ_4$. We want to ask: Can we determine directly from the \SymTO{} which of the two options is realized? Relatedly, even though the set of \oned{} fermionic SPTs are the same as the set of \oned{} bosonic SPTs, the stacking rule (group structure) for \oned{} fermionic SPTs is not the same as the stacking rule of \oned{} bosonic SPTs, which is determined by the group structure $H^2[G^F,U(1)]$~\cite{Wang_2018,turzillo2023duality,Aksoy,Ren2023}. In this section, we demonstrate that the stacking rules between fermionic and bosonic SPTs with isomorphic symmetry groups can indeed be differentiated using the techniques of \SymTO{}.


 
\subsubsection{Review: Bosonic SPT stacking rules from \SymTO{}}
Let us first focus on bosonic SPT phases with symmetry group $G$.
In \SymTO{} different $G$ SPTs correspond to distinct fully-gapped  boundary conditions (Lagrangian condensations) of $D(G)$. 
Stacking of gapped boson SPTs then translates to a product structure on the set of Lagrangian condensations of $D(G)$.  To derive this product, imagine stacking two phases, each represented by a \SymTO{} thin-slab. Let $\A^{1,2}$ be the two Lagrangian condensations of $D(G^B)$ that produce phases $\EF^{1,2}$ when put on the physical boundary $\sf{B_{phys}}$. The stacked \SymTO{} sandwich now has the bulk theory $D(G\times G)$, and the reference boundary has the product condensation algebra $\A_{\sf{ref}}\boxtimes \A_{\sf{ref}}$. Similarly the physical boundary now has the product condensation algebra $\A^1\boxtimes\A^2$. This sandwich now describes a \oned{} system with symmetry $G\times G$. Following the second step of the stacking operation, We then break the symmetry down to the diagonal subgroup. This can be done by Higgsing the bulk gauge theory to the diagonal subgroup, which is achieved by condensing the following diagonal algebra
\begin{align}
    \E_b=\bigoplus_{R\in \rep(G)}e_R\otimes e_R^*.
\end{align}
in the \SymTO{} sandwich. After this diagonal condensation, the bulk theory reduces to the theory $D(G^d)$. The fluxes of the theory $D(G^d)$ are related to the original $D(G\times G)$ theory via $m^d_g\cong m^1_gm^2_g$, therefore correspond to the diagonal symmetry when acting on the reference boundary. The boundary conditions on $\sf{B_{ref}}$ and $\sf{B_{phys}}$ are now boundary conditions of the bulk $G^d$-gauge theory.  The condensation on the reference boundary will reduce from $\A_{\sf{ref}}\otimes \A_{\sf{ref}}$ to $\A_{\sf{ref}}$. This is because the condensation $\E_b$ identifies $e_R$ in the first copy of $D(G)$ with $e_R$ in the second copy. In other words the reference boundary condition is invariant under stacking. The new physical boundary is obtained as follows. For simplicity we consider abelian group $G$ for now. The physical boundary before condensing $\E_b$ is given by the Lagrangian algebra $\A^1\times \A^2$, which is a sum of anyons in $D(G\times G)$. The diagonal condensation will confine certain anyons of $D(G\times G)$, therefore we remove the confined anyons in the product $A^1\times A^2$ and identify the deconfined anyons with anyons in $D(G^d)$. This gives us a condensation algebra in $D(G^{d})$ and defines the new physical boundary condition for $D(G^d)$. This procedure defines a product between $\A^1$ and $\A^2$, which we denote as $\A^1\boxtimes_b\A^2$.

\begin{widetext}
\subsubsection{Example: $G=\bZ_2^2$ }
We illustrate this product structure with an example. Consider the \SymTO{} for $G=\bZ_2^A\times \bZ_2^B$. Condensing $\A^1=1+e_Am_B+e_Bm_A+f_Af_B$ at $\sf{B_{phys}}$ gives the non-trivial SPT phase protected by $G$. The reference boundary has the condensation $\A_{\sf{ref}}=1+e_A+e_B+e_Ae_B$. Now let us try to stack two $\A^1$s.
The diagonal condensation is
\begin{align}
    \E_b=1+e_Ae_A'+e_Be_B'+e_Ae_Be'_Ae'_B.
\end{align}
 The deconfined anyons are generated by $e_A\E_b, e_B\E_b, m_Am'_A\E_b, m_Bm'_B\E_b$. If we denote anyons of $D(G^{d})$ by $a^*$, then anyons of $D(G^d)$ are related to those of $D(G\times G)$ via $e^*_{A}=e_{A}\E_b, e^*_{B}=e_B\E_b, m^*_A=m_Am'_A\E_b, m^*_B=m_Bm'_B\E_b$. As promised, the new fluxes are diagonal products of the fluxes of the two copies. Using this identification, we see that the reference boundary condition becomes 
\begin{align}
    \A_{\sf{ref}}\boxtimes \A_{\sf{ref}}=(1+e_A+e_B+e_Ae_B)(1+e'_A+e'_B+e'_Ae'_B)= 1^*+e^*_A+e^*_B+e^*_Ae^*_B.
\end{align}
Therefore we have $\A_{\sf{ref}}\boxtimes_b \A_{\sf{ref}}=\A_{\sf{ref}}$, as claimed. The physical boundary condition is
\begin{align}
    \A^1\boxtimes \A^1=(1+e_Am_B+e_Bm_A+f_Af_B)(1'+e'_Am'_B+e'_Bm'_A+f'_Af'_B).
\end{align}
Expand this product and collect the deconfined anyons, we have
\begin{align}
    \A^1\boxtimes \A^1\to 1+e_Am_Be'_Am'_B+e_Bm_Ae'_Bm'_A+f_Af_Bf'_Af'_B=1^*+m^*_B+m^*_A+m^*_Am^*_B.
\end{align}
We therefore have $\A^1\boxtimes_b\A^1=1+m_A+m_B+m_Am_B$, which is exactly the Lagrangian condensation that represents the trivial SPT phase. We have recovered the $\bZ_2$-stacking rule of $\bZ_2^2$-SPTs.
\end{widetext}

The stacking of two Lagrangian condensations in other Abelian quantum doubles can be dealt with in a similar fashion, and it can be shown that the stacking of Lagrangian algebras defined in this way agrees with the stacking of bosonic  $G$-SPTs. This construction can also be extended naturally to generic non-abelian symmetry $G$, the details of which will appear elsewhere. 
\subsubsection{Fermionic SPT stacking rules in \SymTO{}}
Similar to the bosonic case, fermionic SPTs form a group under stacking. By the \SymTO{} dictionary this means there is another product structure on Lagrangian condensations of $D(G^F)$, distinct from $\boxtimes_b$, that matches with the fermionic SPT stacking rule. To motivate the new product, let us revisit the \SymTO{} for $\bZ_2^F$. The reference boundary has condensation $1+f$. If we stack two $\bZ_2^F$-systems, the reference boundary becomes $\A_{\sf{ref}}\boxtimes\A_{\sf{ref}}=(1+f)(1+f')$. If we continue to use the diagonal condensation $\E_b=1+ee'$, then, the deconfined anyons in the product $\A_{\sf{ref}}\boxtimes\A_{\sf{ref}}=1+f+f'+ff'$  are $1+ff'\cong 1+m^*$. We see the reference boundary is no longer invariant under the bosonic stacking rule.  Consider another diagonal condensation $\E_f=1+ff'$. The deconfined anyons are generated by $mm'\E_f, f\E_f$, which form a single copy of toric code under the identification $m^*:=mm'\E_f, f^*:=f\E_f, e^*:=em'\E_f$. Here the identification is chosen so that $mm'$ is mapped to the new $m^*$. This is because $m$-line is identified with the symmetry generator, therefore we wish the diagonal symmetry to be represented by $mm'$. The reference boundary now changes to $(1+f)(1+f')=(1+ff')+(f+f')=1^*+f^*$. The reference boundary is thus invariant under the new diagonal condensation $\E_f$.  For a general fermionic symmetry $G^F$, we take the diagonal condensation that breaks $G^F\times G^F$ to the diagonal $G^{F,d}$ to be
\begin{align}
    \E_f=&\left(\bigoplus_{R,|R|=0}e_R\otimes e_R^*\right)
\nonumber\\
&\bigoplus\left(\bigoplus_{R,|R|=1} e_Rm^F\otimes e_R^*m^F\right).
\end{align}
This is the proper modifification of $\E_b$, because odd charges of $G^F$ are really the dyons $e_Rm^F$. For Abelian $G^F$, the stacking of two Lagrangian condensations $\A^1, \A^2$ at $\sf{B_{phys}}$ is defined similar to the bosonic case, namely one first forms the product $\A^1\boxtimes \A^2$, then deletes the anyons that are confined after condensing $\E_f$. The remaining anyons define a Lagrangian condensation of $D(G^{F,d})$. This product is denoted by $\A^1\boxtimes_f \A^2$. Next we use an example to illustrate how the new diagonal condensation $\E_f$ captures the difference between bosonic and fermionic SPT stacking rules.
\subsubsection{Example: stacking of Kitaev chains}
The simplest non-trivial fermionic invertible phase is the Kitaev chain, represented by the condensation $\A=1+e$ in the $\bZ_2^F$-\SymTO{}. The stacking of two Kitaev chain should be the trivial phase, represented by the condensation $\A_0=1+m$ in the $\bZ_2^F$-\SymTO{}.

Now we apply the general construction illustrated above to derive this. The product algebra is initially 
\begin{align}
    \A\times \A'=1+e+e'+ee'.
\end{align}
The anyons in the sum that stay deconfined after the diagonal condensation $1+ff'$ are $1, ee'$.
Using the relation $m^*=mm'\cong ee', f^*:=f\cong f, e^*:=em'\cong e'm$, we have 
\begin{align}
    \A\boxtimes_f \A\cong 1+ee'\cong 1+m^*=\A_0,
\end{align}
reproducing the stacking rule of the Kitaev chain.
\subsubsection{Example: $G^F=\bZ_2\times \bZ_2\times \bZ_2^F$ and $G^B=\bZ_2\times \bZ_2\times \bZ_2$}
The reason for considering this symmetry group is that this is the smallest group for which bosonic and fermionic SPTs have different stacking rules.  This means given elements $\omega_1, \omega_2\in H^2[G^F\cong G^B,U(1)]$ , and fermionic phases determined by them $\EF_{\omega_1}, \EF_{\omega_2}$, we have $\EF_{\omega_1}\boxtimes_f \EF_{\omega_2}\neq \EF_{\omega_1\omega_2}$, where $\boxtimes_f$ means stacking of fermionic phases.

There are various ways of deriving the stacking rule, here we provide a method based on  symmetry fractionalization of gapped symmetries. For a fermionic phase $\EF_{\omega}$ in the class $[\omega]\in H^2[G^F,U(1)]$, the symmetry actions $U_g,~g\in G^F$ form a projective representation at the edge of the system: $\widetilde{U}_g \widetilde{U}_h=\omega(g,h)\widetilde{U}_{gh}$. Here $\widetilde{U}$ stands for the localized version of the symmetry action that only acts on one of the edges of the system.  For the symmetry group $\bZ^A_2\times \bZ^B_2\times \bZ_2^F$, since there is no projective representation of $\bZ_2$ alone, the projective representations of $\bZ^A_2\times \bZ^B_2\times \bZ_2^F$ can be labelled by which pairs of $\bZ_2$-generators anti-commute. Consider a phase $\EF_1$ with anti-commuting $\bZ_2^A$ and $\bZ_2^F$ generators. Denote the localized unitaries by $\widetilde{U}^{\EF_1}_{g^A}$ and $\widetilde{U}^{\EF_1}_{P_f}$, then we have $\widetilde{U}^{\EF_1}_{g^A}\widetilde{U}^{\EF_1}_{P_f}=-\widetilde{U}^{\EF_1}_{P_f}\widetilde{U}^{\EF_1}_{g^A}$. Similarly consider another phase $\EF^2$ with anti-commuting $\bZ_2^B$ and $\bZ_2^F$ generators, we have $\widetilde{U}^{\EF_2}_{g^B}\widetilde{U}^{\EF_2}_{P_f}=-\widetilde{U}^{\EF_2}_{P_f}\widetilde{U}^{\EF_2}_{g^B}$. Notice being charged under fermion parity means the operator is fermionic. In the current case this means $\widetilde{U}^{\EF_1}_{g^A}$ and $\widetilde{U}^{\EF_2}_{g^B}$ are both fermionic operators. Next we stack $\EF^1$ with $\EF^2$, the generator of the $\bZ_2^A, \bZ_2^B, \bZ_2^F$ factors in $\EF^1\boxtimes \EF^2$ are $\widetilde{U}^{\EF_1}_{g^A}, \widetilde{U}^{\EF_2}_{g^B}$ and $\widetilde{U}^{\EF_1}_{P_f}\widetilde{U}^{\EF_2}_{P_f}$ respectively. We see that now the pairs $(\bZ_2^A,\bZ_2^F)$ and $(\bZ_2^B,\bZ_2^F)$ anti-commute. This is expected. However, since $\widetilde{U}^{\EF_1}_{g^A}$ and $\widetilde{U}^{\EF_2}_{g^B}$ are both fermionic operators, they also anti-commute due to fermion statistics. 

If we label a $\bZ_2^A\times \bZ_2^B\times \bZ_2^F$ projective representation by $(i,j,k)$, where $i=0,1$ labels whether $\bZ^A$ and $\bZ_2^F$ generators anti-commute, $j=0,1$ labels whether $\bZ^B$ and $\bZ_2^F$ generators anti-commute, and $k=0,1$ labels whether $\bZ_2^A$ and $\bZ_2^B$ generators anti-commute. Then the stacking rule we just derived states $(1,0,0)\boxtimes (0,1,0)=(1,1,1)$. However, for a bosonic symmetry with $\bZ_2^F$ replaced by another bosonic $\bZ_2^C$ factor, we would have $(1,0,0)\boxtimes (0,1,0)=(1,1,0)$: the $\bZ_2^A$ and $\bZ_2^B$ generators in the stacked system would commute instead of anti-commute. How do we see this difference between bosonic and fermionic stacking rules in \SymTO{}? Let us first construct the \SymTO{} for $\EF^1$ and $\EF^2$. 

The \SymTO{} for $\bZ_2^A\times \bZ_2^B\times \bZ_2^F$ is the quantum double $D(\bZ_2^A\times \bZ_2^B\times \bZ_2^F)$. The reference boundary condition is $\A_{\sf{ref}}=\<e_1,e_2,f_3\>$. Consider the two choices of gapped physical boundary condition $\A^{1}=\<e_1m_3,e_3m_1, m_2\>, \A^{2}=\<e_2m_3,e_3m_2,m_1\>$. By studying their edge modes one can verify $\A^{1,2}$ describe the fermionic phases $(1,0,0)$ and $(0,1,0)$ respectively.  

\begin{widetext}   
The diagonal condensation is $\E_f=1+e_Ae'_A+e_Be'_B+f_Cf'_C$, which leaves $m_A^*:=m_Am_A', m_B^*:=m_Bm_B', m_C^*=m_Cm_C', e_A^*:=e_A, e_B^*:=e_B, e_C^*:=e_Cm_C'$ deconfined. In the product $\A^1\boxtimes \A^2$, the deconfined anyons are generated by
\begin{align}
    e_1e_2'm_3m_3'=e_1^*e_2^*m_3^*,~e_1e_3'm_2m_2'm_3=e_1^*e_3^*m_2^*,~e_3e_2'm_1m_1'm_3'=e_3^*e_2^*m_1^*.
\end{align}
Therefore $\A^1\boxtimes_f \A^2=\<e_1e_2m_3, e_1e_3m_2,e_3e_2m_1\>$. Analyzing the edge mode structure of the RHS, we see that all three pairs of the $\bZ_2$ generators anti-commute. We have reproduced the correct stacking rule $(1,0,0)\boxtimes_f (0,1,0)=(1,1,1)$. Had we used the diagonal condensation $\E_b$, the stacking rule would become $\A^1\boxtimes_b \A^2=\<e_1e_2m_3, e_3m_2,e_3m_1\>$, which matches with the bosonic stacking rule $(1,0,0)\boxtimes_b (0,1,0)=(1,1,0)$. Therefore the new diagonal condensation $\E_f$ is crucial in obtaining the correct fermionic SPT stacking rule.
\end{widetext}

In general, for two fermionic phases given by $[\omega_1], [\omega_2]\in H^2[G^F,U(1)]$, their stacking is 
\begin{align}
    [\omega_1]\boxtimes_f [\omega_2]=[\omega_1\omega_2 \nu_1\cup \nu_2],\label{eq: F-stacking}
\end{align}
where 
\begin{align}
    \nu_i(g)=\frac{\omega_i(g,P_f)}{\omega_i(P_g,g)}=\pm 1
\end{align}
measures fermion parity of the edge operator $\widetilde{U}_g$. The extra factor $\nu_1\cup \nu_2$ accounts for the anti-commutation between edge operators due to fermion statistics. The discussion of the $\bZ_2\times\bZ_2\times \bZ_2^F$-SPT stacking rule can be generalized to generic $G^F$-SPT, and the stacking rule \eqref{eq: F-stacking} can be reproduced by the stacking of Lagrangian algebras in $D(G^F)$. The details of this computation will appear elsewhere.

\section{Bosonization in \SymTO{}\label{sec: bz}}

The crucial difference between \SymTO{} for bosonic and fermion symmetries is the choice of reference boundary condition. If we fix the boundary condition at $\sf{B_{phys}}$, choosing different boundary conditions amounts to some topological manipulations acting on the original 1+1D system. A natural question to consider is the following: if we fix the boundary condition at $\sf{B_{phys}}$, is there any relation between the two phases represented by reference boundaries $\A_{\sf{ref}}^f$ and $\A_{\sf{ref}}^b$ (given by Eq.~\eqref{eq:bref})? We show in this section that the relation is exactly bosonization, also known as Jordan-Wigner transformation in \oned{}. More precisely, if the \SymTO{} sandwich with $A_{\sf{ref}}^f$ reference boundary represents a \oned{} fermionic phase $\sf{T}$ with symmetry $G^F$, then by changing the reference boundary to $A_{\sf{ref}}^b$ we obtain exactly the Jordan-Wigner transformation of $\sf{T}$, which may have a different (non-isomorphic) bosonic symmetry $G^B$.

Bosonization \cite{thorngren2019anomalies,Inamura2022,Karch_2019} is a general correspondence between bosonic systems (whose fundamental degrees of freedom are bosonic) and fermionic systems (whose fundamental degrees of freedom
are fermionic). It offers a unique and effective way to deal with complex fermionic problems by transforming them into more manageable bosonic problems. The \SymTO{} framework we have developed for fermionic symmetries offers a fresh perspective on bosonization within the broader context of managing general topological manipulations, as we now explore.

\subsection{Bosonization of an anomaly-free symmetry} 

 
Before we initiate the proof let us review bosonization/Jordan-Wigner transformation in the language of topological field theory. Formally, bosonization is the process of ``summing over spin structures'' for the partition function of fermionic systems. When no gravitational anomaly is present, bosonization can also be refered to as  ``gauging fermion parity''.\footnote{As discussed in \cite{smith2024backfiring}, for \oned{} fermionic systems, when the gravitational anomaly corresponds to a multiple of 16, there is still a well-defined procedure of ``summing over spin structures'', which coincides with a properly defined version of ``gauging fermion parity''. When the gravitation anomaly is a multiple of 8 but not 16, this process of ``gauging fermion parity'' can still be defined but will generate a fermionic theory instead of a bosonic theory. Hence, there is a subtle difference between the two concepts that is fortunately absent in our context.} One can also stack the system with an invertible fermion topological order (e.g. the Kitaev chain in 1+1D) before summing over spin structures. These two options result in phases related by Kramers-Wannier (KW) duality. The standard Jordan-Wigner transformation, where the trivial fermionic phase is mapped to the trivial bosonic phase, corresponds to the second choice.  

Following our discussion of spin structure in Section~\ref{sec:GFspin}, the partition function of a \oned{} fermionic theory coupled to a $G^F$-spin structure can be written as $\Z_{\sf{T}}(\eta,a_b)$. The bosonized theory has an isomorphic symmetry $G\cong G^F$, therefore it couples to $G$-gauge fields. We can split a $G\cong G^F$ gauge field into a $\bZ^p_2$-gauge field $a_p$, and a $G^B$-gauge field $a_b$, with relation $da_p=a_b^*(\rho)$. Here $\bZ_2^p$ is a  subgroup of the bosonic symmetry $G^B$, which comes from the fermion parity subgroup $\bZ_2^F$. The partition function of the bosonized theory can then be written as $\Z_{\JW^{-1}(\sf{T})}[a_b,a_p]$. The relation between the two partition functions is
\begin{align}\label{eq:bosonization}
    \Z_{\JW^{-1}(\sf{T})}[a_b,a_p]=\sum_{\eta}\frac{\arf(\eta+\tau+a_p)}{\sqrt{|H^1[M,\bZ_2]|}}\Z_{\sf{T}}[a_b,\eta].
\end{align}

Now we derive this result purely within our construction of \SymTO{}. The fermionic reference boundary state with $G^F$-spin structure $(\eta,a_b)$ is given by~\eqref{eq:f-bdry-2}

\begin{align}\label{eq:arfinterface}
    &|\EF(\eta,a_b)\>:=\nonumber\\
    &\sum_{A_{\bZ_2},dA_{\bZ_2}=a_b^*(\rho)}\frac{\arf(\eta+A_{\bZ_2}+\tau)}{\sqrt{|H^1[M,\bZ_2]|}}|A_{\bZ_2},A_b=a_b\>.
\end{align}
Summing over $\eta$ weighted by the $\arf$ invariant, we have 
\begin{widetext}
\begin{align}\label{eq: boundary state relation}
    \sum_{\eta,d\eta=a_b^*(\rho)+w_2}\frac{\arf(\eta+a_p+\tau)}{\sqrt{|H^1[M,\bZ_2]|}}|\eta,a_b\>&=\sum_{\eta,A_{\bZ_2}}\frac{\arf(\eta+a_p+\tau)}{\sqrt{|H^1[M,\bZ_2]|}} \frac{\arf(\eta+A_{\bZ_2}+\tau)}{\sqrt{|H^1[M,\bZ_2]|}}|A_{\bZ_2},A_b=a_b\>\\
    &=\sum_{\eta,A_{\bZ_2}}\frac{q_{\eta+\tau+a_p}(a_p+A_{\bZ_2})}{|H^1[M,\bZ_2]|}|A_{\bZ_2},A_b=a_b\>\\
    &=\sum_{A_{\bZ_2}}\delta_{a_p+A_{\bZ_2}}|A_{\bZ_2},A_b=a_b\>\\
    &=|A_{\bZ_2}=a_p,A_b=a_b\>,
\end{align}
where we used relation 
\begin{align}
    \arf(\tau+x)=\arf(\tau)q_\tau(x),~\sum_{\tau}q_\tau(x)=|H^1[M,\bZ_2]|\delta_x
\end{align}
for a spin structure $\tau$ and an $x\in H^1[M,\bZ_2]$.
The resulting state $|A_{\bZ_2}=a_p,A_b=a_b\>$ is exactly the Dirichlet boundary with $G^B$ gauge field fixed to the value $(a_p,a_b)$. Doing the inner product of Eq.~\eqref{eq: boundary state relation} with the state $|\omega\rangle$ from the physical boundary, we obtain Eq.~\eqref{eq:bosonization}. Therefore the bosonic reference boundary with $\A^b_{\sf{ref}}$ condensed is exactly tha bosonization of the fermionic boundary with $\A^f_{\sf{ref}}$ condensed.
\end{widetext}

Some examples we have studied in previous sections can be understood from the perspective of bosonization. For example, the Majorana CFT is related to the Ising CFT by bosonization, and we have seen that the two phases are indeed related by changing the reference boundary from $\A_{\sf{ref}}^f=1+f$ to $\A_{\sf{ref}}^b=1+e$. Another example is the $e$-condensed reference boundary, this is the Kitaev chain with $\A_{\sf{ref}}^f=1+f$ and the SSB phase with $\A_{\sf{ref}}^b=1+e$. Indeed the SSB phase of $\bZ_2$ is the bosonization of the Kitaev chain phase.

Before the end of this subsection, we mention that the fermionic reference boundary Eq.~\eqref{eq:arfinterface} can also be interpreted as an invertible interface (called an Arf interface in \cite{smith2024backfiring}, and generally a bimodule in higher dimensions \cite{debray2023bosonization}) fused with the bosonic reference boundary. It is possible to understand various aspects of bosonization from the interplay of this interface with other invertible defects of the \SymTO{}, as initiated in \cite{smith2024backfiring}. We defer it to future study.

\subsection{Bosonization of anomalous symmetry\label{sec: bza}}

So far the examples studied have non-anomalous symmetries. In the following we study bosonization of some anomalous symmetries. What is interesting is that in the presence of anomaly the bosonization does not preserve the structure of the symmetry group. The elementary example is when the fermionic symmetry is $\bZ_2\times \bZ_2^F$, this symmetry has a $\bZ_8$-classification of anomaly. The bosonization of the $\nu=2$ level anomaly is  a $\bZ_4$ symmetry with $k=2$ level anomaly. What is even more interesting is the $\nu=1$ level anomaly, this is mapped to the non-invertible Ising symmetry under bosonization. These facts once again show that the \SymTO{} is what is intrinsic in defining a symmetry instead of the group structure. We will show how these dualities can be derived from simple \SymTO{} considerations. 

\subsubsection{The $\nu=2$ anomaly of $\bZ_2\times \bZ_2^F$}
The \SymTO{} for the $\nu=2$ anomalous $\bZ_2\times \bZ_2^F$ symmetry has been derived in Section~\ref{sec:anomalous SymTFT}. Recall that the theory $\Z[\bZ_2\times\bZ_2^F,\nu=2]$ is generated by two order-4 anyons $\phi, \phi_F$, which correspond to gauge fluxes of $\bZ_2$ and $\bZ_2^F$ respectively. They have quantum spins $\theta_{\phi_F}=1, \theta_{\phi}=e^{i\pi/4}$ and mutual statistics $\theta_{\phi,\phi_F}=e^{i\pi/2}$.

Since $\Z[\bZ_2\times \bZ_2^F,\nu=2]$ is a non-chiral Abelian topological order, it must be equivalent to a twisted Abelian quantum double. Indeed it is equivalent to the $\bZ_4$ twisted quantum double $D_{k=2}(\bZ_4)$, where $k\in H^3[\bZ_4,U(1)]$ labels the twists of $\bZ_4$. The theory $D_{k=2}(\bZ_4)$ is generated by a bosonic gauge charge $e$ and a flux $m$ with quantum spin $\theta_m=e^{i\pi /4}$. The mutual statistics is $\theta_{e,m}=e^{i\pi /2}$.  We see that an isomorphism to the theory $\Z[\bZ_2\times \bZ_2^F,\nu=2]$ can be achieved by the mapping $\phi\simeq m, \phi_F\simeq e$. 
Consider a bosonic gapped reference boundary of $\Z[\bZ_2\times\bZ_2^F,\nu=2]\cong D_{k=2}(\bZ_4)$ obtained by condensing $e$. The non-trivial defect lines on this boundary are generated by $m\simeq \phi$, which has a $\bZ_4$ fusion rule and quantum spin $\theta_m=e^{i\pi /4}$. Therefore if we choose to condense $e$ on the reference boundary, then the \SymTO{} describes a bosonic $\bZ_4$ symmetry, with level $k=2$ anomaly. This anomaly can also be seen from the $F$-symbols of the $m$-lines. This leads to the conclusion 
\begin{mdframed}
    The fermionic symmetry $\bZ_2\times\bZ_2^F$ with anomaly $\nu=2$ is dual to the bosonic symmetry $\bZ_4$ with anomaly $k=2$.
\end{mdframed}
This result has been derived by direct bosonization in~\cite{thorngren2019anomalies}, here we see the \SymTO{} provides a very simple derivation. 

Moreover, we may generalize the ``equivalent symmetry" principle in~\cite{Kong_2020,Chatterjee_2023}to include both bosonic and fermionic symmetries. We state that, two symmetries, either bosonic or fermionic, are equivalent, if they provide the same set of constraint on local symmetric operators. Then, the sets of local symmetric operators under two equivalent symmetries have an one-to-one correspondence and the same algebraic relations. The Hamiltonians as sums of those symmetric local operators also have an one-to-one correspondence, and the corresponding Hamiltonians have the same spectrum.  Based on the \SymTO{} framework we developed for fermionic symmetries, we claim that two symmetries, either bosonic or fermionic, are equivalent, if they have isomorphism \SymTO{}. For instance, the fermionic symmetry $\bZ_2\times \bZ_2^F$, with $\nu=2$ anomaly, is equivalent to the bosonic symmetry $\bZ_4$ with anomaly $k=2$.

\subsubsection{The $\nu=1$ anomaly of $\bZ_2\times \bZ_2^F$}
Let us now consider the more exotic case: the root anomaly of $\bZ_2\times \bZ_2^F$. This anomaly is realized on the boundary of the root phase of 2+1D $\bZ_2$-symmetric topological superconductor. This 2+1D phase is the stacking $p+ip\boxtimes p-ip$, with the $p+ip$ layer charged under $\bZ_2$. The \SymTO{} for the $\nu=1$ anomalous $\bZ_2\times \bZ_2^F$ is then the gauged $p+ip\boxtimes p-ip$. This theory, denoted by $\Z[\bZ_2\times\bZ_2^F,\nu=1]$, is just $\text{Ising}\boxtimes\overline{\text{Ising}}$. Denote the Ising anyons by $\sigma, \psi$, and the time-reversal by $\overline{\sigma},\overline{\psi}$, with fusion rules $\sigma\times \sigma=1+\psi$. Then the flux of $\bZ_2$ is the $\pi$-flux in the $p+ip$ layer, $\sigma$. The $\bZ_2^F$ flux is the $\pi$-flux in both layers, $\sigma\overline{\sigma}$. $\psi,\overline{\psi}$ are charged and uncharged fermions respectively, and $\psi\overline{\psi}$ is the charged boson. On the reference boundary we should condense all gauge charges, in this case $\psi$ and $\overline{\psi}$. The non-trivial defect lines on the reference boundary is then generated by $\sigma,\overline{\sigma}$. With $\psi,\overline{\psi}$ condensed, the flux lines on the reference boundary indeed have the $\bZ_2\times \bZ_2^F$ fusion rule: $\sigma^2\simeq 1, \overline{\sigma}^2\simeq 1$. 

To obtain the bosonized reference boundary we make the following observation: Recall in the non-anomalous case the fermionic reference boundary condenses the fermionic anyons $e_Rm^F$, which is the bosonic gauge charge $e_R$, bound with fermion parity flux $m^F$. Reverse this relation, we may say the bosonic gauge charge $e_R$, is obtained by binding a fermion parity flux $m^F$ to $e_Rm^F$. Generalize this observation to the current case, we see that the fermionic charges condensed on the reference boundary are $\psi,\overline{\psi}$. If we bind a fermion parity flux $\sigma\overline{\sigma}$ to them, we obtain $\psi\times \sigma\overline{\sigma}=\sigma\overline{\sigma}$ and $\overline{\psi}\times \sigma\overline{\sigma}=\sigma\overline{\sigma}$. Therefore, the bosonized reference boundary condenses $\sigma\overline{\sigma}$ and $\psi\overline{\psi}$.
Notice that the theory $\Z[\bZ_2\times \bZ_2^F,\nu=1]$ is equivalent to $\text{Ising}\boxtimes \overline{\text{Ising}}$, we may also view it as the \SymTO{} for the Ising symmetry. The bosonized boundary has $\sigma\overline{\sigma}, \psi\overline{\psi}$ condensed. This is exactly the Lagrangian condensation that describes the Ising non-invertible symmetry. Namely, this condensation leaves non-trivial defect lines generated by $\sigma\cong \overline{\sigma}, \psi\cong \overline{\psi}$, with Ising fusion rules $\sigma\times\sigma=1+\psi,~\sigma\times\psi=\sigma,~\psi\times \psi=1$. Therefore  we have the conclusion the conclusion
\begin{mdframed}
    The fermionic symmetry $\bZ_2\times\bZ_2^F$ with anomaly $\nu=1$ is equivalent to the non-invertible Ising symmetry.
\end{mdframed}
We see that the \SymTO{} provides us a systematic and convenient way to identify equivalent symmetries, including both bosonic and fermionic symmetries.
\section{Summary and outlook}\label{sec: summary}
In this work we developed a framework of topological holograph for fermionic symmetries of fermionic systems. For a group-like symmetry $G^F$ of \oned{} systems, the \SymTO{} is the \twod{} bosonic topological order $D(G^F)$ together with a fermionic reference boundary, with a fermionic condensation $\A_{\sf{ref}}^f$. The \SymTO{} sandwich describes an effective \oned{} fermionic system with $G^F$ symmetry. We showed how various aspects of fermionic systems, such as the non-trivial invertible phase modeled by the Kitaev chain, dependence on spin structure, etc. are described in the \SymTO{} framework. We analyzed gapped fermionic phases, including the Kitaev chain and fermionic SPTs via the \SymTO{}. We showed that the \SymTO{} gives full characterization of fermionic invertible phases, including their partition function, edge modes, and their stacking rules. 

We also studied fermionic gapless phases and critical points in \SymTO{}. We reproduced the partition function of the Majorana CFT via \SymTO{}. Moreover, we used the \SymTO{} framework to construct an example of an intrinisically-fermionic, intrinsically gapless SPT with $\bZ_8$ symmetry, uncover its edge-mode structure, and showed how the \SymTO{} structure can be directly used to construct a field theory description of this phase. This demonstrated that beyond reproducing known properties of phases, the \SymTO{} also has proved to be a \emph{useful} theoretical tool for designing novel gapless phases. 


Finally, we addressed the issue of bosonization in the context of \SymTO{}. We showed that bosonization in \SymTO{} amounts to changing the reference boundary from the fermionic condensation $\A^f_{\sf{ref}}$ to the bosonic condensation $\A^b_{\sf{ref}}$. With this understanding, we showed that the $\nu=1$ anomalous $\bZ_2\times \bZ_2^F$ symmetry bosonizes to the non-invertible Ising symmetry and the $\nu=2$ anomalous $\bZ_2\times \bZ_2^F$ symmetry bosonizes to the $k=2$ bosonic $\bZ_4$ symmetry. 

We end this paper with a few comments. 

\begin{enumerate}
\item There are other works discussing the construction of \SymTO{} for fermionic systems, including \cite{Gaiotto2020,freed2023topological,KantaroLecture}. In particular, in \cite{KantaroLecture}, the authors used a 4d-3d-2d construction to build the \SymTO{} for a \oned{} fermionic system. The whole system is constructed from a rigid symmetric monoidal 4-category called \textbf{BrFus} according to the cobordism hypothesis \cite{Baez_1995,lurie2009classification}. The 4d bulk is a Crane-Yetter like theory, which is ``Morita equivalent'' to the trivial theory. After trivializing the 4d bulk, we believe that their construction matches with our construction. 

\item These references also suggest that it may be possible to construct the \SymTO{} for \oned{} fermionic systems from a \twod{} \emph{fermionic} topological order. It is interesting to work out the details and rederive our results using this approach. 

\item It is also interesting to extend our approach to \oned{} systems with gravitational anomaly/nonzero chiral central charge. We expect that the \twod{} \SymTO{} should correspond to a chiral topological order. See \cite{smith2024backfiring,KantaroLecture} for some construction in this direction.

\item In this work we focused mostly on invertible fermionic symmetries in \oned. We expect that the framework established in this work can be naturally generalized to non-invertible fermionic symmetries. A non-invertible fermionic symmetry is described by a super-fusion category with a distinguished order 2 invertible element as the fermion parity symmetry. In particular, we expect that if the fermion parity $\bZ_2^F$ is non-anomalous, then the \SymTO{} for a super-fusion categorical symmetry $\EF$ is the bosonic topological order $\Z[\EF_b]$, where $\EF_b$ is the bosonization (also known as the modular extension \cite{Bruillard_2017}) of the super-fusion category $\EF$. However, on the reference boundary of $\Z[\EF_b]$, there should be a canonical fermionic condensation, to ensure the \SymTO{} sandwich describes the original fermionic symmetry $\EF$.  In particular, this fermionic reference boundary should be related to the standard bosonic reference boundary by ``binding fermion parity flux", similar to the relation between $\A^b_{\sf{ref}}$ and $\A^f_{\sf{ref}}$ we discussed in Section~\ref{sec:general condensation}. 

\item In higher dimensions, one may attempt to use the tools of higher fusion categories and its center to formulate the \SymTO{}~\cite{Kong_2020,Chatterjee_2023}. From a more physical point of view, the \SymTO{} is simply the ``gauged SPT" in one dimension higher. For instance, for the fermionic parity $\bZ_2^F$ in \twod{}, the \SymTO{} is simply the gauged $3+1D$ trivial insulator, also known as the $3+1D$ toric code with a fermionic charge. The theory of boson condensation in $3+1D$ has been developed in ~\cite{Zhao_2023,kong2024higher}.  It would be interesting to develop the theory of fermionic condensation in $3+1D$ and apply it to the \SymTO{} theory. For instance, non-Lagrangian fermionic condensation in the \SymTO{} of a \twod{} fermionic symmetry would correspond to fermionic \twod{} gSPTs. 

\end{enumerate}

\textit{Acknowledgements} -- We thank Fiona Burnell, Julio Parra-Martinez, Joseph Sullivan, Kantaro Omori, Nathanan Tantivasadakarn for insightful
discussions. This work is supported by the Natural Sciences
and Engineering Research Council of Canada (NSERC), and
by an Alfred P. Sloan Foundation Fellowship (A.C.P.).

\textit{Note} -- Near the end of this work, we learn that Inamura-Ohmori group is also working on the construction of \SymTO{} for \oned{} fermionic systems \cite{KantaroLecture}. As far as we understand, our construction matches with their construction when no gravitational anomaly is present. See the discussion of Section~\ref{sec: summary} for more details. We also learn that there are some discussions about \SymTO{} for \oned{} fermionic systems in \cite{smith2024backfiring}, mostly focusing on bosonization.

\appendix

\section{Quantum double and its condensable algebra\label{app:DG algbera}}
We review the structure of the braided fusion category $D(G)\cong\sf{Z}[\text{Vec}_G]$, the category that describes the IR limit of the Kitaev quantum double lattice model. All linear spaces are over $\bC$.

An object of $D(G)$ is a finite dimensional $G$-graded linear space $X=\oplus_{g\in G} X_g$ together with a $G$-action $\phi: G\to \text{End}(X)$, such that $\phi(g)(X_h)\subset X_{ghg^{-1}}$. Such a $G$-action is called a \textit{compatible} one. We will abbreviate the $G$-action $\phi(g)(v)$ as $g\star v,~\forall v\in X$. The grading of a vector $v$ will be denoted by $|v|\in G$. The support of a graded vector space is $\text{supp}(X):=\{g\in G|X_g\neq 0\}$. 

A morphism between two objects $X,Y$ is a grading preserving linear map $f: X_g\to Y_g$, that preserves the $G$-action, $h\star\circ f=f\circ h\star$.  The monoidal product of two objects is the tensor product of vector spaces with grading
\begin{align}
    (X\otimes Y)_{g}:=\bigoplus_{h}X_h\otimes Y_{h^{-1}g},
\end{align}
and $G$-action 
\begin{align}
    g\star (x\otimes y)=g\star x\otimes g\star y.
\end{align}
The braiding is defined by 
\begin{align}
    R_{X,Y}:& X\otimes Y\to Y\otimes X,\\
&x\otimes y\mapsto (g\star y) \otimes x, \text{if}~|x|=g.
\end{align}
And the associator is trivial
\begin{align}
    F_{X,Y,Z}: (x\otimes y)\otimes z\mapsto x\otimes (y\otimes z).
\end{align}
The quantum dimension of an object is its dimension as a linear space.

Let us look at some examples of objects in this category.
\paragraph{Objects with trivial support.}
If an object is supported only at the identity element, i.e. $X=X_e$. Then the $G$-action maps $X_e\to X_e$, and form a representation of $G$. This kind of objects, labelled by representations of $G$, are called gauge charges. They form a fusion sub-category of $D(G)$, namely $\rep(G)$.

\paragraph{Objects with trivial $G$-action.}
The $G$-action by $h$ always maps the $g$-component of an object to its $hgh^{-1}$-component. Therefore the support of an object is always closed under conjugation. Take a conjugacy class $\sigma$, and consider the $G$-graded vector space
\begin{align}
    X=\oplus_{g\in \sigma}\bC_g
\end{align}
Denote the basis of the $g$-component as $e_g$. Define a  compatible $G$-action on $X$ as $h* e_g:=e_{hgh^{-1}}$. This family of objects are labelled by conjugacy classes of $G$, and are called gauge fluxes. They form a fusion sub-category of $D(G)$, namely $\text{Vec}_G$.

\paragraph{General simple objects}
The support of a general simple object of $D(G)$ still needs to be closed under conjugation. Therefore a simple object has its support in a conjugacy class of $G$, $\sigma$. Take a $g\in\sigma$. The $g$-component of the object, $X_g$, carries an representation of the centralizer $\C_G(g)$, and the $G$-action on $X$ is induced from this representation of $\C_G(g)$. Therefore, a general simple object of $D(G)$ is labelled as $(\sigma, r)$, where $\sigma$ is a conjugacy class of $G$, and $r\in \rep(\C_G(\sigma)$. If both $\sigma$ and $r$ are non-trivial, the object is called a dyon.

\section{Fermionic boundary of toric code and spin structures\label{app: f-bdry lattice}}
In this appendix we provide a lattice model for the $f$-condensed boundary of \twod{} toric code and show that the spin structure of the physical fermions is related to the expectation value of the $f$-loop on the $f$-condensed boundary. 
\subsection{Setup}
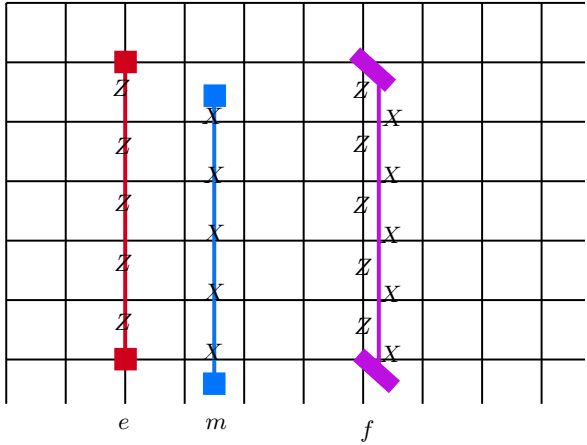
\begin{figure}

\tikzset{every picture/.style={line width=0.75pt}} 

\begin{tikzpicture}[x=0.75pt,y=0.75pt,yscale=-1,xscale=1]

\draw  [draw opacity=0] (67,101) -- (358.91,101) -- (358.91,303.37) -- (67,303.37) -- cycle ; \draw   (67,101) -- (67,303.37)(97,101) -- (97,303.37)(127,101) -- (127,303.37)(157,101) -- (157,303.37)(187,101) -- (187,303.37)(217,101) -- (217,303.37)(247,101) -- (247,303.37)(277,101) -- (277,303.37)(307,101) -- (307,303.37)(337,101) -- (337,303.37) ; \draw   (67,101) -- (358.91,101)(67,131) -- (358.91,131)(67,161) -- (358.91,161)(67,191) -- (358.91,191)(67,221) -- (358.91,221)(67,251) -- (358.91,251)(67,281) -- (358.91,281) ; \draw    ;
\draw [color={rgb, 255:red, 208; green, 2; blue, 27 }  ,draw opacity=1 ][fill={rgb, 255:red, 254; green, 26; blue, 55 }  ,fill opacity=1 ][line width=1.5]    (127,131) -- (127,281) ;
\draw  [color={rgb, 255:red, 208; green, 2; blue, 27 }  ,draw opacity=1 ][fill={rgb, 255:red, 208; green, 2; blue, 27 }  ,fill opacity=1 ] (122,125.64) -- (132.36,125.64) -- (132.36,136) -- (122,136) -- cycle ;
\draw  [color={rgb, 255:red, 208; green, 2; blue, 27 }  ,draw opacity=1 ][fill={rgb, 255:red, 208; green, 2; blue, 27 }  ,fill opacity=1 ] (122,275.64) -- (132.36,275.64) -- (132.36,286) -- (122,286) -- cycle ;
\draw [color={rgb, 255:red, 13; green, 114; blue, 237 }  ,draw opacity=1 ][fill={rgb, 255:red, 254; green, 26; blue, 55 }  ,fill opacity=1 ][line width=1.5]    (171.91,147.28) -- (171.91,293.19) ;
\draw  [color={rgb, 255:red, 0; green, 118; blue, 255 }  ,draw opacity=1 ][fill={rgb, 255:red, 1; green, 116; blue, 253 }  ,fill opacity=1 ] (167,142.64) -- (177.36,142.64) -- (177.36,153) -- (167,153) -- cycle ;
\draw  [color={rgb, 255:red, 0; green, 118; blue, 255 }  ,draw opacity=1 ][fill={rgb, 255:red, 1; green, 116; blue, 253 }  ,fill opacity=1 ] (166.73,288.01) -- (177.09,288.01) -- (177.09,298.37) -- (166.73,298.37) -- cycle ;
\draw [color={rgb, 255:red, 189; green, 16; blue, 224 }  ,draw opacity=1 ][line width=1.5]    (254.91,142.46) -- (254.91,291.73) ;
\draw  [color={rgb, 255:red, 189; green, 16; blue, 224 }  ,draw opacity=1 ][fill={rgb, 255:red, 189; green, 16; blue, 224 }  ,fill opacity=1 ] (245.54,124.13) -- (262.91,139.73) -- (257.96,145.25) -- (240.59,129.64) -- cycle ;
\draw  [color={rgb, 255:red, 189; green, 16; blue, 224 }  ,draw opacity=1 ][fill={rgb, 255:red, 189; green, 16; blue, 224 }  ,fill opacity=1 ] (247.54,276.13) -- (264.91,291.73) -- (259.96,297.25) -- (242.59,281.64) -- cycle ;

\draw (119,138.4) node [anchor=north west][inner sep=0.75pt]    {$Z$};
\draw (120,167.4) node [anchor=north west][inner sep=0.75pt]    {$Z$};
\draw (120,196.4) node [anchor=north west][inner sep=0.75pt]    {$Z$};
\draw (120,226.4) node [anchor=north west][inner sep=0.75pt]    {$Z$};
\draw (120,256.4) node [anchor=north west][inner sep=0.75pt]    {$Z$};
\draw (164,152.4) node [anchor=north west][inner sep=0.75pt]    {$X$};
\draw (165,182.4) node [anchor=north west][inner sep=0.75pt]    {$X$};
\draw (165,211.4) node [anchor=north west][inner sep=0.75pt]    {$X$};
\draw (165,241.4) node [anchor=north west][inner sep=0.75pt]    {$X$};
\draw (164,271.4) node [anchor=north west][inner sep=0.75pt]    {$X$};
\draw (240,139.4) node [anchor=north west][inner sep=0.75pt]    {$Z$};
\draw (240,166.4) node [anchor=north west][inner sep=0.75pt]    {$Z$};
\draw (240,197.4) node [anchor=north west][inner sep=0.75pt]    {$Z$};
\draw (241,228.4) node [anchor=north west][inner sep=0.75pt]    {$Z$};
\draw (241,258.4) node [anchor=north west][inner sep=0.75pt]    {$Z$};
\draw (255,153.4) node [anchor=north west][inner sep=0.75pt]    {$X$};
\draw (254,182.4) node [anchor=north west][inner sep=0.75pt]    {$X$};
\draw (254,212.4) node [anchor=north west][inner sep=0.75pt]    {$X$};
\draw (254,242.4) node [anchor=north west][inner sep=0.75pt]    {$X$};
\draw (254,272.4) node [anchor=north west][inner sep=0.75pt]    {$X$};
\draw (122,309.41) node [anchor=north west][inner sep=0.75pt]    {$e$};
\draw (166,309.41) node [anchor=north west][inner sep=0.75pt]    {$m$};
\draw (244,310.41) node [anchor=north west][inner sep=0.75pt]    {$f$};

\end{tikzpicture}
   
    \caption{The toric code lattice model. Representatives of the three types of string operators are shown. The excitations created at their ends are shown as red/blue/purple boxes.}
    \label{fig:tc lattice}
\end{figure}

The toric code is defined on a sqaure lattice. Every edge of the lattice has a two dimensional Hilbert space $\bC^2_e, e\in E$. The Pauli operators on an edge $e$ is denoted by $X_e,Z_e$. We will also denote the edge connecting two vertices $v_1,v_2$ as $v_1v_2$. The bulk Hamiltonian consists of two types of terms. The first type is vertex terms, defined for every vertex $v$ as 
\begin{align}
    A_v:=\prod_{vw\in E} X_{vw}
\end{align}
The second type is plaqutte terms, defined for every plaqutte $p$ of the lattice as
\begin{align}
    B_p:=\prod_{e\in \partial p} Z_e.
\end{align}
The bulk Hamiltonian is then sum of all vertex and plaqutte terms, 
\begin{align}
    H=-\sum_{v\in V}A_v-\sum_{p\in F} B_p
\end{align}
Since the Hamiltonian is a commuting projector one, the ground state is characterized by all vertex and plaqutte terms having eigenvalue $+1$.The excitations of this model can be labelled by the violations of vertex and plaquette terms. Violation of a vertex term will be called an $e$ particle. Violation of a plaqutte term will be called an $m$ particle. An adjacent pair of $e$ and $m$ is called an $f$  particle. String operators can be used to create pairs of excitations. The string operators associated with particles $e,m,f$ will be called $e/m/f$-string respectively. An $e$-string is defined on a path of the direct lattice as follows. Let $l=v_0v_1\cdots v_n$ be a path on the direct lattice. Then 
\begin{align}
    S_l^e:=\prod_{i=0}^{n-1}Z_{v_iv_{i+1}}
\end{align}
which creates a pair of $e$ particles at vertices $v_0, v_n$. An $m$-string is defined on a path of the dual lattice as follows. Let $\tilde{l}=\tilde{v}_0\tilde{v}_1\cdots \tilde{v}_n$ be a path of the dual lattice. For every edge $\tilde{v}_i\tilde{v}_{i+1}$ of the dual lattice, it intersects a unique edge of the direct lattice, denoted by $e_{i}$. Then 
\begin{align}
    S^m_{\tilde{l}}:=\prod_{i=0}^{n-1}X_{e_i}
\end{align}
An $f$-string is defined by an $e$-string and an adjacent $m$-string. The three types of string operators are depicted in Fig~\ref{fig:tc lattice}. 

\subsection{The $f$-boundary of toric code}
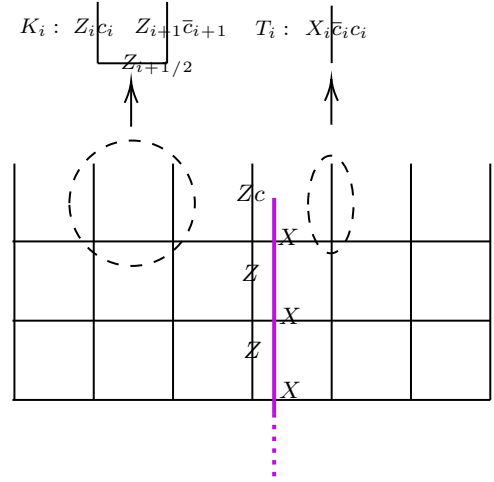
\begin{figure}

\tikzset{every picture/.style={line width=0.75pt}} 

\begin{tikzpicture}[x=0.75pt,y=0.75pt,yscale=-1,xscale=1]

\draw  [draw opacity=0] (318.91,220.28) -- (77.91,220.28) -- (77.91,101) -- (318.91,101) -- cycle ; \draw   (318.91,220.28) -- (318.91,101)(278.91,220.28) -- (278.91,101)(238.91,220.28) -- (238.91,101)(198.91,220.28) -- (198.91,101)(158.91,220.28) -- (158.91,101)(118.91,220.28) -- (118.91,101)(78.91,220.28) -- (78.91,101) ; \draw   (318.91,220.28) -- (77.91,220.28)(318.91,180.28) -- (77.91,180.28)(318.91,140.28) -- (77.91,140.28) ; \draw    ;
\draw  [dash pattern={on 4.5pt off 4.5pt}] (106.63,121.01) .. controls (106.63,103.54) and (120.8,89.37) .. (138.27,89.37) .. controls (155.75,89.37) and (169.91,103.54) .. (169.91,121.01) .. controls (169.91,138.49) and (155.75,152.65) .. (138.27,152.65) .. controls (120.8,152.65) and (106.63,138.49) .. (106.63,121.01) -- cycle ;
\draw    (137.77,82.37) -- (137.77,61.37) ;
\draw [shift={(137.77,59.37)}, rotate = 90] [color={rgb, 255:red, 0; green, 0; blue, 0 }  ][line width=0.75]    (10.93,-3.29) .. controls (6.95,-1.4) and (3.31,-0.3) .. (0,0) .. controls (3.31,0.3) and (6.95,1.4) .. (10.93,3.29)   ;
\draw    (120.91,50.37) -- (155.91,50.37) ;
\draw    (120.91,50.37) -- (120.91,19.38) ;
\draw    (155.91,50.37) -- (155.91,19.38) ;
\draw  [dash pattern={on 4.5pt off 4.5pt}] (227,121.64) .. controls (227,108.03) and (232.13,97) .. (238.46,97) .. controls (244.78,97) and (249.91,108.03) .. (249.91,121.64) .. controls (249.91,135.25) and (244.78,146.28) .. (238.46,146.28) .. controls (232.13,146.28) and (227,135.25) .. (227,121.64) -- cycle ;
\draw    (238.77,81.37) -- (238.77,60.37) ;
\draw [shift={(238.77,58.37)}, rotate = 90] [color={rgb, 255:red, 0; green, 0; blue, 0 }  ][line width=0.75]    (10.93,-3.29) .. controls (6.95,-1.4) and (3.31,-0.3) .. (0,0) .. controls (3.31,0.3) and (6.95,1.4) .. (10.93,3.29)   ;
\draw    (238.91,21.28) -- (238.91,50.28) ;
\draw [color={rgb, 255:red, 189; green, 16; blue, 224 }  ,draw opacity=1 ][line width=1.5]    (209.91,118.28) -- (209.91,226.92) ;
\draw [color={rgb, 255:red, 189; green, 16; blue, 224 }  ,draw opacity=1 ][line width=1.5]  [dash pattern={on 1.69pt off 2.76pt}]  (209.91,226.92) -- (209.91,260.56) ;

\draw (80,26.4) node [anchor=north west][inner sep=0.75pt]  [font=\footnotesize]  {$K_{i} :$};
\draw (107,26.4) node [anchor=north west][inner sep=0.75pt]  [font=\footnotesize]  {$Z_{i} c_{i}$};
\draw (138,26.4) node [anchor=north west][inner sep=0.75pt]  [font=\footnotesize]  {$Z_{i+1}\overline{c}_{i}{}_{+}{}_{1} \ $};
\draw (131,44.4) node [anchor=north west][inner sep=0.75pt]  [font=\footnotesize]  {$Z_{i}{}_{+}{}_{1}{}_{/}{}_{2}$};
\draw (224,27.4) node [anchor=north west][inner sep=0.75pt]  [font=\footnotesize]  {$X_{i}\overline{c}_{i} c_{i}$};
\draw (199,27.4) node [anchor=north west][inner sep=0.75pt]  [font=\footnotesize]  {$T_{i} :$};
\draw (192,150.4) node [anchor=north west][inner sep=0.75pt]    {$Z$};
\draw (193,189.4) node [anchor=north west][inner sep=0.75pt]    {$Z$};
\draw (211,172.4) node [anchor=north west][inner sep=0.75pt]    {$X$};
\draw (211,209.4) node [anchor=north west][inner sep=0.75pt]    {$X$};
\draw (210,132.4) node [anchor=north west][inner sep=0.75pt]    {$X$};
\draw (189,110.4) node [anchor=north west][inner sep=0.75pt]    {$Zc$};
\end{tikzpicture}

    \caption{Top: Two types of boundary Hamiltonian terms for the $f$-boundary. Purple string: an $f$-string that ends on the boundary. The endpoint has an operator $c$ attached. The purple string commutes with all bulk and boundary Hamiltonian terms.}
    \label{fig:f-bdry}
\end{figure}

Now consider the square lattice with a rough boundary as shown in Fig~\ref{app: f-bdry lattice}. The top array of edges are called boundary edges. The set of boundary edges is denoted by $E_{\partial}$. In addition to the $\bC^2$ spin degrees of freedom, we put a local fermion at every boundary edge, with a pair of Majorana operators $c_i, \bc_i, \forall e_i\in E_{\partial}$. Next we define the boundary Hamiltonian terms. There are two types of terms. The first type is defined on every boundary edge, denoted by $T_i$,
\begin{align}
    T_i=-X_i \bc_i c_i.
\end{align}
The second is defined on every ``horseshoe" plaquette as follows. For every adjacent boundary edges $e_i, e_{i+1}$, there is a unique bulk edge $e_{i+1/2}$ connecting the two boundary edges. Then the horseshoe term is defined as
\begin{align}
    K_i=-(Z_{i} c_{i}) Z_{i+1/2}(Z_{i+1} \bc_{i+1})
\end{align}
The two types of boundary terms are depicted in Fig~\ref{fig:f-bdry}. It is straightforward to see that the two types of boundary terms commute with each other and also commute with all bulk Hamiltonian terms. We then define the boundary Hamiltonian to be sum of all $K_i$ and $T_i$ terms. On every boundary edge there are two degrees of freedom, one spin and one fermion. Since there are also two boundary terms per boundary edge, the bulk+boundary Hamiltonian defines toric code with a gapped boundary. 

We now verify that this gapped boundary corresponds to the $fc$-condensed boundary of toric code. To show the anyon $f$ is condensed on the boundary, observe the string operator depicted in Fig.~\ref{app: f-bdry lattice}. The bulk part of this string operator is an $f$-string. However, there is a fermion operator $c$ attached to its end on the boundary. It is straightforward to verify that this string operator does not violate any bulk or boundary terms. Therefore, an $f$-particle can be moved from the bulk to the boundary and annihilated via this type of string operators. This shows $f$ is indeed condensed on the boundary. 
\subsection{Lattice version of $\bZ_2^F$-\SymTO{}}
\begin{figure}
    \centering
    \includegraphics{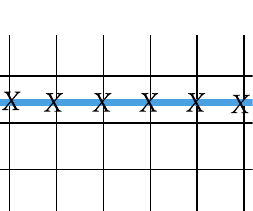}
    \caption{A horizontal $m$-string in the $D(\bZ_2)$ sandwich.}
    \label{fig: m-string}
\end{figure}

We now consider the $\bZ_2^F$-\SymTO{} setup. Namely, we consider the lattice to be extensive in the horizontal direction but finite in the vertical direction. The top boundary is the $f$-condensed boundary as we constructed above. The bottom boundary does not need to be specified for now. Recall that in the construction of \SymTO{} in Section~\ref{sec:Z2FSymTO} we used  horizontal $m$-strings to represent the fermion parity operator. We now verify this in the lattice model. Consider the horizontal $m$-string as shown in Fig~\ref{fig: m-string}. Using the bulk vertex terms, we may move the $m$-string up towards the $f$-boundary. Therefore on the ground space of the system, the bulk $m$-string is equivalent to the boundary $m$-string
\begin{align}
  S^m_{\tilde{l}}\cong  \prod_{e_i\in E_\partial} X_{i}.
\end{align}
Now use the boundary term $T_i=X_i\bc_i c_i$, we see that on the ground space the $m$-string is equivalent to 
\begin{align}
   S^m_{\tilde{l}}\cong \prod_{i}\bc_{i}c_{i},
\end{align}
which is indeed the fermion parity operator. 

Notice that we can  re-define the local Hilbert spaces of the boundary fermions to be formed by the pairs $c_i,\bc_{i+1}$, instead of The pairs $c_i,\bc_i$. If we make this choice,  the horizontal $e$-strings will become the  fermion parity operator, instead of $m$-strings. This confirms that the choice of representative of fermion parity in the $\bZ_2^F$-\SymTO{} is ambiguous, and this ambiguity exactly corresponds to the ambiguity of defining local Hilbert spaces of a closed \oned{} fermion chain.

\begin{figure}
    \centering
    \includegraphics{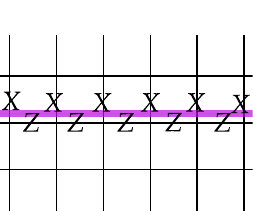}
    \caption{A horizontal $f$-string in the $D(\bZ_2)$ sandwich.}
    \label{fig: f-string}
\end{figure}

Next we confirm the relation between the expectation value of $f$-string on the $f$-boundary and spin structures of the local fermions in the lattice model. Consider the case where all $T_i$ terms in the boundary Hamiltonian  have the same sign(-1 for now), corresponding to spatially periodic boundary condition for the fermions. Now let us try to absorb an $f$-string onto the $f$-condensed boundary. Let $S^f$ be a horizontal $f$-string, as shown in Fig.~\ref{fig: f-string}. By using the bulk vertex and plaquette terms, we may move the $f$-string up towards the $f$-boundary. Then this $f$-string is equivalent to the boundary $f$-string
\begin{align}
    S^f=\prod_i Z_{i+1/2}\prod_i X_{i}.
\end{align}

Now using the boundary termss $K_i,T_i$ we may rewrite $Z_{i+1/2}$ as $Z_i Z_{i+1}c_i \bc_{i+1}$, and rewrite $X_i$ as $\bc_i c_i$. We obtain
\begin{align}
    S^f\cong \prod_i c_i\bc_{i+1} \prod_i \bc_i c_i=-1.
\end{align}
This confirms that for periodic boundary conditions, the $f$ string on the $f$-condensed boundary has expectation value $-1$. On the other hand, the anti-periodic boundary condition can be achieved by flipping the sign of one(or any odd number) of the boundary $K_i$ terms. This leads to $S^f\cong 1$ on ground space. This matches with the result of the diagrammatic argument we presented in~\ref{sec:spin1}.

\bibliography{FSymTFT}

\end{document}